\pdfoutput=1
\documentclass[12pt]{article}
\usepackage{jheppub}
\usepackage{subfigure}
\def\dd{{\rm d}}
\def\df{\varphi}
\def\vl{f_2}
\def\sp{\;\;\;,\;\;\;}
\title{Thermalization in a Holographic Confining Gauge Theory}
\author[a]{Takaaki Ishii,}
\author[a,b]{Elias Kiritsis,}
\author[a]{and Christopher Rosen}
\affiliation[a]{\href{http://hep.physics.uoc.gr}{Crete Center for Theoretical Physics},
Department of Physics, University of Crete, 71003 Heraklion, Greece.}
\affiliation[b]{Univ Paris Diderot, Sorbonne Paris Cit\'e, \href{http://www.apc.univ-paris7.fr/APC_CS/}{APC},
UMR 7164 CNRS, F-75205 Paris, France.}
\emailAdd{ishii@physics.uoc.gr}
\emailAdd{hep.physics.uoc.gr/$\sim$kiritsis}
\emailAdd{rosen@physics.uoc.gr}
\abstract{Time dependent perturbations of states in the holographic dual of a 3+1 dimensional confining theory are considered. The perturbations are induced by varying the coupling to the theory's most relevant operator. The dual gravitational theory belongs to a class of Einstein-dilaton theories which exhibit a mass gap at zero temperature and a first order deconfining phase transition at finite temperature. The perturbation is realized in various thermal bulk solutions by specifying time dependent boundary conditions on the scalar, and we solve the fully backreacted Einstein-dilaton equations of motion subject to these boundary conditions.  We compute the characteristic time scale of many thermalization processes, noting that in every case we examine, this time scale is determined by the imaginary part of the lowest lying quasi-normal mode of the final state black brane. We quantify the dependence of this final state on parameters of the quench, and construct a dynamical phase diagram. Further support for a universal scaling regime in the abrupt quench limit is provided.
}
\preprint{CCTP-2015-08 \\ \hspace*{\fill} CCQCN-2015-67}

\begin{document}
\maketitle
\section{Overview}
\subsection{Holographic Thermalization}
To date,  holographic theories of strongly interacting matter have been deformed, probed,  perturbed and otherwise studied in an enormous number of theoretical experiments. The results of these investigations have provided many important lessons about the nature of strongly coupled field theories, even in applications that fall outside the well established examples of AdS/CFT duality. While many earlier studies of holographic matter focused on the system's linear response to the insertion of various operators, quite a bit of attention has recently turned towards understanding the full non-linear dynamics of these holographic field theories.

Broadly, the aim of these applications is to uncover results about the processes by which strongly coupled gauge theories respond to arbitrary time dependent perturbations. Often, as in the case of the ground-breaking early examples of numerical holography \cite{Chesler:2008hg, Chesler:2009cy,Heller:2011ju,Bizon:2011gg}, the gauge theory responds by ``thermalizing" the perturbation. This means that after some characteristic time scale, typically set by the Hawking temperature of a black hole in the dual gravity solution, the field theory arrives at a final static state in thermodynamic equilibrium. In this equilibrium state, all correlation functions assume their thermal values, which is to say the trace is taken with respect to the thermal probability distribution appropriate to the ensemble. In other examples \cite{Maliborski:2013jca,Buchel:2013uba,Craps:2014eba}, perturbations of the gauge theory  have been found which apparently never thermalize. These ``islands of stability" correspond to initial gravitational data that never leads to gravitational collapse and horizon formation.

A conceptually simple means of perturbing a gauge theory is to turn on a time dependent source for some operator and evaluate the system's response. Beyond the linear regime, such a source can be used to continuously drive the system, or to ``quench" it. By the latter, one typically means that a coupling in the gauge theory is varied over a compact timescale $\tilde{\tau}$, where $\tilde{\tau}$ is sometimes chosen to be small compared to other scales in the theory.

When quenched in this way, it is clear that the system's response will depend on the properties of the operator to which the source couples as well as the parameters which specify the quench. In the weak field approach pioneered in \cite{Bhattacharyya:2009uu}, perturbative results have been obtained for the quench of a marginal operator in the case where the perturbation's amplitude and characteristic timescale are both sufficiently small. This analysis allowed the authors to study the formation of black holes and black branes in asymptotically Anti-de Sitter spacetimes.

Specifically, they constructed the limiting behavior of the dynamical phase diagram for the outcome of massless scalar collapse in global and Poincare patch AdS. Notably, they found that arbitrarily small perturbations of the Poincare patch solution always resulted in the formation of a black brane, whereas finite volume effects in the global solution prevented horizon formation under certain conditions.  Interestingly, they also observe a crossover between final states with small and large black holes, as well as Choptuik behavior in the vicinity of the transition between final states with a horizon and those without. This scaling behavior strongly suggests the presence of a second order line separating these phases. More recently, the weak field approach has been extended to provide interesting insights into thermalization in non-isotropic quenches \cite{Balasubramanian:2013rva},  in finite density states of a gauge theory  \cite{Caceres:2014pda}, and in a simple model of strongly coupled matter with a mass gap  \cite{Craps:2013iaa}.

The holographic toolkit makes it conceptually straightforward to move beyond the weak field perturbative scheme as well as to generalize the quench to perturbations by relevant operators.  Along the first direction, calculating beyond the reach of perturbative methods clearly implies grappling with the full non-linearity of the Einstein equations. In general, this is an exercise in ``numerical holography", loosely defined as the set of all holographic calculations that require numerical techniques more involved than a call to Mathematica's {\tt NDSOLVE}. Instead, one typically appeals to any one of an assortment of numerical methods specifically tailored to the problem of gravitational in-fall.

A recent example of holographic thermalization which connects perturbative weak field calculations to the full numerical evolution of a marginal perturbation appears in \cite{Craps:2014eba,Craps:2013iaa}. In this setup, the authors consider the response of a confining gauge theory to time dependent perturbations by studying the dynamical evolution of a massless scalar quench in the AdS hardwall geometry. This gravitational theory is characterized by a length scale $z_0 = 1/\Lambda$ at which the geometry is artificially terminated. In turn, this length scale gives rise to a mass gap $\sim \Lambda$ in the dual gauge theory, as well as a temperature above which a large black brane can appear in the bulk. Accordingly, the AdS hardwall serves as a crude model for a confining gauge theory with a deconfined phase dual to the large black brane solution.

Weak field calculations in the hardwall background indicated that the dynamical phase diagram ought to include a transition between perturbations which result in black brane formation and those that remain in a horizon-less scattering state. This should be contrasted with the situation in \cite{Bhattacharyya:2009uu} in which such a distinction between final states was only possible when the dual field theory is defined on a sphere. Unfortunately, the weak field approach was unable to answer questions about the late time fate of the scattering solutions, which is of particular importance for the field theory interpretation. To resolve the late time behavior of the scattering solutions, a numerical method was introduced to solve the Einstein equations and evolve the system arbitrarily far forward in time.

A particularly noteworthy result of the numerical investigation was the observation that there exist perturbations such that the subsequent scattering solutions never undergo gravitational collapse. Since no horizon is formed in such a process, the implication is that the holographically dual strongly coupled matter does not achieve thermodynamic equilibrium at late times. This finding differs in important ways from the islands of stability identified in scalar collapse in global AdS. In particular, it demonstrates that a perturbation initiated by explicitly sourcing a marginal operator need not {\it necessarily} thermalize in the infinite volume gauge theory. To rephrase this result in a way that will be more directly applicable to our present work, the introduction of a mass gap in the dual gauge theory can strongly effect the thermalization time for a certain class of perturbations.

A related, but logically distinct line of research involves the deformation of strongly coupled matter by time dependent perturbations of a relevant scalar operator. Numerical investigations into this situation appear in both systems with a finite density of charge carriers (for example \cite{Bhaseen:2012gg}) and without \cite{Buchel:2012gw,Buchel:2013lla,Buchel:2014gta}. We will focus on the latter scenario, in which an uncharged black brane solution is perturbed by varying the non-normalizable boundary mode of a massive bulk scalar in time. Holographically, the dual picture is that an initial state of the gauge theory in thermodynamic equilibrium is deformed by turning on a relevant operator over some timescale $\tilde{\tau}$. The strongly coupled matter generically responds by passing through a non-linear regime before settling once more into thermodynamic equilibrium, albeit in a different thermodynamic macrostate.

One of the most interesting results to emerge from the numerics of  \cite{Buchel:2012gw,Buchel:2013lla,Buchel:2014gta} has been the appearance of a ``universal fast quench regime" in which the change in energy density $\mathcal{E}$ after the quench scaled as a power law in the quench width $\tilde{\tau}$,
\begin{equation}\label{eq:uniFQ}
\mathcal{E}_\textrm{FINAL}-\mathcal{E}_\textrm{INIT}\sim \frac{\tilde{\delta}^2}{\tilde{\tau}^{2\Delta-d}}.
\end{equation}
This scaling is argued to manifest when the quench width is small compared to any other scales in the theory, such as the characteristic amplitude of the perturbation $\tilde{\delta}$. Moreover, the scaling was shown to be robust for many choices of the relevant operator's conformal dimension $\Delta$, and its appearance was further elucidated analytically in a recent series of papers \cite{Buchel:2013gba,Das:2014jna,Das:2014hqa}.

Given the non-trivial dynamical phase structure and strong dependence of the thermalization time on perturbations in the hardwall model, as well as the appearance of universal regimes in the space of quench parameters, it is natural to wonder how these properties might manifest in a holographic dual to a gauge theory that is more similar to $SU(3)$ Yang-Mills. In a sense we make more precise in the following subsection, this means we are interested in finding a natural way to ``soften" the hardwall while retaining the mass gap $\sim \Lambda$ that was responsible for the interesting dynamical features of the quench.  The construction and subsequent perturbation of such a holographic model will be the focus of this work.

\subsection{Confining Gauge Theories and Holography}

Holographic models of confining gauge theories generically involve a gravitational theory whose solutions include those which explicitly break scale invariance in the bulk. This condition is necessary but not  sufficient to allow for  non-trivial temperature dependence of states in the dual gauge theory, as well as a mass gap at zero temperature. These models also include solutions whose metric breaks bulk conformal invariance, the most relevant of which will be branches of black hole solutions which are holographically dual to thermal states with non-zero stress-energy.  The deconfinement transition will be a thermodynamic phase transition from solutions without a black hole, to solutions with one.

In what follows, we will focus on a particular class of holographic models for confining gauge theories which are in the Einstein-Dilaton family. As this title implies, our model will involve a non-trivial scalar whose profile explicitly breaks conformal invariance in the radial direction, and can be tuned to produce gravitational solutions dual to states on either side of a deconfinement transition. Within this rather simple class of models there exists a remarkable wealth of phenomenological possibilities, which include sharp deconfining transitions of first or second order, crossover behavior between phases, and theories driven from their UV fixed point by relevant operators of arbitrary dimension.

When the gravitational theory is dimensionally reduced to five dimensions, all such bulk solutions as described above (with the exception of flows to a non-trivial IR CFT  which are not relevant for holographic models of Yang-Mills) will have a naked singularity in the IR \cite{Gursoy:2007cb,Gursoy:2007er}. The presence of this IR singularity is a signal that something is missing in the effective holographic description. In known examples, typically the missing ingredients are an assortment of other bulk fields that will necessarily obtain vevs in the ground state solution. If such fields are reintroduced, the singularity is expected to be resolved.
In \cite{Gubser:2000nd}, Gubser provides a simple criterion  for a resolvable singularity which states that for a  singularity to be resolvable it should be possible to construct a solution with an infinitesimal regular horizon surrounding the singularity. Such singularities are called ``good" or resolvable singularities.

Even in such ``good" solutions, holographic calculations of correlators may be ill-defined in the presence of the naked IR singularity. This happens when in the neighborhood of the singularity the Sturm-Liouville problem that governs the bulk fluctuation has two normalizable solutions. In such a case an extra boundary condition is needed to determine the solution, and this is the signal that the correct solution is sensitive to the details of the singularity's resolution. Such cases are called ``holographically unreliable" because without knowing the details of the singularity's resolution holographic computations are inherently ambiguous. There are, however,  numerous cases where the Sturm-Liouville problem near the naked singularity has only a single normalizable solution. In that case the solution is unique and {\em does not} depend on the resolution of the singularity. When this occurs, the singularity is said to be repulsive, \cite{Gursoy:2007cb,Gursoy:2007er} and boundary theory correlation functions can be reliably calculated from holography in such backgrounds.

In confining states of a gauge theory, the expectation value of Wilson loops exhibit area law scaling. Holographically, the Wilson loop has a natural definition as a minimal surface in the bulk affixed  to the edges of a Wilson loop on the boundary. In \cite{Kinar:1998vq}, it was shown that holographic Wilson loops will give area law behavior providing that the string frame metric scale factor has a minimum, and that the scale factor at this minimum is non-zero. In Einstein-Dilaton theories, this requirement can be reformulated as a constraint on the IR behavior of the dilaton potential, as we will discuss in more detail below.

In our model, the zero temperature, zero entropy\footnote{Strictly speaking the entropy is $O(1)$ and is therefore due to one loop effects in the bulk.} ground state of the gauge theory will be dual to a bulk solution with a singularity in the far IR, but no event horizon. This singularity allows the bulk scalar to diverge in the IR, as well as the spatial metric scale factor to vanish.\footnote{It has been observed in \cite{Gursoy:2007cb,Gursoy:2007er} that in Einstein-Dilaton theories which have genuine breaking of scale invariance, a discrete spectrum and a mass gap, the string frame metric becomes flat in the IR. Therefore, in the string frame the bulk ``singularity" is due to a diverging dilaton.}  The latter is the feature responsible for vanishing entropy density, and is a familiar aspect of many candidate holographic ground states. As discussed above, the IR singularity can be thought of as a rather innocuous consequence of working in a limit of non-critical  string theory, providing the singularity satisfies several criteria. Our singularity is of this ``good" type.

As the temperature increases in the gauge theory, no new solutions are immediately available in the dual gravitational theory. Instead,  the dual bulk solutions are given by the zero-temperature (confining) solution with its time coordinate compactified into the thermal circle. The temperature $T$ associated to this solution is governed by the size of the thermal circle in the familiar way,
\begin{equation}
t \to i\tau \qquad \textrm{where} \qquad \tau = \tau + \beta\qquad \textrm{with} \qquad \beta = \frac{1}{T},
\end{equation}
and such a state is often called the ``thermal gas" saddle point. As it lacks a black hole horizon, it is obvious that the thermal gas is characterized by thermodynamics subleading in the large $N_c$ (where $N_c\gg1$ is roughly the rank of the dual  gauge theory) expansion inherent to our holographic setup.

Increasing the temperature further, one eventually arrives at a special temperature $T=T_0$ which marks the appearance of a new branch of solutions in the bulk theory. These solutions are of the black brane type, with a non-compact horizon of planar topology.
As seen in the left plot of figure \ref{fig:TnSd3} for any $T>T_0$, there are two black hole solutions. One is in the left branch of the curve while the other is in the right branch of the curve. The black holes on the left branch are called large black holes as their horizon size is larger than any of those in the right-branch. They have positive specific heat and are therefore locally thermodynamically stable. On the contrary, the small black hole branch on the right has negative specific heat and the associated black holes are thermodynamically unstable. Qualitatively the diagram is similar to that of global AdS space, although here the space is flat and the volume infinite. In the case of global AdS, small black holes asymptote to Schwarzschild black holes in flat space in the limit of small horizon size (with $T\to\infty$). Here also, small black holes in the limit of small horizon size have $T\to \infty$ but they are very different from Schwarzschild black holes, \cite{Gursoy:2007er,Kiritsis:2011qv,Kiritsis:2011yn}. Finally, the unique black hole that is at $T=T_0$ is very special, as here the specific heat diverges.

The thermodynamics of such black holes is governed by the standard black-brane relations, which implies they have an entropy density proportional to $L^3/\kappa^2$ where $L$ is the characteristic length scale of the metric (for example the AdS radius) and $\kappa$ is the five-dimensional gravitational constant. A standard application of the holographic dictionary for known five-dimensional gravitational duals gives $L^3/\kappa^2\propto N_c^2$. The proportionality  constant depends on the details of the duality under consideration. In holographic applications that do not descend from a known string theory solution, the proportionality constant is an undetermined parameter of the theory, but the $N_c^2$ scaling of various thermodynamic quantities is expected to be robust.

As mentioned above, by studying the susceptibilities of the small black hole branch, for example the specific heat, it is straightforward to show that these black brane solutions are thermodynamically unstable. This is a local statement, independent of the global thermodynamic (in)stability of these solutions.  In light of the Correlated Stability Conjecture (CSC) \cite{Gubser:2000mm}, one might worry that these solutions are also dynamically unstable, which is to say their fluctuation spectrum may contain a mode in the upper-half complex frequency plane. Such an instability obviously leads to an exponential growth in time, and the conjecture posits that the existence of such a mode indicates a Gregory-Laflamme type instability towards a ``lumpy" horizon. This dynamical instability will play no role in the present work as we will study s-wave perturbations of spatially homogeneous solutions.

\begin{figure}
\centering
\includegraphics[height=6cm]{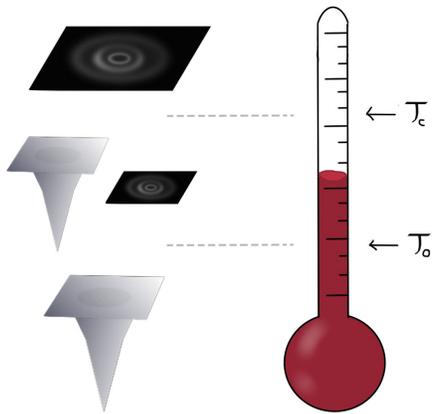}
\caption{The phase diagram of our model. The thermal gas solution (grey) is the only solution below $T=T_0$. Above $T_0$ the black hole branches appear (black plane), but the thermal gas is thermodynamically preferred. At $T=T_c$ there is a first order deconfinement transition from the thermal gas to large black brane solutions (large black plane). Cartoon from \cite{DeWolfe:2013cua}.\label{fig:pd}}
\end{figure}

Above $T_0$, the thermal gas still provides the thermodynamically-dominant solution, as it has lower free energy than the black brane solutions at the same temperature. This remains true until $T=T_c$, at which point the model predicts a first order phase transition from the thermal gas to the black-brane solutions. Above $T_c$, the system is described by the large black brane solutions that are thermodynamically stable.
This first order phase transition is a holographic realization of a deconfining transition to states with an entropy density that scales like $N_c^2$ and perimeter law behavior of the Wilson loop. The full phase structure of our model is illustrated in figure \ref{fig:pd}.

\subsection{An Einstein-Dilaton Model of Thermalization in a Confining Theory}

Our goal in this work is to continue the investigation of the effects of confinement on thermalization processes in holographic gauge theories. We accomplish this by perturbing a state of a confining gauge theory with a relevant operator that we loosely associate with\footnote{In realistic bottom up holographic models of QCD, the scalar is dual to a marginally relevant operator \cite{Gursoy:2007cb,Gursoy:2007er}. However,  we do not expect qualitatively important changes in the IR if the UV dimension of that operator is three (as in the case studied here). The reason is that in both cases, in Einstein dilaton gravity the scalar becomes strongly relevant in the IR.}     $\textrm{Tr}F^2$, and then studying the system's response. From the bulk perspective, we are choosing a solution of our model and deforming it briefly through time-dependent boundary conditions on the dilaton.  The deformation will generically throw the system out of equilibrium, which will then undergo non-linear evolution governed by the Einstein-Dilaton equations.

An important question is whether or not an arbitrary perturbation of a confined state in the strongly interacting gauge theory will thermalize. In this context, ``to thermalize" means that the late time behavior of the state is in thermodynamic equilibrium, and all correlation functions are time independent and assume their thermal values. The gravitational formulation of this question is whether or not an arbitrary perturbation of the horizon-less thermal gas solutions of our model necessarily results in black brane formation. Put another way, is the exponentially diverging dilaton potential of our model sufficiently similar to the hard wall of \cite{Craps:2013iaa,Craps:2014eba} to encourage scattering solutions and prohibit horizon formation under certain conditions?

At present, the answers to these questions remain beyond our grasp. In the absence of a horizon in the initial state, the diverging scalar in the IR leads to a number of challenging numerical obstacles. To address these issues, one should regulate the bulk solutions in the IR through the introduction of a numerical cut-off, and study the effects of the position of this cut-off on the computation's results.  A related approach, which we will adopt, forfeits the ability to comment on black hole formation while hopefully retaining qualitative effects of the confining dilaton potential. This scheme replaces the hard numerical cut-off with a small black brane horizon, which then acts as a natural IR regulator.

In our model, this necessarily implies that our initial bulk solution will not be holographically dual to the ground state of the gauge theory. At first glance it actually looks much worse, since these small black brane solutions aren't the thermodynamically preferred state at sufficiently low temperature,  nor are they necessarily dynamically stable.  Nonetheless, we argue that they provide a sensible starting point for our calculation, as in the limit of vanishing horizon size these black holes coincide with the ground state of the system.

The small black-brane solutions asymptotically approach the zero-temperature solution in the limit where the horizon recedes infinitely far from the boundary into the IR. Thus, provided we accept the concession that a horizon has already been formed in the bulk, for sufficiently large separation between the UV boundary  and the position of the horizon the gravitational solution shares many important features with the ground state solution. The most important of these are a scalar profile which is rapidly growing towards the IR, and a rapidly vanishing warp factor in the Einstein frame metric. We expect that these features will be responsible for most of the interesting dynamics in our system, such as the timescale with which thermalization takes place, and accordingly that the small black holes can provide us with a qualitative understanding of the consequences of these features.

Moreover, the thermodynamic instability of these black branes and the possible presence of a dynamical instability in their fluctuation spectrum posited by the CSC is not expected to have much relevance to our discussion. Gregory-Laflamme type instabilities are related to a ``clumping" of mass or charge at the horizon, and should thus involve finite wave-number perturbations in the directions transverse to the brane. We study translationally invariant perturbations in the non-critical (five dimensional) theory, and hence do not anticipate the possibility of exciting such instabilities. Indeed, we have not yet seen any direct evidence for a dynamical instability in the small black hole branch of our model.

The main results of our study are a portion of the dynamical phase diagram, and the scaling properties of various equilibration processes in our model. The former is realized as a map between perturbation parameters and the final state achieved after equilibration. Our perturbation will be controlled by two parameters corresponding to the amplitude of the perturbation and its duration. Thus, the dynamical phase diagram is a two dimensional plot with a curve marking the boundary between perturbations that result in a small black hole and those that thermalize into large black holes.

By focusing on the details of the final state, we are also able to make claims about the dependence of the characteristic thermalization time in the boundary gauge theory on the quench parameters. Not surprisingly, this timescale turns out to be quantified by the quasi-normal mode spectrum of the final state solution. Moreover, by quantifying the relationship between the features of the quench and the change in energy density induced by the perturbation, we demonstrate explicitly that the simple scaling anticipated in (\ref{eq:uniFQ}) appears in our model as well.

The remainder of this work is devoted to detailing the setup, implementation, and interpretation of time dependent quenches of a holographic confining gauge theory. In section \ref{sec:Model} we introduce the specifics of the bulk theory we are interested in perturbing, as well as the equations of motion that govern its behavior. The behavior of these solutions near the conformal boundary are of particular importance for the application of holographic methods, and thus we provide these near boundary solutions in some detail.

The numerical construction and thermodynamic properties of the static, equilibrium solutions of our model are discussed in section \ref{sec:InitState}. This section primarily serves as an orientation to our holographic gauge theory. In it one finds the dependence of the free energy and entropy density of our strongly coupled matter on temperature, as well as a comparison of the temperature dependence of the speed of sound in our model to that of $SU(3)$ Yang-Mills. It is complemented by the results of section \ref{sec:onepts} which further characterize the static states in terms of their thermodynamic one-point functions.

Section \ref{sec:strategy} details the computational approach we adopt to evolve our gravitational system in time. This approach is based on an assortment of well known numerical techniques that we carefully tune to accommodate the specifics of our model. Most notably, we explain how we handle the copious logarithmic fall-offs in the near boundary behavior of bulk fields. These fall offs are generic to gravitational theories in odd bulk dimensions, and present obstacles to the accurate determination of the normalizable and non-normalizable UV coefficients of the various bulk fields. This section will be primarily interesting to those who would like to numerically study dynamical quenches in related models.

The output from our numerical method and the interpretation of this output is contained in sections \ref{sec:EX} and \ref{phase}. These sections constitute the primary results of our study. They include examples of the response our model to various classes of quench by a relevant scalar operator, as well as a dynamical phase diagram for the outcome of these quenches when performed in a particular initial state. We examine the dependence of the final equilibrium state on the quench parameters, and comment on the appearance of an anticipated universal scaling regime in the fast quench limit. The connection and applicability of these results to similar processes in other theories is discussed in section \ref{sec:diss}. Specifically, we comment on the implications of our calculations for the thermalization of probes in the strongly coupled matter produced in heavy ion collisions, as well as future directions one might wish to pursue.

\section{The Model}\label{sec:Model}
\subsection{The Action}
We will be interested in Einstein-Dilaton theories that are tuned to qualitatively reproduce some important properties of QCD. These theories can be described by an action of the form
\begin{equation}\label{eq:action}
S = \frac{1}{2\kappa^2}\int\dd^5x\sqrt{-g}\left(R-\frac{4}{3}(\partial\df)^2+V(\df)\right) -\frac{1}{\kappa^2}\int_\partial \dd^4x\sqrt{-\gamma}\,\mathcal{K}
\end{equation}
where $V(\df)$ is the dilaton potential which will govern the dynamics of the system, and $\mathcal{K}$ is the Gibbons-Hawking-York term necessary to help render the variational problem well defined on the boundary. The model we study can be classified as a ``bottom-up" description of a confining gauge theory, in the sense that $V(\df)$ is not derived from a known supergravity theory. Instead, this potential is manufactured to satisfy certain criteria in the IR and/or UV limits of the gravitational theory so as to induce desirable features in the dual boundary theory. For example, these asymptotics determine whether or not there is a mass gap, while other asymptotics determine the scaling dimension of the dual scalar operator. These criteria have been elucidated and categorized in a series of papers beginning with \cite{Gursoy:2007cb,Gursoy:2007er}. Importantly, the asymptotic behaviors of the scalar potential that we choose for the current work {\it are} present in known solutions of gauged supergravity. An example of this which is closely related to the model investigated in this work is described in \cite{Girardello:1999hj}. As many qualitative features of holographic models are dominated by these asymptotic properties, it is reasonable to expect that predictions from our model may apply to a broad class of holographic systems, including those with a more esteemed string theory lineage.

While there are a variety of scalar potentials whose distinct IR asymptotics lead to confining theories, it was found in \cite{Gursoy:2007cb,Gursoy:2007er} that a particular IR behavior also  produces linear radial trajectories for glueballs, as well as the structure of the Yang-Mills phase diagram just above the first order deconfinement transition. This behavior requires

\begin{equation}
V_{\mathrm{IR}}\sim e^{\frac{4}{3}\df}\sqrt{\df}
\end{equation}
which, apart from the subleading $\sqrt{\df}$,  is the non-critical string theory dilaton potential in five dimensions in the Einstein frame.

At high energies, the UV properties of the gauge theory depend on the nature of the scalar operator which perturbs it. In \cite{Gursoy:2007cb,Gursoy:2007er} a marginal operator was used, and the potential chosen such that the UV fixed point was at $\df\to -\infty$. The potential is of the form
\begin{equation}
V_{\mathrm{UV}}=\sum_{n=0}^{\infty} V_{(n)}~\lambda^{n}\qquad\textrm{with}\qquad\lambda=e^{\df},
\label{1}\end{equation}
so that $\lambda\to 0$ where $\lambda$ in the UV is identified with the 't Hooft coupling constant. This fixed point is extremely shallow as all finite derivatives of the potential vanish  at the fixed point.

This potential is tuned to match the thermodynamics of $SU(3)$ Yang-Mills, and the resulting theory, dubbed Improved Holographic QCD (IHQCD), is capable of quantitatively reproducing many features of QCD across all energy scales. Clearly it would be desirable to study dynamical processes in IHQCD, beyond the linearized level. However, the presence of a marginally relevant operator in the UV creates serious computational difficulties. One consequence of the marginal deformation inherent to IHQCD is that the scalar $\lambda$ vanishes logarithmically near the UV boundary. This mild falloff is very difficult to handle numerically, as it requires retaining a very high level of numerical precision to correctly identify the leading and subleading coefficients governing the near boundary behavior of the solution. To circumvent this issue, we will sacrifice the quantitative match to Yang-Mills theory provided by IHQCD in favor of computational convenience.

One can do this without departing drastically from the coarse features of Yang-Mills by following the approach advocated in \cite{Gubser:2008yx}. In this approach, the marginal scalar operator dual to $\lambda$ is traded for a relevant operator with dimension not far from 4, which one hopes to roughly identify with a boundary operator of the form $\textrm{Tr}F^2$ where $F$ is the Yang-Mills field strength. The fact that this operator is no longer marginal is meant to capture the anomalous dimension that the operator acquires after running some ways towards the IR, reminiscent of what happens in Yang-Mills.

We will therefore assume that there is a regular UV fixed point at $\df=0$, without loss of generality,  and the potential near this fixed point takes the standard form
\begin{equation}
V_{\mathrm{UV}}\sim V^{(0)}+\frac{1}{2}V^{(2)}\df^2+\ldots\sp \df\to 0.
\end{equation}
The various coefficients $V^{(i)}$ are fixed by the symmetries of the gauge theory and were shown in \cite{Gursoy:2007cb,Gursoy:2007er,Bourdier:2013axa} to be in one to one correspondence to the coefficients of the holographic $\beta$-function.
The AdS scale $L$ of the UV AdS space is determined by  $L^2V^{(0)}=12$. The relationship between the mass of the scalar and the conformal dimension of the dual gauge theory operator $\Delta$ can be made precise. In the standard quantization $\Delta$ is defined to be the larger of the two roots of
\begin{equation}
\label{eq:confdim}
m^2 L^2\equiv -\frac{3}{8}L^2\,V^{(2)}= \Delta(\Delta-4).
\end{equation}

An example of a potential for which these properties are realized first appeared in \cite{Gubser:2008yx}:
\begin{equation}\label{eq:Vd3}
V(\df) = \frac{12\left(1+a\df^2 \right)^{1/4}\cosh\frac{4}{3}\df-b\df^2}{L^2}
\end{equation}
and  a parameter set which produces a bulk theory dual to a confining gauge theory deformed by a dimension $\Delta=3$ operator is $(a,b)=(1/500,10009/1500)$. This theory is decidedly not real world YM. Unlike the more refined IHQCD models of  \cite{Gursoy:2010fj}, it fails to produce a very good  match to the thermodynamics of real world $SU(3)$ Yang-Mills (see section \ref{sec:thermo}).  Nevertheless, this holographic theory does exhibit a mass gap, the thermodynamic properties determined by its potential are qualitatively reminiscent  of $SU(3)$ glue, and it renders the bulk system relatively amenable to numerical investigation. Accordingly, in what follows we will primarily be concerned with solutions to the equations of motion derived from (\ref{eq:action}) with the potential written in (\ref{eq:Vd3}).

\subsection{Solutions}
Among the solutions to this model's equations of motion are the planar geometries (with and without a horizon) of the form
\begin{equation}\label{eq:gans}
\dd s^2=-A\,\dd v^2-\frac{2}{z^2}\dd v\dd z+\Sigma^2 \dd\vec{x}^2\qquad\mathrm{and}\qquad \df = \df(v,z)
\end{equation}
where $v$ is an ingoing null-coordinate, $z$ is the radial direction, and $\vec{x}$ describe the planar $\mathbb{R}^3$. The various metric functions appearing in this ansatz, as well as the scalar, are taken to be functions of both $v$ and $z$.
The solutions satisfy the equations of motion
\begin{align}\label{eq:eom1}
0 =& -\frac{1}{z^2}\frac{3}{8}\partial_\df V+2(\dd_+\df)' + 3\,\dd_+\df\,\frac{\Sigma'}{\Sigma}+3\,\dd_+\Sigma\,\frac{\df'}{\Sigma},\\\label{eq:eom2}
0=& \frac{1}{6z^2}V+\frac{(\dd_+\Sigma)'}{\Sigma}+2\,\dd_+\Sigma\,\frac{\Sigma'}{\Sigma^2},\\\label{eq:eom3}
0=& \Sigma''+\frac{2}{z}\Sigma'+\frac{4}{9}\Sigma\,\df'^2,\\\label{eq:eom4}
0=& \frac{1}{3z^4}V+\frac{12}{z^2}\,\dd_+\Sigma\,\frac{\Sigma'}{\Sigma^2}-\frac{1}{z^2}\frac{8}{3}\df'\,\dd_+\df+ \frac{2}{z}A'+A'',\\\label{eq:eom5}
0=& \dd_+^2\Sigma+\frac{4}{9}(\dd_+\df)^2\,\Sigma+\frac{1}{2}z^2\,\dd_+\Sigma\,A',
\end{align}
where the prime ($'$) denotes differentiation with respect to $z$, and $\dd_+$ is a modified derivative defined by
\begin{equation}\label{eq:ddp}
\dd_+\equiv \partial_v -\frac{z^2}{2}A\,\partial_z,
\end{equation}
which is the derivative along the outgoing null vector.

\subsubsection{Boundary Analysis}\label{sec:BA}
Near the UV boundary, the various metric functions and a bulk scalar with $m^2=-3/L^2$ can be expanded like
\begin{align}
A(v,z) &= \sum_{n=0}\left[a_n(v)+ \right.\sum_{m=1}\left.\alpha_{nm}(v)\log^m z\right] z^{n-2},\\
\Sigma(v,z) &=  \sum_{n=0}\left[s_n(v)+\right.\sum_{m=1}\left.\sigma_{nm}(v)\log^m z\right] z^{n-1},\\
\varphi(v,z) &= \sum_{n=0}\left[f_n(v)+\right.\sum_{m=1}\left.\phi_{nm}(v)\log^m z\right] z^{n+1},
\end{align}
and we adopt coordinates in which the asymptotically $AdS_5$ boundary has $a_0(v) = s_0(v)=1$ and $\alpha_{0m}(v)=\sigma_{0m}(v)=0$. In general, logarithmic terms are expected in even boundary theory dimensions, $d$, and when the scalar operator has dimension $\Delta$ such that $\Delta-d/2$ is an integer. The scalar's source, $f_0(v)$, remains a boundary condition to be implemented.

Inserting these expansions into the Einstein equations and expanding near the boundary at $z=0$ provides the  asymptotic behaviors of the various fields. In what follows, we leave the potential largely unspecified, requiring only that it gives rise to a dual conformal gauge theory which is deformed by a parity invariant relevant operator of dimension $\Delta=3$. This constrains a few terms in an expansion of the dilaton potential about the AdS fixed point like
\begin{equation}
V\Big|_{\df=0} = \frac{12}{L^2}, \qquad \partial^{2n+1}_{\df }V\Big|_{\df=0} = 0,\qquad \mathrm{and}\qquad \partial^2_{\df }V\Big|_{\df=0} = \frac{8}{L^2}
\end{equation}
for positive integers, $n$. Accordingly, the fields behave like this:
\begin{align}
A =& \left(\frac{1}{z}+\zeta \right)^2-2\dot{\zeta}-\frac{4}{9}f_0^2+a_4\, z^2-\alpha_4 \,z^2\log z+\ldots ,\label{eq:AUV}\\
\Sigma = &  \frac{1}{z}+\zeta-\frac{2}{9}f_0^2\,z+\frac{2}{27}f_0\left(3\zeta f_0-4\dot{f_0} \right)z^2+s_4\,z^3-\sigma_4 z^3\log z+\ldots,\label{eq:SUV}\\
\df = & f_0\,z +\left(\dot{f_0}-\zeta f_0 \right)z^2 + \vl\,z^3-\phi_2\,z^3\log z+\ldots\label{eq:FUV},
\end{align}
where
\begin{align}
\alpha_4 = &-\frac{4}{9}\dot{f_0}^2+\frac{4}{9}f_0\ddot{f_0}+f_0^4\left(\frac{16}{81}-\frac{1}{72}V^{(4)} \right),\\
s_4 = &-\frac{2}{9}f_0\vl-\frac{4}{27} \dot{f_0}^2+\frac{1}{108}f_0\left(16\zeta\dot{f_0} +3\ddot{f_0}\right)+f_0^4\left(\frac{8}{243}-\frac{1}{576}V^{(4)}\right),\\
\sigma_4 =&\frac{1}{9}f_0\ddot{f_0} +f_0^4\left( \frac{8}{81}-\frac{1}{144}V^{(4)}\right),\\
\phi_2 = &-\frac{1}{2}\ddot{f_0}+f_0^3\left(\frac{1}{32}V^{(4)}-\frac{4}{9}\right),
\end{align}
in which $V^{(4)} \equiv \partial^4_{\df }V|_{\df=0}$.  The functions $\zeta$, $a_4$, $f_0$ and $f_2$ all depend on time, and are not determined by the asymptotic series expansion. The function $\zeta$ is completely unfixed, and reflects residual reparametrization invariance in $z$. The coefficient $a_4$ is related to the background energy density, and $f_0$ and $f_2$ can be related to the source for and response of the dual scalar operator, respectively. The AdS radius $L$ has been (and will continue to be) set to one, and can be reinstated in any  formula by dimensional analysis. In order for the full set of Einstein equations to be consistently solved, the time derivative of $a_4$ is subject to a constraint,
\begin{align}\label{eq:Econs}
\dot{a_4} =&\frac{8}{9}\left(\vl\dot{f_0}-f_0\dot{\vl}\right)+\frac{16}{9}f_0\dot{\zeta}\left( f_0\zeta-\dot{f_0}\right)+\frac{4}{27}\left(4f_0\dddot{f_0}-\dot{f_0}\ddot{f_0}\right)\nonumber\\
&\,\,\,\qquad\qquad\qquad\qquad+ \frac{16}{9}\zeta\left(\dot{f_0}^2-f_0\ddot{f_0}\right)+\dot{f_0}f_0^3\left(\frac{160}{243}-\frac{V^{(4)}}{36} \right).
\end{align}
This constraint is related to the Ward identity governing the divergence of the stress tensor, which we revisit in section \ref{sec:onepts}.

From these expressions it is simple to work out the asymptotic behavior of the modified derivatives $\dd_+\Sigma$ and $\dd_+\df$. They are
\begin{align}
\dd_+\Sigma = & \frac{1}{2}\left(\frac{1}{z}+\zeta\right)^2-\frac{1}{9}f_0^2-\frac{4}{27}f_0\dot{f_0}\,z+\mathcal{C}_\Sigma \,z^2+\ldots,\label{eq:dpSUV}\\
\dd_+\df = &-\frac{1}{2}f_0 +\left(\frac{3}{2}\zeta^2f_0-3\zeta\dot{f_0}+\frac{3}{4}\ddot{f_0}+\frac{1}{64}f_0^3V^{(4)}-\frac{3}{2}\vl\right)z^2+\ldots,\label{eq:dpFUV}
\end{align}
where $C_\Sigma$ is the exhausting constant
\begin{equation}\label{eq:cterm}
\mathcal{C}_\Sigma = \frac{1}{2}a_4+\frac{1}{3}f_0\vl-\frac{1}{3}\zeta^2 f_0^2+\frac{2}{27}\dot{f_0}\left(11\zeta f_0-\dot{f_0} \right)-\frac{61}{216}f_0\ddot{f_0}-f_0^4\left( \frac{4}{81}+\frac{V^{(4)}}{1152}\right).
\end{equation}
For later convenience, we write these asymptotic expansions in the form
\begin{align}
\mathrm{d}_{+}\Sigma(v,z) &=  \sum_{n=0}\left[(Ds)_n(v)+\right.\sum_{m=1}\left.(D\sigma)_{nm}(v)\log^m z\right] z^{n-2},\\
\mathrm{d}_{+} \varphi(v,z) &= \sum_{n=0}\left[(Df)_n(v)+\right.\sum_{m=1}\left.(D\phi)_{nm}(v)\log^m z\right] z^{n}.
\end{align}

Solutions with non-zero $f_0$ correspond to adding to the boundary Lagrangian a term like
\begin{equation}
\delta\mathcal{L} = f_0\, \mathcal{O}, \qquad \mathrm{where}\qquad \left[f_0\right] = \Lambda^{4-\Delta}
\end{equation}
for mass scale $\Lambda$. For the model of interest (\ref{eq:Vd3}), we have $\Delta=3$ and thus the source for the relevant perturbation has dimensions of energy. Regularity of the solutions in the IR implies one constraint between the  UV coefficients $f_0, f_2$ and $a_4$,  and thus the UV deformations are specified by a single dimensionless quantity which we can take to be $\langle T_{tt}\rangle/f_0^4$. In other words,  we anticipate a one parameter family of black hole solutions characterized by their energy density in units of $f_0$.  Often times it will be more instructive to parametrize various solutions by the value their scalar obtains in the IR, a method introduced in the following section.

\section{The Initial State}\label{sec:InitState}
We eventually wish to thermalize the scalar perturbation in an initially static state of the gauge theory dual to the bulk theory described by (\ref{eq:action}) and  (\ref{eq:Vd3}), at finite temperature. These states are constructed by restricting the background functions in the ansatz to vary only radially, that is removing $v$-dependent terms in (\ref{eq:eom1}-\ref{eq:eom5}), and then integrating the resulting system of ordinary differential equations to obtain solutions for $A(z), \Sigma(z)$ and $\df(z)$. These bulk solutions then provide the starting point for the subsequent deformation and evolution.

To construct numerical solutions to the equations of motion (\ref{eq:eom1}-\ref{eq:eom5}) describing the initial state, we exploit the fact that we desire a regular horizon at $z=z_H$ and thus expect that the background functions can be expanded around the horizon like
\begin{equation}
F(z) = F_0+F_1(z_H-z)+F_2(z_H-z)^2+\ldots,
\end{equation}
where $F$ is any of  $\{A,\Sigma,\df\}$. The coefficients $F_i$  can be partially fixed by appealing to symmetries of the underlying gravity theory. For example, $A_0=0$ by assumption of a regular horizon, and $\Sigma_0 = 1$ can be enforced by rescalings of the spatial coordinates. As mentioned previously, the metric ansatz (\ref{eq:gans}) does not fully fix the gauge freedoms of the gravitational theory, and this is reflected in horizon data as the ability to choose $A_1$ arbitrarily. In practice, we have found it convenient to fix this residual reparametrization invariance by requiring that the boundary coefficient $\zeta$ of the previous section vanishes. The value of the scalar at the horizon, $\df_0\equiv \df_H$ thus parametrizes the available solutions in this model. All the higher order coefficients are fixed in terms of these first few coefficients, and accordingly one can use the expansion to sufficiently high order to provide IR data at a radial location near the horizon with which to seed a numerical routine.
\subsection{Thermodynamics}\label{sec:thermo}
The solutions to the equations of motion are asymptotically AdS$_5$ near the boundary at $z=0$, with a scalar that vanishes linearly. However, as our integration strategy relies on fixing the values of various parameters at the horizon, it will often happen that the numerical solutions obtained are characterized by different energy scales, $f_0$ and appear in different coordinates. More explicitly, static solutions near the boundary generically behave as
\begin{align}
\dd s^2 = & -\left(\frac{1}{z^2}+\ldots\right)\dd v^2 -\frac{2}{z^2}\dd v \dd z + \left(\frac{\Sigma_F^2}{z^2}+\ldots\right)\dd\vec{x}^2,\\
\df = & \tilde{f_0} z+\ldots
\end{align}
for constants $\Sigma_F$ and $\tilde{f_0}$. As the aim is to compare states of different temperature in the same theory, we should arrange that the solutions take a canonical form at the boundary, with identical metric and energy scale. This can be accomplished by performing a coordinate transformation to a new radial coordinate in which $\tilde{f_0} = 1$, as well as simple rescalings of the other coordinates. To wit, the transformation
\begin{align}\label{eq:canV}
\tilde{v} = &\tilde{f_0}\, v,\\
\vec{\tilde{x}}=& \tilde{f_0}\Sigma_F\,\vec{x},\\
\tilde{z} = & \tilde{f_0}\, z,\label{eq:canZ}
\end{align}
ensures a near boundary solution of the form
\begin{align}
\dd \tilde{s}^2 = & -\left(\frac{1}{\tilde{z}^2}+\ldots\right)\dd \tilde{v}^2 -\frac{2}{\tilde{z}^2}\dd \tilde{v} \dd \tilde{z} + \left(\frac{1}{\tilde{z}^2}+\ldots\right)\dd\vec{\tilde{x}}^2,\\
\tilde{\df} = &\, z+\ldots,
\end{align}
which is then suitable for comparing thermodynamic properties between solutions.
\begin{figure}
\centering
\includegraphics[height=5.2cm]{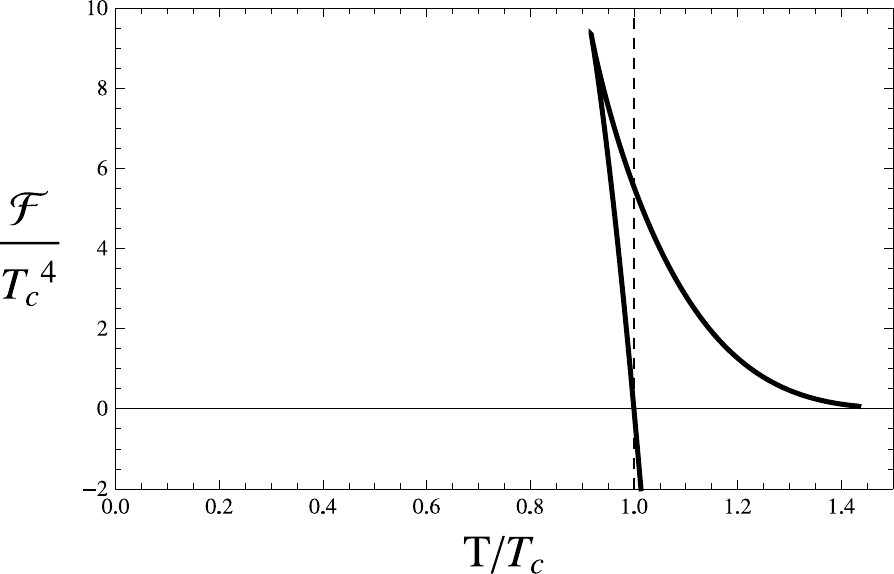}
\caption{The free energy density in units of the critical temperature $T_c$ as a function of $T/T_c$ for the model described by (\ref{eq:Vd3}).\label{fig:FreeFd3}}
\end{figure}
Following the standard prescriptions for black brane thermodynamics, temperatures and entropies can be readily extracted from the near horizon geometry. In the ``canonical" coordinates of (\ref{eq:canV}-\ref{eq:canZ}), the temperature is easily computed from the surface gravity $\hat{\kappa}$ at the horizon since
\begin{equation}
T = \frac{\hat{\kappa}}{2\pi}\qquad \textrm{where} \qquad\hat{\kappa}^2 = -\frac{1}{2}\nabla^\mu\xi^\nu\nabla_\mu\xi_\nu\Big|_{\tilde{z}_H}
\end{equation}
 and $\xi$ is a unit Killing vector, timelike outside the Killing horizon at $\tilde{z}=\tilde{z}_H$. Meanwhile, the entropy density is proportional to the area of the horizon\footnote{For the sake of simplicity, we will often refer to the ``area" of a horizon when we more precisely mean the ``area density". All of the black brane solutions we discuss have planar horizons and thus, strictly speaking, infinite horizon area.}, so that
\begin{equation}\label{eq:thermBB}
T = \frac{1}{4\pi}\tilde{z}_H^2\, \tilde{A}'\Big|_{\tilde{z}_H}\qquad \mathrm{and} \qquad s = \frac{2\pi}{\kappa^2}\sqrt{\tilde{\gamma}},
\end{equation}
where $s$ is the entropy density and $\tilde{\gamma}$ is the determinant of the spatial part of the metric at the horizon. For the solutions we consider, inserting the near horizon form of the canonical metric into (\ref{eq:thermBB}) gives
\begin{equation}
T = \frac{1}{4\pi \tilde{f_0}}z_H^2\,A_1 \qquad \mathrm{and}\qquad  s = \frac{2\pi}{\kappa^2 \tilde{f}_0^3\Sigma_F^3}.
\end{equation}

As different solutions are fully characterized by the value of their scalar at the horizon, it is often useful to think of their thermodynamic properties as functions of $\lambda_H = \exp\df_H$. In this spirit, once the temperature and entropy of a given state can be reliably computed, one can measure the free energy of the solution from the integrated first law:
\begin{equation}\label{eq:fzh}
\mathcal{F}(\lambda_H) = \int_{\lambda_H}^\infty\dd\bar{\lambda}_H\,s(\bar{\lambda}_H)\frac{\dd T(\bar{\lambda}_H)}{\dd\bar{\lambda}_H}.
\end{equation}
For the model considered here, the free energy density is plotted as a function of $T$ in figure \ref{fig:FreeFd3}.
\begin{figure}
\centering
\includegraphics[height=4.5cm]{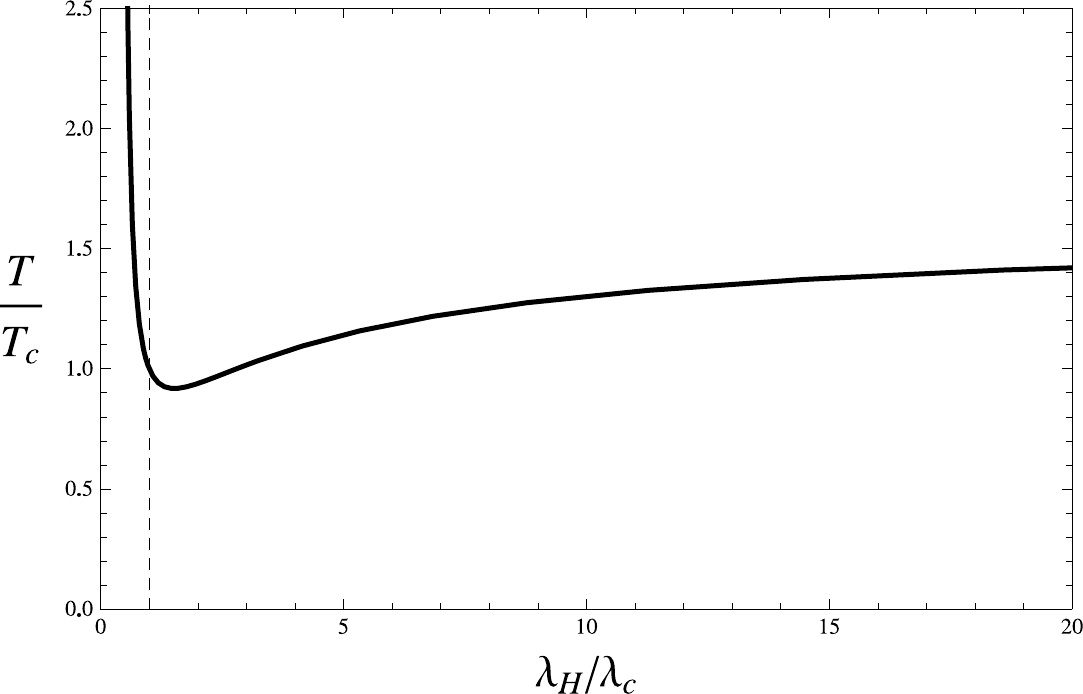}
\includegraphics[height=4.5cm]{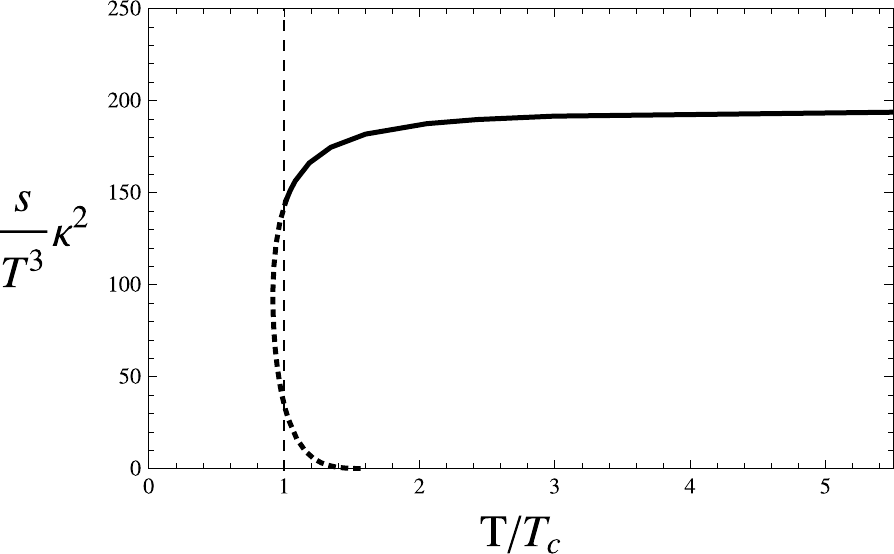}
\caption{Plots of the temperature scaled by the critical temperature as a function of $\lambda_H/\lambda_c$ (left) and the entropy density scaled by the third power of the temperature as a function of $T/T_c$ (right). The rightmost plot becomes ``dotted" as one passes through the phase transition by lowering the temperature from above. This is meant to indicate that in the field theory, this low temperature phase is governed by the thermal gas solutions, whose entropy is subleading in the number of colors $N_c$.\label{fig:TnSd3}}
\end{figure}
\begin{figure}
\centering
\includegraphics[height=4.5cm]{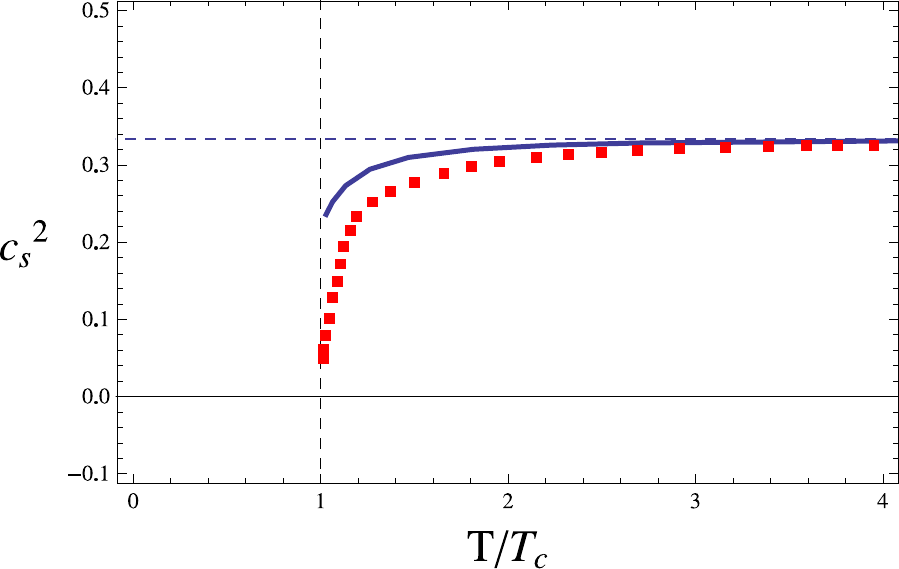}
\caption{Plot of the temperature dependence of the speed of sound. The blue curve is the gravity prediction for the model described by (\ref{eq:Vd3}), while the red squares are lattice data for $SU(3)$ gauge theory from \cite{Boyd:1996bx}. The blue dashed line indicates the conformal value $c_s^2 = 1/3$.\label{fig:cs2d3}}
\end{figure}

From these plots, one learns that there exists a critical $\lambda_H\equiv \lambda_c$, located where the free energy changes sign, at which there is a first order phase transition from the thermal gas to the black brane phase. One also notes the presence of two black hole branches in the plot of $\mathcal{F}$ as a function of $T$. Since the specific heat is given by
\begin{equation}
C_v = -T\frac{\partial^2\mathcal{F}}{\partial T^2},
\end{equation}
wherein $T$ is manifestly positive, one finds that the ``upper" branch of black hole solutions, with positive curvature, have $C_v <0$ and are thus thermodynamically unstable. Accordingly we will anoint this the ``small black hole" branch in analogy to similar solutions in global AdS.
In figure \ref{fig:TnSd3} the temperature is plotted as a function of $\lambda_H$, and the entropy density is plotted against the temperature. From the former one notes the existence of a minimum temperature, which we denote $T_0$, while from the latter one finds that this model is characterized by a large jump in the entropy density at the first order phase transition. Moreover, as the squared speed of sound is just
\begin{equation}
c_s^2 = \frac{\dd\log T}{\dd\log s},
\end{equation}
this model attains the high-$T$ conformal value of $c_s^2 = 1/3$ more rapidly than anticipated by $SU(3)$ glue on the lattice. The results are shown in figure \ref{fig:cs2d3}.

In what follows, we will focus our attention on finite temperature initial states in thermal equilibrium. This is a subset of thermalization processes which excludes the possibility of commenting on a variety of interesting questions related to the conditions required for the formation of a black hole in this theory. Included in this subset, however, are processes which probe the non-trivial gravitational phase structure of the bulk theory. Generically, one expects that this phase structure will result in an interplay between physics on the large and small black hole branches, which in turn might better inform our understanding of thermalization in confining gauge theories.

\section{More Boundary Theory Observables}\label{sec:onepts}
The dual gauge theory data is encoded in correlation functions of the boundary theory operators. In this work, we will primarily be interested in the one-point functions of the stress-energy tensor $T^{ij}$ and the dimension three operator $\mathcal{O}$. Holographically, the values of these one point functions are related to the boundary coefficients of the normalizable modes of the metric and bulk scalar, $a_4$ and $f_2$ respectively.
\subsection{Renormalized One-Point Functions}
The precise relationship, however, requires a careful analysis of the near boundary onshell bulk action, which generically has both power law and logarithmic divergences at $z=0$. These divergences can be regularized and renormalized following the standard dogma of Holographic Renormalization \cite{Bianchi:2001kw,Papadimitriou:2011qb}, the end result being a (possibly scheme dependent) identification of bulk falloffs with boundary theory correlation functions. This procedure is by now a familiar aspect of many holographic calculations, and accordingly the details will be left to appendix \ref{sec:AppHR}.

The main results of this analysis are the set of local counterterms required to regulate the on shell action, and the one point functions of the stress energy tensor and the scalar operator as functions of the bulk boundary data. In the present system, with flat boundary metric and $\df$ dual to a dimension three operator, the relevant counterterms turn out to be
\begin{equation}
S_{\mathrm{ct}} = - \frac{1}{2\kappa^2}\int_{\epsilon}\dd^4x\,\sqrt{-\gamma}\Bigg[6+\frac{4}{3}\varphi^2+\log\epsilon\bigg(F_4\,\varphi^4
-\frac{4}{3}\varphi\,\Box_\gamma\varphi\bigg)+\mathcal{A}\big[ \gamma,\varphi \big]\Bigg],
\end{equation}
where $\gamma$ is the pull back of the metric to the radial cutoff at $\epsilon$, $\mathcal{A}$ is a set of finite counterterms which define the renormalization scheme, and $F_4 = 16/27-V^{(4)}/24$. Absent supersymmetry, or another guiding principle with which to fix the coefficients of the finite counterterms, we shall henceforth adopt a holographic minimal subtraction scheme and simply ignore them. The corresponding one point functions are then given by
\begin{align}\label{eq:1pts1}
\kappa^2\langle\mathcal{O}\rangle = &\, \frac{2}{3}\left(4 f_2-\ddot{f}_0 \right)+f_0^3\left(\frac{8}{27}-\frac{V^{(4)}}{24} \right),\\
\label{eq:1pts2}
\kappa^2\langle T_{tt} \rangle = &\, -\frac{3}{2}a_4 -\frac{4}{3}\left( f_2-\frac{2}{3}\ddot{f}_0\right)f_0-\frac{2}{9}\dot{f}_0^2+\frac{14}{81}f_0^4,\\
\label{eq:1pts3}
\kappa^2\langle T_{xx} \rangle = &\,-\frac{1}{2}a_4 +\frac{2}{9}\left(2 f_2+\frac{1}{3}\ddot{f}_0\right)f_0+\frac{4}{27}\dot{f}_0^2+\frac{1}{9}f_0^4\left(\frac{14}{27} -\frac{V^{(4)}}{16}\right),
\end{align}
where we have used (\ref{eq:c2c}) to express the correlation functions in terms of the coefficients in (\ref{eq:FUV}). As expected, these correlation functions are constrained by the presence of Ward identities. In terms of these near boundary expansion coefficients, the conformal Ward identity reads
\begin{equation}
\langle T^i{}_i\rangle = \frac{2}{3}\left( 4f_2-\ddot{f}_0\right)f_0+\frac{2}{3}\dot{f}_0^2-\frac{1}{48}V^{(4)}f_0^4 = f_0\langle \mathcal{O}\rangle+\frac{2}{3}\dot{f}_0^2-\frac{1}{2}f_0^4 F_4,
\end{equation}
while the divergence of the stress tensor yields
\begin{equation}\label{eq:dW}
\nabla^t\langle T_{tt}\rangle = \dot{f}_0\langle \mathcal{O}\rangle,
\end{equation}
which illustrates the non-conservation of the system's energy in the presence of time dependent sources.

To better interpret the values of these various one-point functions, it is convenient to define subtracted correlators which effectively measure deviations of the energy, pressure, and/or scalar expectation value from the static zero temperature solution obtained in the limit $\df_H\to\infty$. One way to do this is to modify directly the renormalized on shell action by addition of appropriate finite local counterterms. For example,  a non-zero $\mathcal{A}= c_{\df}\,\df^4$ introduces a parameter $c_\df$ which can be tuned such that the energy density of the zero temperature solution vanishes. Since the counterterms are properties of the theory and not a specific solution, this subtraction will be manifest in all one-point functions computed in the boundary theory.
\begin{figure}[t]
\centering
\includegraphics[height=4cm]{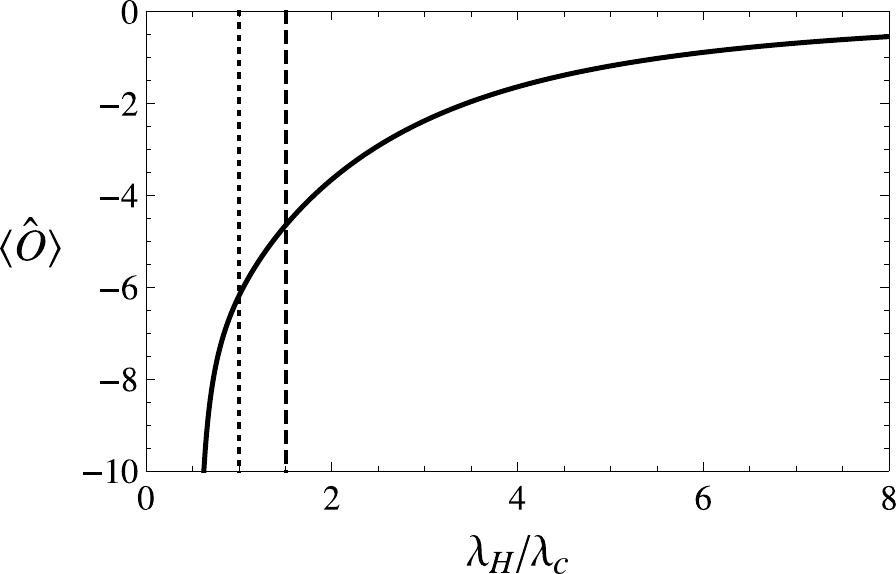}
\includegraphics[height=4cm]{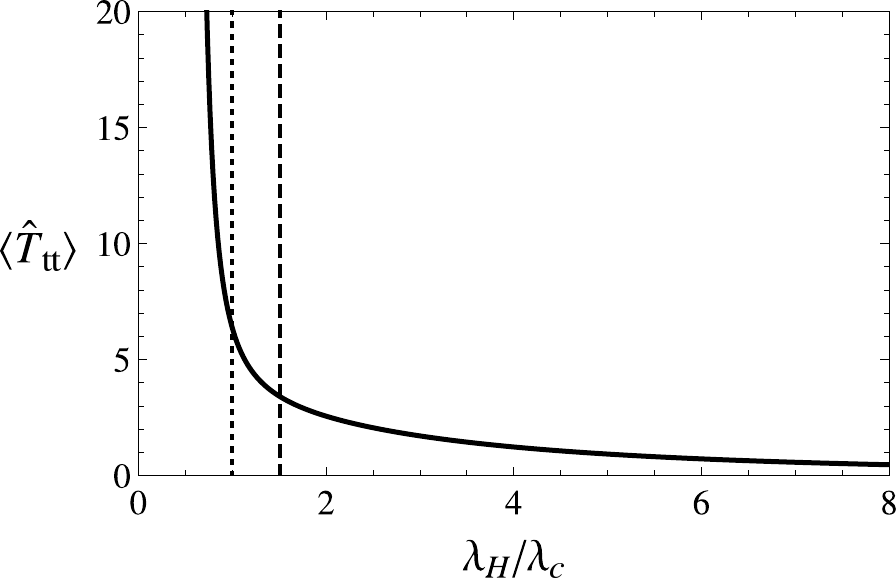}
\includegraphics[height=4cm]{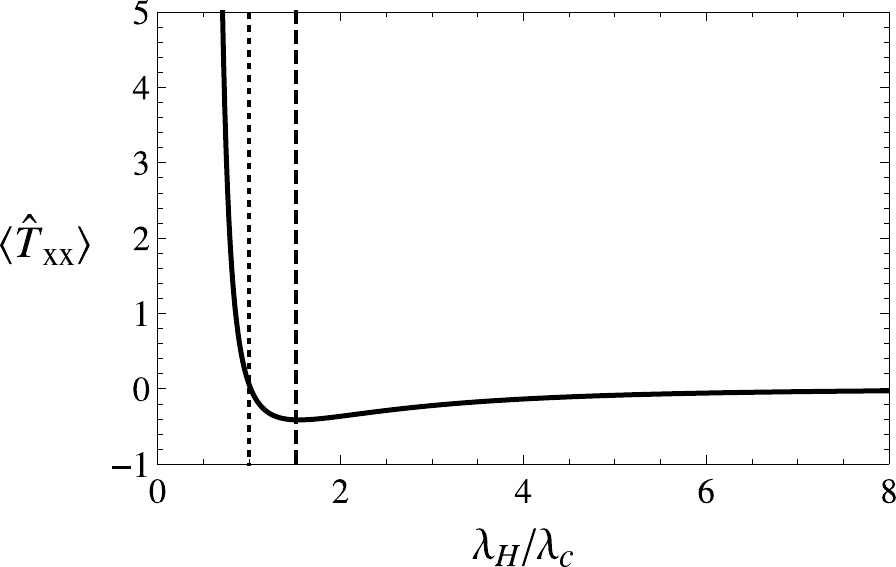}
\caption{``Hatted'' one point functions characterizing the static states of the boundary gauge theory. The plots are given in units of $f_0$ with $\kappa^2=1$. The dotted (left) and the dashed (right) lines correspond to $T=T_c$ and $T=T_0$, respectively.}
\label{fig:renVEV}
\end{figure}

 A related, and perhaps more practical prescription is to simply define new correlators with the limiting $\df_H\to\infty$ values explicitly removed. Accordingly, in the examples that follow we often construct ``hatted" one-point functions defined such that
\begin{equation}
\langle \hat{\Omega} \rangle \equiv \langle \Omega \rangle-\langle \Omega \rangle_{\df_H=\infty}
\end{equation}
for any boundary theory operator $\Omega$. In figure~\ref{fig:renVEV} we plot the hatted correlators as functions of the value the scalar obtains in the IR for future reference.

The numerical procedure we adopt is initialized by the boundary coefficients $a_4, f_2$ and $f_0$, which are in turn extracted from numerically generated initial states as described in the previous section. For this reason, the evolution's stability depends crucially on the accuracy of these values. It is thus reassuring that we find excellent agreement between the free energy density computed from (\ref{eq:fzh}), and from boundary data using $\mathcal{F} = -p \equiv -\langle \hat{T}_{xx}\rangle$.

\begin{figure}[t]
\centering
\includegraphics[height=5.5cm]{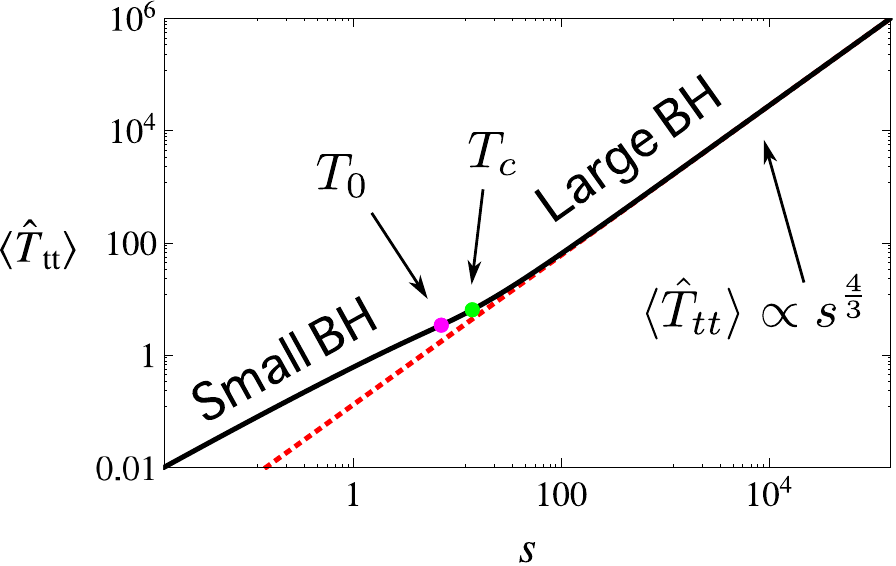}
\caption{The energy density $\langle \hat{T}_{tt} \rangle$ as a function of entropy $s$, in units of $f_0$ and with $\kappa=1$. The asymptotic behaviors are $\langle \hat{T}_{tt} \rangle \propto s^{4/3}$ and $\langle \hat{T}_{tt} \rangle \propto s\sqrt{-\ln s}$ in the limits of very large and very small black holes, and a fit for the former is plotted with a red dotted line. The green and magenta dots mark the locations of the first order phase transition at $T=T_c$ and the division between small and large black holes at $T=T_0$, respectively.}
\label{fig:s43Ttt}
\end{figure}

We also compute the dependence of the energy density $\langle \hat{T}_{tt} \rangle$  on the system's entropy density $s$. The former is calculated from the UV boundary coefficients extracted from our numerical solutions, while the latter is computed from horizon data. The result appears in figure~\ref{fig:s43Ttt}, in units of $\tilde{f}_0$ and $\kappa=1$. At large energy density, our model correctly reproduces the expectation for a conformal theory, $\langle \hat{T}_{tt} \rangle \propto s^{4/3}$. As the energy density decreases, the dimensionful source's explicit breaking of the UV theory's conformal invariance becomes increasingly important. In the extremal limit,  the small black hole branch is characterized by a logarithmic dependence of the energy density on entropy density, $\langle \hat{T}_{tt} \rangle \propto s\sqrt{-\ln s}$.  From the point of view of the boundary gauge theory, the latter behavior does not manifest as the thermodynamically preferred solution is the thermal gas whose entropy density is subleading in $N_c$.

\section{Numerical Strategy}
\label{sec:strategy}

The primary goal of the present work is to discover how an initial state described by a solution from section \ref{sec:thermo} responds to time dependent perturbation by a dimension three operator. Holographically, we implement this perturbation by varying the UV boundary condition on the scalar in time. For example, a source whose profile is given by
\begin{equation}\label{eq:gquench}
f_0(v) = \tilde{f}_0-\delta f_0\,e^{-\frac{v^2}{2\tau^2}}
\end{equation}
corresponds to perturbing the initial state (with coupling to $\mathcal{O}$ fixed by $\tilde f_0 \equiv f_0(-\infty)$) by a gaussian deformation with amplitude $-\delta f_0$ and variance $\tau^2$ centered at $v=0$. Accordingly, the boundary theory experiment we have in mind is dialing down the coupling to the relevant operator by an amount $\delta f_0$ and for a time $\tau$.

In the remainder of this section, a recipe for computing the time evolution of an initial state in the presence of time dependent scalar source $f_0(v)$ is explained from both generic and practical viewpoints.

\subsection{Integration Routine}
\label{sec:routine}
The equations of motion (\ref{eq:eom1}-\ref{eq:eom5})  have been arranged into a ``nested" structure convenient for numerical study. We employ and consequently review the method detailed in \cite{Chesler:2013lia}, which exploits this nested structure to reduce the partial differential equations to a sequence of ordinary differential equations that can be solved time step by time step.

Before addressing the task of solving the Einstein equations describing this system, it is important to understand the boundary information needed to fully specify a given solution. For this, one may turn to the near boundary behavior of the {\it homogeneous} Einstein equations. For example, in the UV equation (\ref{eq:eom3}) is
\begin{equation}
\Sigma''+\frac{2}{z}\Sigma'+\frac{4}{9}f_0^2\,\Sigma=0,
\end{equation}
which has the general solution
\begin{equation}
\Sigma = \mathcal{C}_1\frac{1}{z}\cos\frac{2f_0\,z}{3}+\mathcal{C}_2\frac{1}{z}\sin\frac{2f_0\,z}{3}.
\end{equation}
Thus, near the boundary linearly independent solutions behave like $z^{-1}$ and $z^0$. Comparing to (\ref{eq:SUV}) it is clear that $\Sigma$ can be fully specified by the first two terms in the UV expansion---the AdS boundary condition and the radial reparametrization artifact $\zeta$. Similarly, from the unsourced (\ref{eq:eom2}) one finds near the boundary
\begin{equation}
(\dd_+\Sigma)'-\frac{2}{z}\,\dd_+\Sigma=0,
\end{equation}
which has the solution $\dd_+\Sigma\sim z^2$. From (\ref{eq:dpSUV}) it is clear that one must specify the value of $C_\Sigma$ in order to determine  $\dd_+\Sigma$, which in turn depends on $f_0$, $\vl$, and $a_4$. Finally, from the unsourced (\ref{eq:eom1}) in the UV one solves
\begin{equation}
(\dd_+\df)'-\frac{3}{2z}\,\dd_+\df=0
\end{equation}
with  $\dd_+\df\sim z^{3/2}$. From (\ref{eq:dpFUV}), evidently the desired solution is the one in which the coefficient of $ z^{3/2}$ vanishes. In principle, the same sort of analysis can be applied to (\ref{eq:eom4}) to explore the boundary behavior of $A$. The homogeneous equation is readily solved by $A\sim z^{-1}+z^0$, which (\ref{eq:AUV}) shows are fixed by $\zeta$ and $\dot{\zeta}$ on the given time slice. Since $\zeta$  parameterizes a residual gauge degree of freedom, one may adopt a gauge where $\dot{\zeta}=0$. Alternatively, it may be desirable to determine $\zeta(v)$ dynamically. In this case one may use the value of $A$ on the apparent horizon instead of $\dot{\zeta}$ as an integration constant.  In practice this can be accomplished by immobilizing the location of the horizon (e.g. requiring $z_H = 1$ always). The condition for the location of an apparent horizon in these backgrounds (see Appendix \ref{sec:AppAH}) is simply
\begin{equation}\label{eq:AppH}
\dd_+\Sigma \big|_{z_H}= 0,
\end{equation}
and the horizon stationarity equation thus requires
\begin{equation}\label{eq:AH}
A(z_H) = -\frac{16}{3}\frac{(\dd_+\df)^2}{V}\Bigg|_{z_H}.
\end{equation}

The upshot of the preceding analysis is that solving the Einstein equations requires knowledge of  $a_4$, $\zeta$ and $\df(v_0,z)$---from which $f_2$ can be extracted---on the initial time slice, together with a choice of the forcing function $f_0(v)$.

After obtaining this initial data, one is well poised to evolve the system. The procedure is as follows:

\begin{enumerate}
\item From $\df(v_0,z)$, the AdS boundary condition $s_0(v_0)=1$, and the value of $\zeta(v_0)$, equation (\ref{eq:eom3}) can be integrated to obtain $\Sigma(v_0,z)$.
\item With this knowledge, and the value of $\mathcal{C}_\Sigma(v_0)$---which depends on $a_4(v_0)$, $f_0(v)$ and $f_2(v_0)$---equation (\ref{eq:eom2}) can be integrated for $\dd_+\Sigma(v_0,z)$.
\item Then equation (\ref{eq:eom1}) can be solved with the boundary condition that $\dd_+\df$ has no fall off like $z^{3/2}$ at the boundary, resulting in the knowledge of $\dd_+\df(v_0,z)$.
\item Finally, we turn to the second order ODE given by equation (\ref{eq:eom4}). As indicated above, this equation can be solved given  the values of $\zeta(v_0)$ and either $\dot{\zeta}(v_0)$ or  $A_H(v_0)$.
\end{enumerate}

Already this routine is enough to evolve the fields $\Sigma$ and $\df$, as well as $\zeta$ in time. More precisely, for any field $F$, knowledge of $\dd_+F$ permits one to write
\begin{equation}\label{eq:euler}
\dot{F} = \dd_+F+\frac{z^2}{2}A\,F'\qquad \mathrm{and\,\,thus} \qquad F(v_0+\Delta v)\approx F(v_0)+\dot{F}(v_0)\Delta v.
\end{equation}
In practice, it is not necessary to evolve $\Sigma$ explicitly, as its profile on the next time slice will be constructed when step 1 repeats.  The value of  $f_2(v_0+\Delta v)$ can be computed by extracting $\dot{\vl}(v_0)$ from  the boundary behavior of $\dot{\df}(v_0,z)$ and integrating, or it can be extracted directly from the boundary behavior of $\varphi(v_0+\Delta v)$. In practice we have found the latter numerically favorable.


At this juncture, all that is needed to update the routine on the next time step is $\dot{a_4}(v_0)$. There are options for computing this. One method is to solve equation (\ref{eq:eom5}), rearrange the $\dd_+$ derivative to obtain $\partial_v\dd_+\Sigma$ at $v_0$, and then study the UV behavior to extract $\dot{a_4}(v_0)$. A better way is to use \eqref{eq:Econs} and directly evolve $a_4(v_0)$ to the next time slice. In this way one arrives at time $v=v_0+\Delta v$ with updated values of $f_2$, $a_4$, $\zeta$, and $\df(v_0+\Delta v,z)$, which is all the information required to begin the integration routine anew.

\subsection{A practical method}

The radial integration of the equations of motion can in principle be performed by any of a number of well worn techniques, including pseudo spectral and finite difference methods. In practice, we found that the logarithmic terms in the near boundary behaviors of the fields rendered pseudo spectral methods unstable unless sufficient care was exercised to explicitly remove these terms (an observation also made by the authors of \cite{Buchel:2014gta}). While one can typically ameliorate this issue through suitable field redefinitions, we have found it more convenient to employ a finite difference discretization method and integrate from the boundary to the IR. This method is similar in spirit to the one that appears in \cite{Murata:2010dx}.

The response of the boundary gauge theory to these time dependent perturbations is encoded in the time evolution of $f_2$ and $a_4$, which show up in higher orders of the boundary series solutions (\ref{eq:AUV}-\ref{eq:FUV}). Therefore, to determine them accurately, we work with redefined fields where $f_2$ and $a_4$ appear in the leading order at the boundary expansions by subtracting source contributions in lower orders. To surmount some of the challenges posed by the logarithmic behavior near the boundary, we also subtract the first few logarithmic terms. Moreover, we rescale by appropriate factors of $z$ so that the redefined fields do not vanish at the boundary. We thus work with the subtracted fields defined by
\begin{align}
\label{eq:sbtA}
z^2\,\bar{A} &= A - \sum_{n=0}^{3} a_n z^{n-2}  - \sum_{n=4}^{6} \alpha_{n1} z^{n-2} \log z - \alpha_{62} z^4 \log^2 z, \\
\label{eq:sbtS}
z^3\,\bar{\Sigma} &=  \Sigma -  \sum_{n=0}^{3} s_n z^{n-1} - \sum_{n=4}^{6} \sigma_{n1} z^{n-1} \log z - \sigma_{62} z^5 \log^2 z , \\
\label{eq:sbtF}
z^3\,\bar{\varphi} &= \varphi - \sum_{n=0}^{1} f_n z^{n+1} - \sum_{n=2}^{4} \phi_{n1} z^{n+1} \log z - \phi_{42} z^5 \log^2 z , \\
\label{eq:sbtDS}
z^2\,\overline{D\Sigma} &= \mathrm{d}_{+} \Sigma - \sum_{n=0}^{3} (Ds)_n z^{n-2} - \sum_{n=4}^{6} (D\sigma)_{n1} z^{n-2} \log z - (D\sigma)_{62} z^4 \log^2 z , \\
\label{eq:sbtDF}
z^2\,\overline{D\varphi} &= \mathrm{d}_{+}\varphi - \sum_{n=0}^{1} (Df)_n z^{n} - \sum_{n=2}^{4} (D\phi)_{n1} z^{n} \log z - (D\phi)_{42} z^4 \log^2 z .
\end{align}
In our choice, the redefined fields are $C^2$ over the radial domain. Substituting these expressions into the equations of motion, we obtain evolution equations of the redefined fields, which are solved with the strategy described in section~\ref{sec:routine}.

The equations of motion are discretized by finite difference grids in the $z$ and $v$ coordinates whose sizes are $\Delta z$ and $\Delta v$, and we arrange our numerical computation so that second order accuracy is supposed to be achieved. We use a fourth-order Runge-Kutta method in $z$-integration, and integrate from the boundary toward the interior. The $z$-derivatives of the fields which are not evolved in each differential equation are computed by a central finite difference scheme. We adopt a gauge where $\zeta(v)=0$ is fixed for all times. This generically implies that the location of the black hole horizon can vary throughout the evolution. We monitor the location of the apparent horizon on each time step and terminate the integration slightly inside it, which is also inside the event horizon. To update $\bar{\df}$ to the next time step, we use an upwind difference scheme for the advection term in \eqref{eq:ddp}. The grid sizes are chosen such that the Courant-Friedrichs-Lewy condition is satisfied, and we use $\Delta v \le \Delta z$. For $v$-evolution, we use a modified Euler's method, where to update $a_4$ we integrate the constraint \eqref{eq:Econs} with a fourth order Adams-Bashforth formula,\footnote{For first few steps, \eqref{eq:euler} is used.}
\begin{align}
a_4(v_{n+1}) = a_4(v_{n}) + \frac{55 \, \dot{a}_4(v_{n}) -59 \, \dot{a}_4(v_{n-1}) + 24 \, \dot{a}_4(v_{n-2}) - 9 \, \dot{a}_4(v_{n-3})}{24} \Delta v.
\end{align}

Once the final static configuration is reached, when the apparent and event horizons coincide, we solve \eqref{eq:zhgeo} backward in time to compute the evolution of the event horizon $z_\mathrm{EH}(v)$. We can then calculate the area density of the event horizon on constant $v$-slices as
\begin{align}
A_\mathrm{EH}(v) = \Sigma(v,z_\mathrm{EH})^3.
\end{align}
The monotonic increase of the event horizon is utilized as a consistency check of our numerical computation. The area density of the apparent horizon is computed analogously, as $A_\mathrm{AH}(v) = \Sigma(v,z_\mathrm{AH})^3$.

\section{Thermalization Examples}\label{sec:EX}
\begin{figure}[t]
\centering
\includegraphics[height=4.5cm]{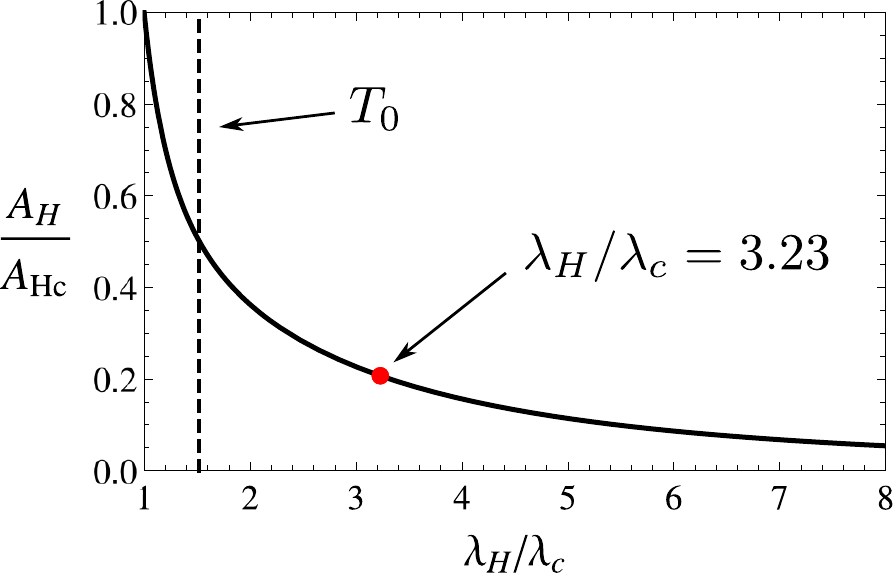}
\caption{The area density of small black holes compared with that at the phase transition, $A_{Hc}$. The red dot marks the location in our space of solutions of the smallest black hole we perturb in this study. The dashed line indicates the location of the small and large black hole transition at $T=T_0$. As these are static black brane solutions, the apparent and event horizons coincide.}
\label{fig:sBHA}
\end{figure}

In this section we provide examples of the thermalization processes produced by our numerical computation. As the scalar source of the static states is nonzero, quenches are not symmetric between positive and negative $\delta f_0$, and we consider quenches with $\delta f_0>0$ for simplicity. Several tests suggest that broad stroke, qualitative features of the resulting time evolution seem not to be significantly altered by the choice of this sign.\footnote{The primary differences occur for very large, very slow quenches but we will primarily be interested in more modest perturbations in what follows.} Throughout this section and the next we will work in units of $\tilde{f}_0$ and henceforth set $\kappa^2=1$. The stability and accuracy of initial data is discussed in Appendix~\ref{app:num}.

An important question in the interpretation of these results is ``how far" our initial states are along a given black hole branch. A practical metric for quantifying this distance is the area density of the black hole horizon. The zero temperature solution, which can be thought of as the $\lambda_H\to\infty$ limit of the small black hole branch, has vanishing horizon area, and the horizon area grows monotonically as the scalar vanishes at $\lambda_H\to 1$. This further justifies our ``large" and ``small" naming conventions for the black hole branches. In the studies which follow, the smallest black hole we perturb has $\lambda_H/\lambda_c = 3.23$. In figure~\ref{fig:sBHA}, the area of the black hole horizons  as a function of $\lambda$ is given, and this smallest black hole is indicated by a red dot. Evidently, the horizon area of $\lambda_H/\lambda_c = 3.23$ in this case is about 1/5 of that at the first order phase transition.

\begin{figure}[t]
\centering
\includegraphics[height=5cm]{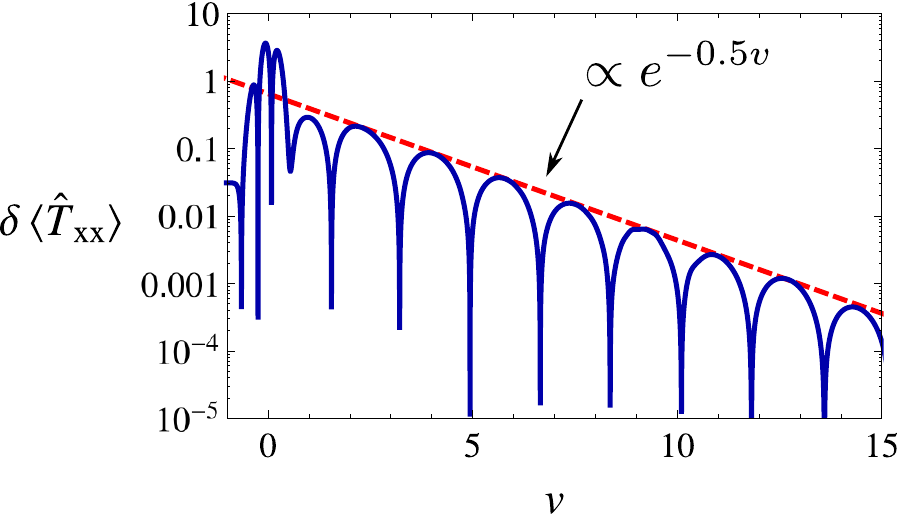}
\caption{Typical behavior of the magnitude of the late time deviation of $\langle T_{xx} \rangle$ from its final value. This particular ring-down corresponds to the late time behavior of the perturbation shown in figure \ref{fig:smVEV}. The red line is a fit to the exponential decay provided by $\delta\langle T_{xx} \rangle\sim0.65\, e^{-0.5 v}$.}
\label{fig:smLog}
\end{figure}

Generically, a time dependent perturbation of a black hole will take a static initial state through a non-linear regime initiated by the details of the quench profile, followed by a linear regime governed by the ``ring-down" to the final steady state configuration (there may also be late time power law tails in some situations, but we will not be concerned with these here).
The ring-down is fully determined by the quasi-normal modes of the final state black hole, and is dominated by the  mode closest to the real axis, $\omega_1$.  In turn, the quasi-normal modes characterize the linear response of the final state to small perturbations in any of several available channels. In the present case, where the perturbations preserve the homogeneity of the spatial $\mathbb{R}^3$,   the gauge invariant perturbations organize themselves  into representations of $SO(3)$ \cite{Kovtun:2005ev,DeWolfe:2011ts} transforming as the transverse-traceless (spin-2), vector, or scalar.

\begin{figure}[t]
\centering
\includegraphics[height=5cm]{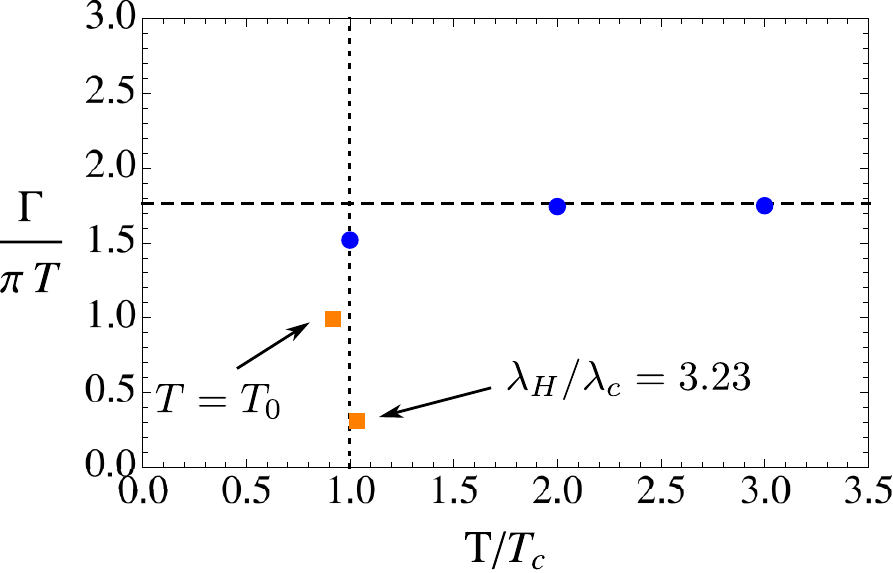}
\caption{The temperature dependence of the decay width $\Gamma$ for the lowest lying scalar quasi-normal mode in several states of our theory. The blue circles are large black branes whose temperature is an integer multiple of $T_c$. The orange squares correspond to the minimum temperature black brane (top) and the smallest black hole we perturb in our study (bottom). The ratio $\Gamma/\pi T$ approaches 1.75953 (the dashed line) at high temperatures, which coincides with the expected value for perturbations of AdS$_5$ Schwarzschild by a dimension 3 scalar operator \cite{Nunez:2003eq}. }
\label{fig:QNM}
\end{figure}

In figure \ref{fig:smLog} the late time ring-down of one particular time dependent perturbation is shown on a logarithmic scale. The figure clearly indicates the presence of an excited mode of the form
\begin{equation}
\delta\langle \hat{T}_{xx} \rangle \sim \textrm{Re}\,Z_1 e^{-i \omega_1 v}\qquad \textrm{with} \qquad \omega_1 = \omega_{*}-i \Gamma
\end{equation}
for some real constants $Z_1$, $\omega_*$, and $\Gamma >0$. Although figure \ref{fig:smLog} illustrates the general late time features of any perturbation in our study, it is important to note that the quasi-normal mode spectrum, and hence the particular value of $\omega_1$, is a property of the final state achieved after the quench. In other words, one should keep in mind that $\omega_1$ varies as one traverses the static state solution space so that $\omega_1 = \omega_1(\lambda_H)$. In figure \ref{fig:QNM} we quantify this by plotting $\Gamma$ as a function of temperature for several thermal states of our holographic theory. As expected, linear scaling of $\Gamma$ with temperature appears in the conformal (small scalar) limit, and deviations from this behavior are already evident at the first order phase transition where $T=T_c$. The qualitative properties of this plot are anticipated by the thermodynamic features of our model shown in figure \ref{fig:cs2d3} and the right plot of figure \ref{fig:TnSd3} which display a similar approach to conformal behavior above $T_c$.

To construct a meaningful measure of the ``thermalization time" in our quench process, it is useful to imagine partitioning the system's response to a time-dependent perturbation into several distinct timescales. The gaussian perturbations that we define in \ref{eq:gquench} are controlled by two dimensionless parameters, $\tilde{\delta}\equiv\delta f_0/\tilde{f}_0$ and $\tilde{\tau}\equiv\tau \tilde{f_0}$. Thus, the timescale  $\tilde{\tau}$ is roughly the amount of time it takes to drive the system out of equilibrium. Once the system is no longer in thermodynamic equilibrium, it may return to a (generically different) static thermal state by passing through any number of additional dynamical regimes, each with its own characteristic timescale, $\mathcal{T}_i$. For example, one might plausibly imagine a regime with timescale $\mathcal{T}_{\mathrm{S}}$ dominated by scattering of the perturbation in the bulk from the rapidly increasing scalar potential, or a regime whose characteristic timescale $\mathcal{T}_{\mathrm{RD}}$ is controlled entirely by the low-lying quasi-normal modes of the final state black brane.  Of course the system could respond in more complicated ways, and the distinction between regimes may not necessarily be crisp. In any case, we define the thermalization time as the sum of these characteristic timescales:
\begin{equation}\label{eq:tthermdef}
\mathcal{T}_\textrm{therm} \equiv\sum_i \mathcal{T}_i.
\end{equation}
An important attribute of this definition is the fact that $\tilde{\tau}$ is excluded from the sum. This feature ensures that the thermalization time we measure is characteristic of the system's response to a given perturbation, and does not explicitly depend on the time it takes us to prepare the non-equilibrium state.

In practice, we find that our quenches are well described by a very rapid transition from the regime driven by the quench to the ring down characterized by linear response. Put another way, in (\ref{eq:tthermdef}) only one term, $\mathcal{T}_{\mathrm{RD}}$, appears in the sum. We will thus be particularly interested in the decay width $\Gamma$, as it provides a natural time scale for all of the equilibration processes we examine.  In other words, throughout the rest of this work we will always find that
\begin{equation}
\mathcal{T}_\textrm{therm} \sim \frac{1}{\Gamma(\tilde{\delta},\tilde{\tau})}.
\end{equation}
Given the preceding discussion, there are two natural questions we would like to address:
\begin{enumerate}
\item For a given initial state, what is the dependence of the thermalization time on $\tilde{\delta}$ and $\tilde{\tau}$?
\item Why does this thermalization time appear to be dominated by $\mathcal{T}_{\mathrm{RD}}$ in our confining gauge theory?
\end{enumerate}
Our answers to these questions  motivate the remainder of this work. The answer to the first question begins in the following section, while we postpone addressing the second until the discussion in section \ref{sec:diss}.

\subsection{Thermalization Between Branches}

\begin{figure}[t]
\centering
\includegraphics[height=4cm]{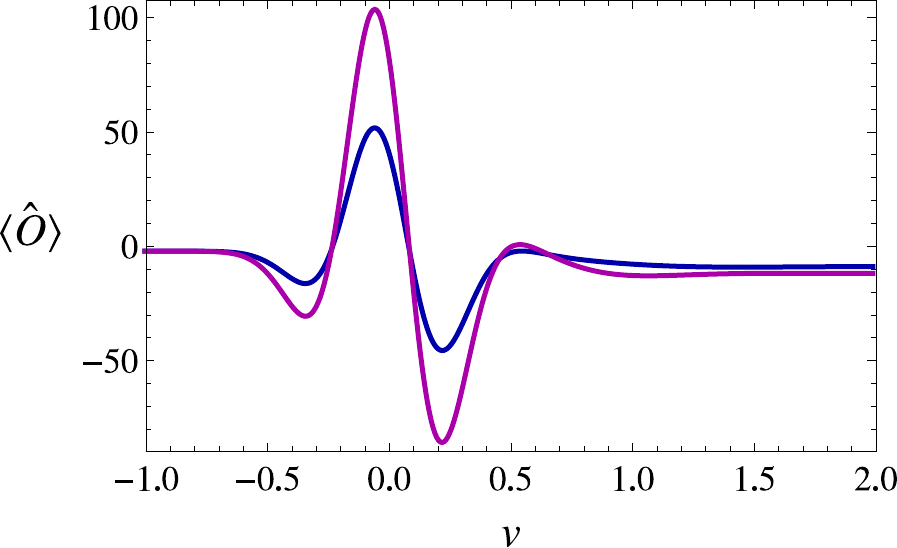}
\includegraphics[height=4cm]{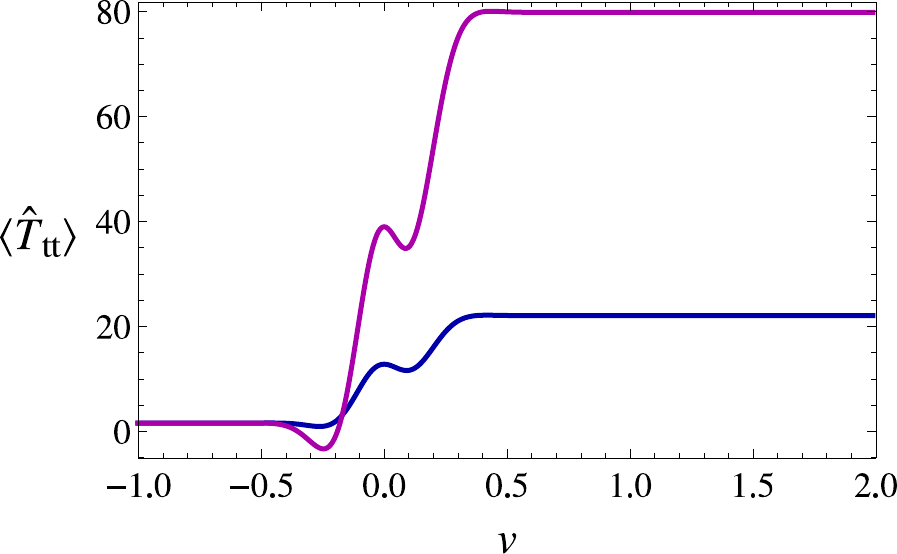}
\includegraphics[height=4cm]{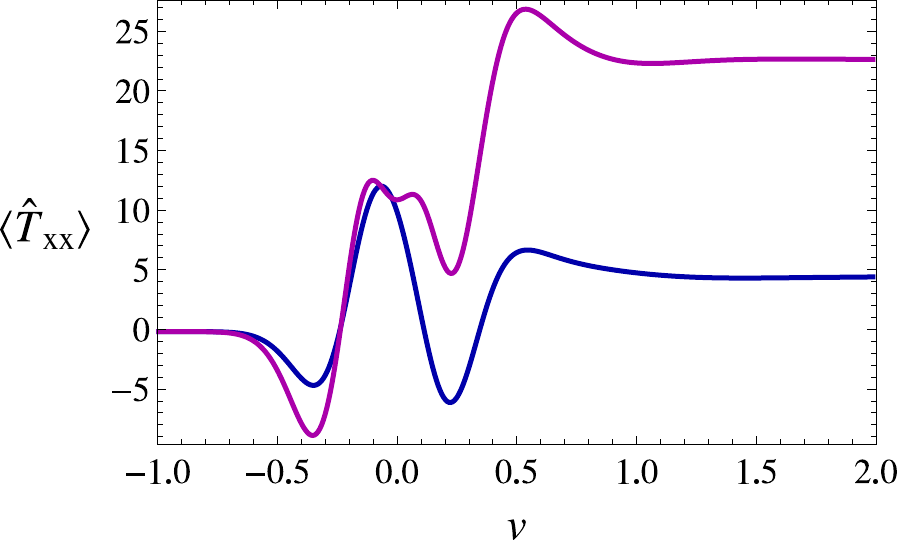}
\caption{Large amplitude quench. The blue and purple lines correspond to $\tilde{\delta} = 0.5$ and $1$, respectively. In the case of the larger amplitude quench, it is interesting to note that the energy density appears to be driven below the ground state energy density (i.e.~negative) in the first moments of the quench.}
\label{fig:laVEV}
\end{figure}

The form of the Ward identity (\ref{eq:dW}) encourages the intuitive expectation that a large, fast quench will result in the most significant increase in the system's energy. In figure~\ref{fig:laVEV}, examples of the evolution of the boundary operators in such a scenario are shown. These computations are performed with a  width $\tilde{\tau}=0.168$ and amplitudes $\tilde{\delta} = 0.5$ and  1. These perturbations basically dominate the dynamics, and can generically be tuned to result in a final configuration characterized by a large black hole independent of the particulars of the initial state. Indeed, we have performed analogous quenches in many different initial states, all of which show qualitatively similar time evolution during and after the quench. In line with our previous discussion, the plots of $\langle \hat{\mathcal{O}} \rangle$ and $\langle \hat{T}_{ii} \rangle$ show late time behavior that is well fit by an exponential decay related to the ring-down of the black hole. As expected, we find that the lowest lying quasi-normal mode for these large black hole final states retreats from the real axis as the final state energy density increases. Accordingly, we observe that these large amplitude perturbations imply rapid thermalization in the dual field theory.

\begin{figure}[t]
\centering
\includegraphics[height=4.5cm]{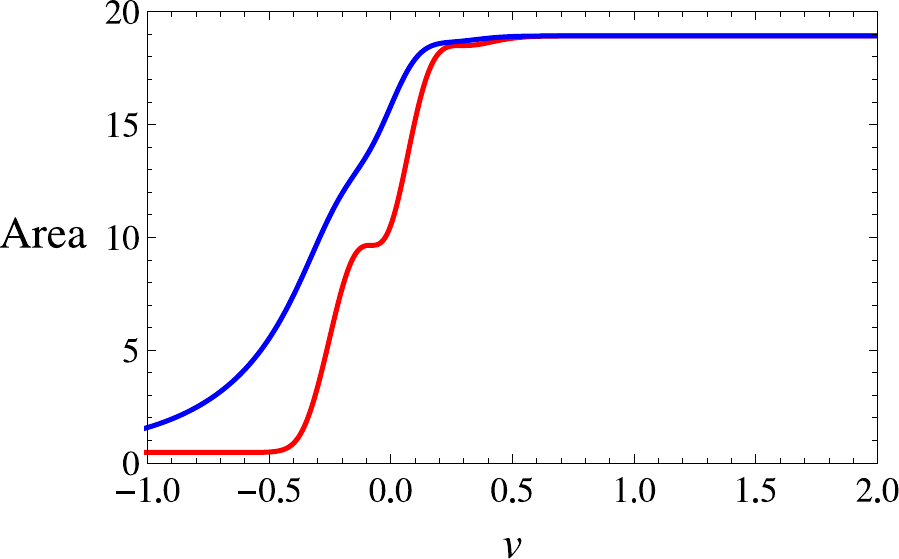}
\caption{Example of the time evolution of the apparent (red) and event (blue) horizons. The event horizon coincides with the apparent horizon when the bulk solution is static, at $v \to \pm\infty$.}
\label{fig:laArea}
\end{figure}

Once the entire evolution between static initial and final states is known, the time evolution of the apparent and event horizons can be computed. This provides a useful view of the equilibration process, as well as a consistency check of the time evolution. The area of the event horizon monotonically increases as shown in figure~\ref{fig:laArea}, and its location should always lie outside (closer to the boundary than) that of the apparent horizon. We verify explicitly that both of these statements hold in each process we study. Additionally, we note that in all perturbations we evolve, the area of the apparent horizon increases monotonically as well.

\subsection{Thermalization Along a Branch}\label{sec:tab}

When the quench amplitude is very small compared to its width, the Ward identity (\ref{eq:dW})  implies that the system's energy will be very nearly conserved, and thus we anticipate final configurations that are very close to the initial state. In particular, an initial state characterized by a small black hole can remain within the small black hole branch.

\begin{figure}[t]
\centering
\includegraphics[height=4cm]{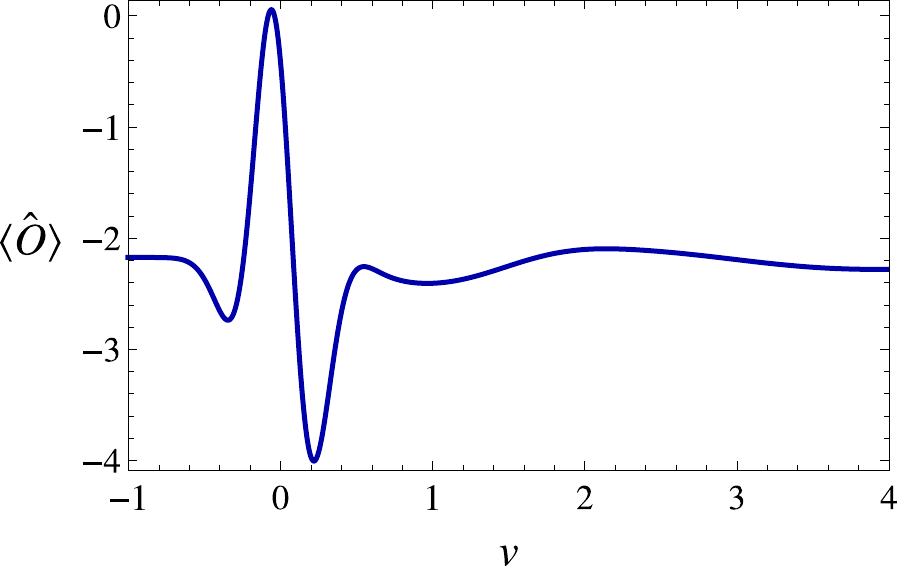}
\includegraphics[height=4cm]{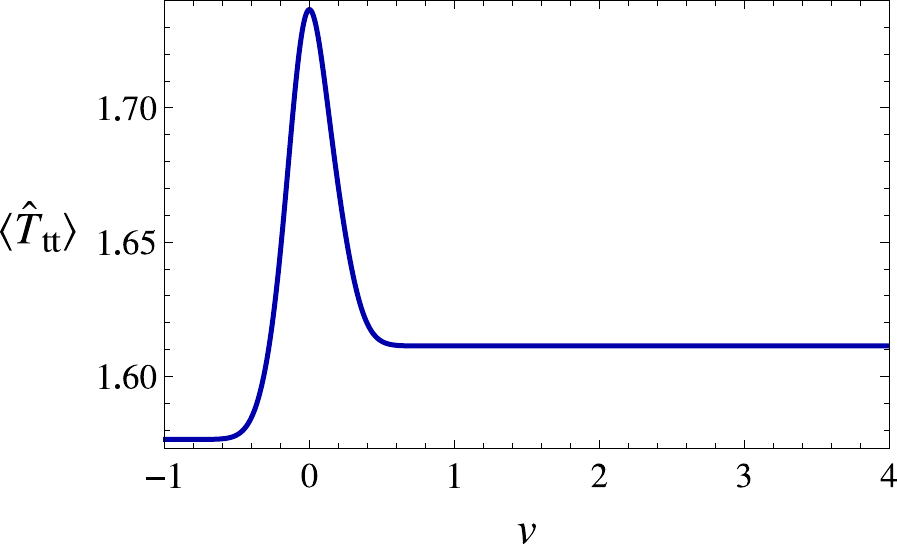}
\includegraphics[height=4cm]{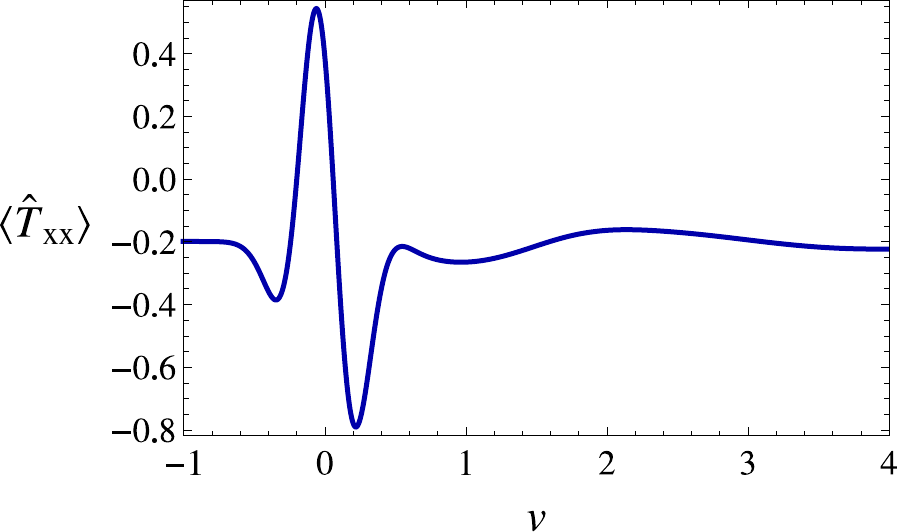}
\caption{A ``small" amplitude quench with $\tilde{\delta} = 0.02$ and $\tilde{\tau}=0.168$.}
\label{fig:smVEV}
\end{figure}

An example of this kind of quench is shown in figure~\ref{fig:smVEV}. There the perturbation width is taken to be the same as in the example of previous section, $\tilde{\tau}=0.168$, but now the amplitude is chosen to be $\tilde{\delta}=0.02$.

With these parameters, the area of the event horizon in the final state is only about 2.5\% larger than that of the initial state. Pictorially, on the scale of figure~\ref{fig:sBHA} this quench results in the red dot which marks the location of the initial state moving imperceptibly to the left after the perturbation.
\begin{figure}[t]
\centering
\includegraphics[height=4cm]{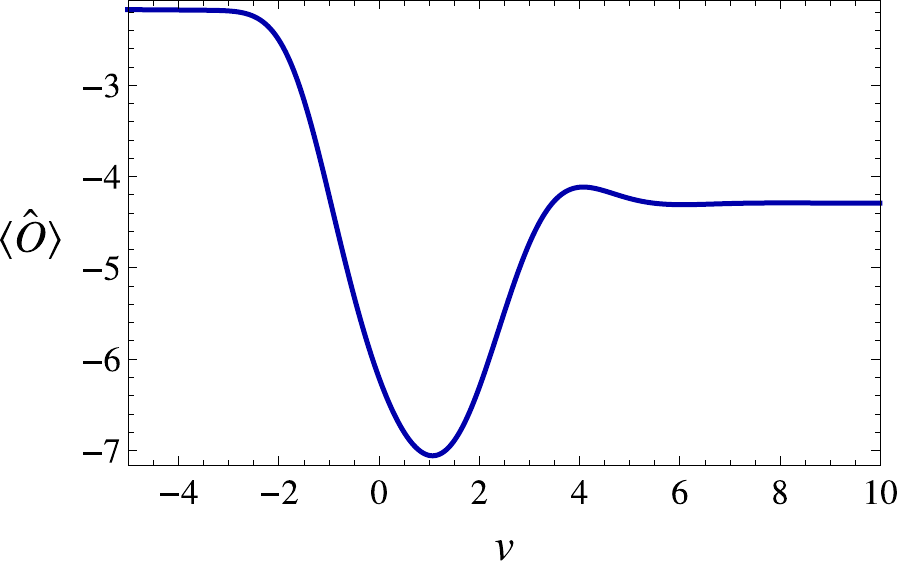}
\includegraphics[height=4cm]{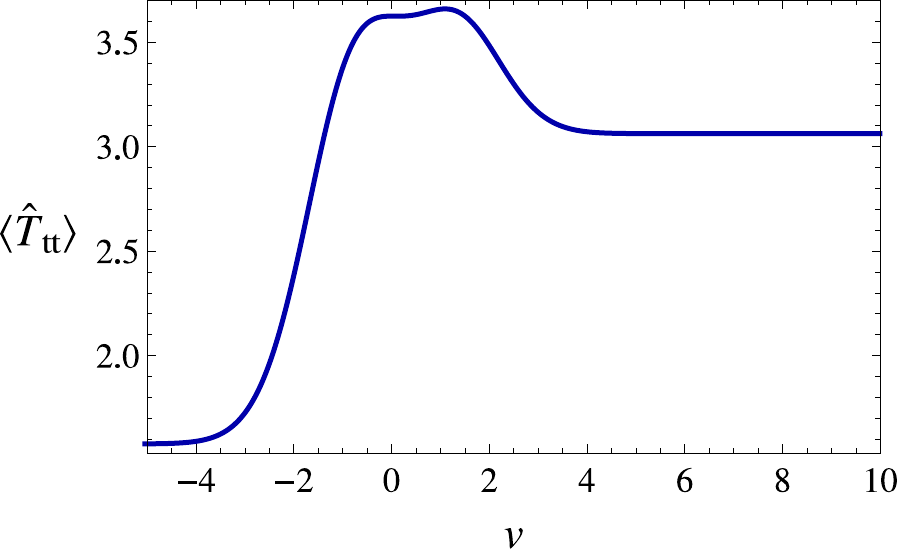}
\includegraphics[height=4cm]{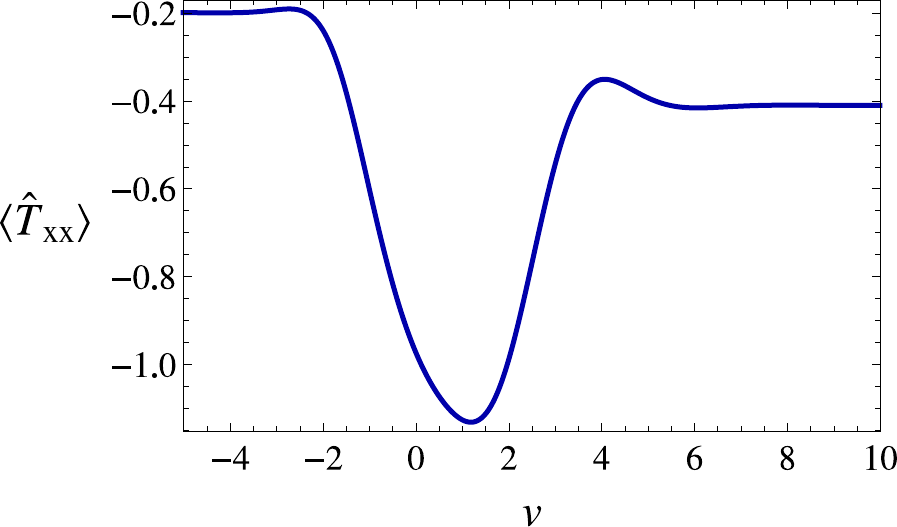}
\caption{A slow quench with large amplitude: $\tilde{\delta}= 0.5$ and $\tilde{\tau} = 1.19$.}
\label{fig:slVEV}
\end{figure}
At late times, the ring-down of $\langle \mathcal{O} \rangle$ and $\langle T_{xx} \rangle$ is far more pronounced in this case, with easily identifiable oscillations shown in figure~\ref{fig:smVEV}. The late time behavior of this quench is the focus of figure \ref{fig:smLog}, which  highlights the role of the dominant quasi-normal mode. The late time behavior of $\langle \mathcal{O} \rangle$ displays very similar behavior. In line with expectations, unlike the large amplitude quenches this class of perturbation is sensitive to the particulars of the initial state. For example, for the quench shown in figure~\ref{fig:smVEV}, we find that $\mathcal{T}_\textrm{therm}\sim 2$ in units of $\tilde{f_0}$---but the same perturbation in an initial state with about 13.5 times the energy density has a thermalization time  $\mathcal{T}_\textrm{therm}\sim 0.3$, more than $6.6$ times faster.

Of course there are a variety of perturbations one can perform which result in a final state thermodynamically similar to the initial state. As mentioned above, these all share the feature that the perturbation amplitude is relatively small compared to its width. We will see in detail below how to quantify relatively small. At present, we turn our attention to a similar example of thermalization along a branch. This is the large amplitude, slow quench shown in figure~\ref{fig:slVEV}. In this case, $\tilde{\delta} = 0.5$ and   $\tilde{\tau} = 1.19$, so that the ratio of the amplitude to width here is about 3.5 times the previous example. Accordingly, we anticipate a greater change in energy density and that this perturbation will thermalize more quickly. From the late time behavior  we find that this is indeed the case, with $\mathcal{T}_\textrm{therm}\sim 0.86$.

Our code is able to evolve even slower perturbations, including those whose widths are many times larger than their amplitude. Predictably, in this case the various one point functions respond by being gradually deformed from (and then nearly returned to) their initial vales.  For these adiabatic perturbations, the ``bumpy" features  in the late time behavior shown in figure~\ref{fig:slVEV} disappear, and the exponential decay at late times is difficult to detect due to numerical accuracy. This can be viewed as a consequence of the fact that for such slow perturbations, the quench width is much larger than the time scale defined by the final state's lowest lying quasi-normal mode.

\section{Perturbation Analysis and Dynamical Phase Structure}
\label{phase}

\subsection{Parameter Dependence and Scaling Regimes}

\begin{figure}[t]
\centering
\subfigure[$\tau$ fixed]{\includegraphics[height=4.8cm]{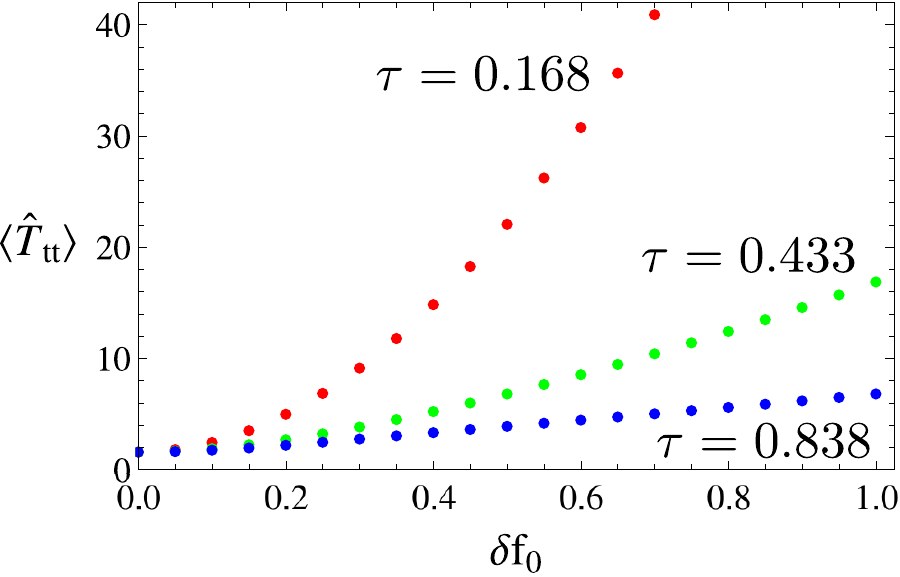}\label{fig:sfTttD}}
\quad
\subfigure[$\delta f_0$ fixed]{\includegraphics[height=4.8cm]{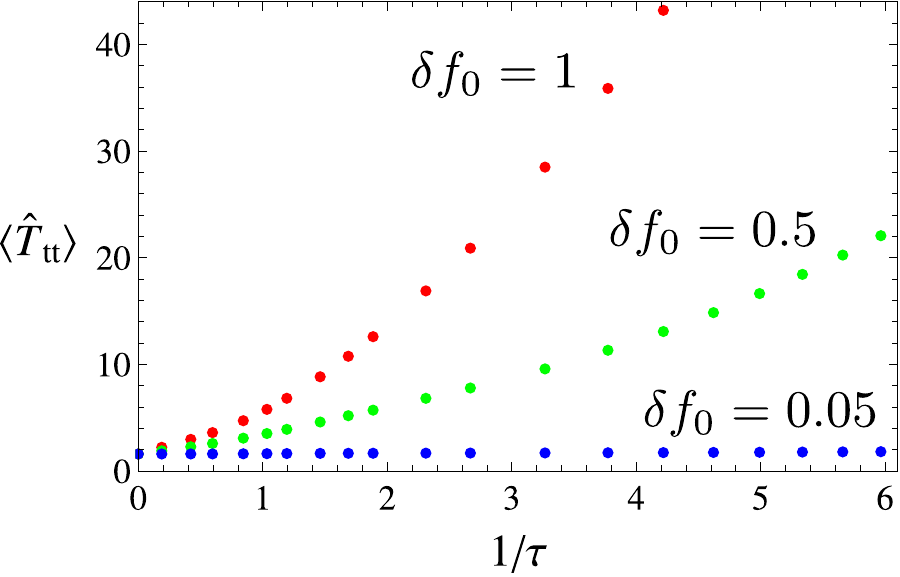}\label{fig:stTttD}}
\caption{The final state energy density for various perturbations with either $\delta f_0$ or $\tau$ fixed. All quantities in this figure (and those that follow) are measured in units of $\tilde{f}_0$}
\label{fig:sTttdat}
\end{figure}

To better understand the features of different quenches, we look in more detail at the dependence of the final state on the perturbation parameters. In figure \ref{fig:sTttdat}, we show several results for the dependence of the final state energy density on $\tilde{\delta} $ for fixed $\tilde{\tau} $ and vice-versa. In figure~\ref{fig:sfTttD}, we compare perturbations with many different amplitudes but one of three fixed quench widths: $\tilde{\tau} =0.168, 0.433$ and 0.838. In figure~\ref{fig:stTttD}, we fix the perturbation amplitude at $\tilde{\delta} =0.05, 0.5$ or 1 and consider quenches with many different widths.

\begin{figure}[t]
\centering
\subfigure[$\tau=0.168$ fixed]{\includegraphics[height=4.5cm]{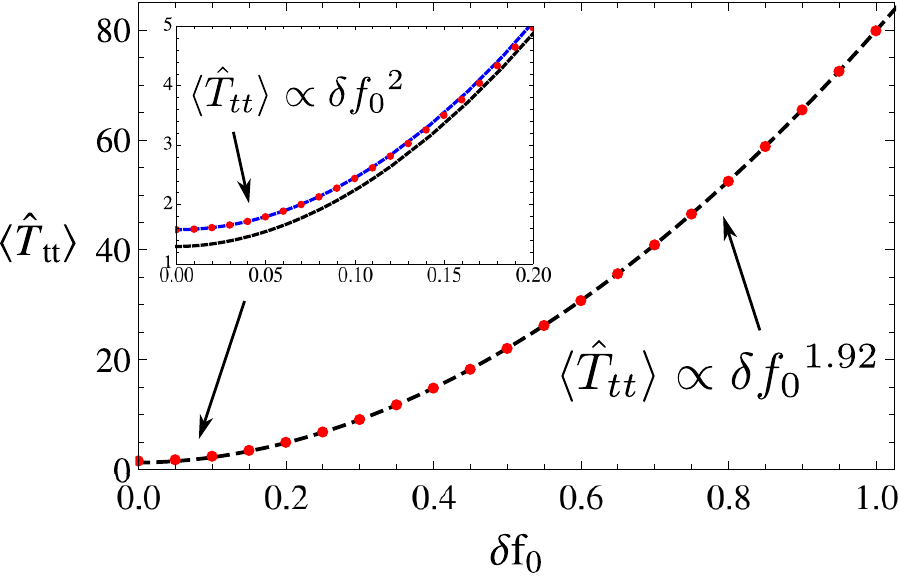}\label{fig:sfTtt1f}}
\quad
\subfigure[$\tau=0.838$ fixed]{\includegraphics[height=4.5cm]{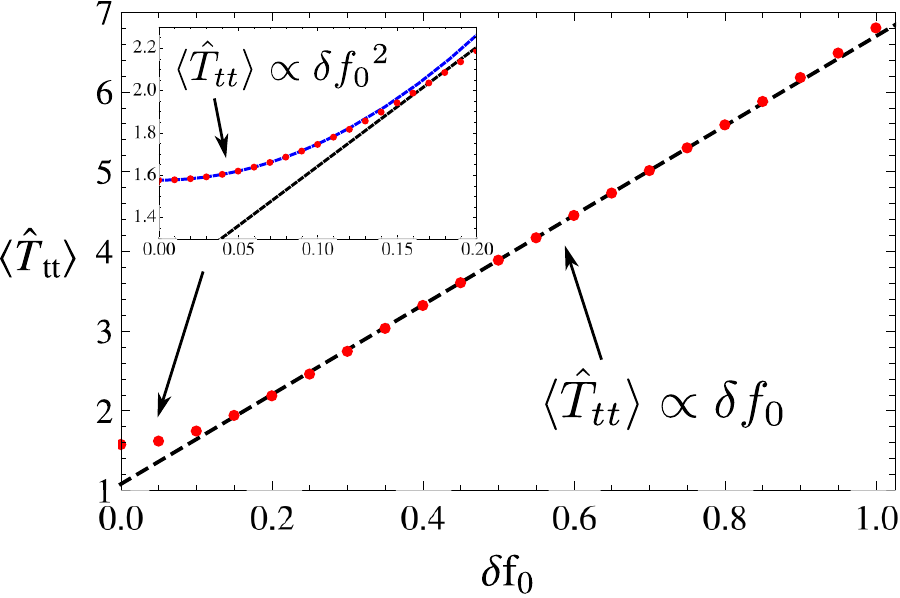}\label{fig:sfTtt1s}}
\caption{Analysis of two fixed $\tau$ cases. In the small $\delta f_0$ limit, both cases can be well fit with quadratic functions of $\delta f_0$ (insets). In \ref{fig:sfTtt1f}, the large $\delta f_0$ region favors a power law fit with $\langle \hat{T}_{tt} \rangle \sim \delta f_0{}^{1.92}$. Around $\delta f_0 \simeq 0.7$ in \ref{fig:sfTtt1s}, a linear behavior $\langle \hat{T}_{tt} \rangle \sim \delta f_0$ appears, while some deviations are seen around $\delta f_0 \simeq 1$ in this slow quench.}
\label{fig:sfTtt}
\end{figure}

The fixed $\tau$ scenario is studied closely in figure~\ref{fig:sfTtt}. For sufficiently small $\tilde{\delta}$ and $\tilde{\tau}$, the change in $\langle \hat{T}_{tt} \rangle$ across the quench is well approximated by a quadratic in the amplitude of the perturbation. In fact, this behavior manifests over a wide range of quench amplitudes providing that the quench is very fast, so that $\tilde{\tau}$ is very small (or $\tau \ll \tilde{f}_0$). As the quench width is increased, however,  deviations from this simple scaling become more pronounced. In figure \ref{fig:sfTtt1s}, in which $\tilde{\tau}$ is no longer small, the quadratic amplitude dependence of the final state energy density is only present for $\tilde{\delta}\lesssim 0.1$. Outside of this small amplitude region we find that in this case $\langle \hat{T}_{tt} \rangle \sim \delta f_0$.

\begin{figure}[t]
\centering
\subfigure[$\delta f_0=1$ fixed]{\includegraphics[height=4.5cm]{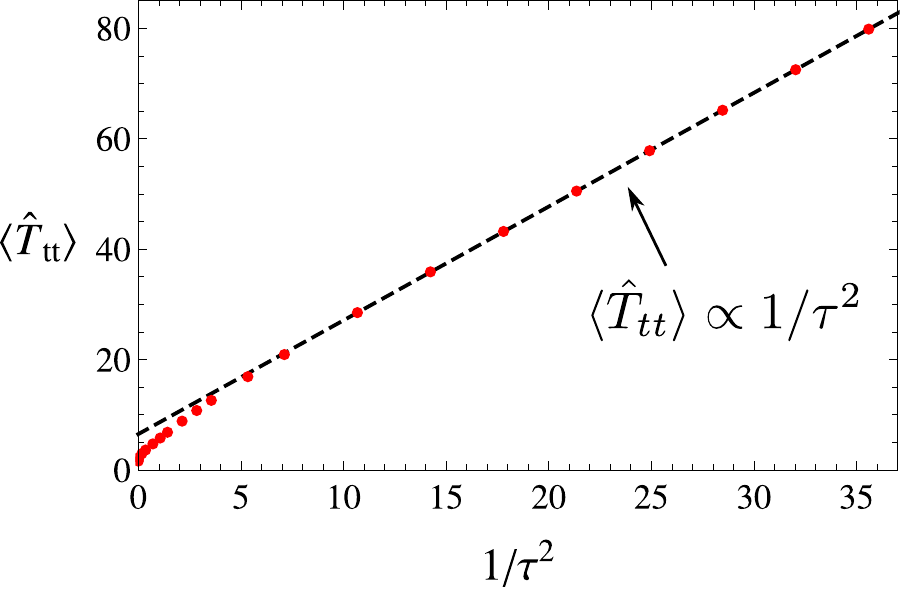}\label{fig:stTtt4h}}
\quad
\subfigure[$\delta f_0=0.05$ fixed]{\includegraphics[height=4.5cm]{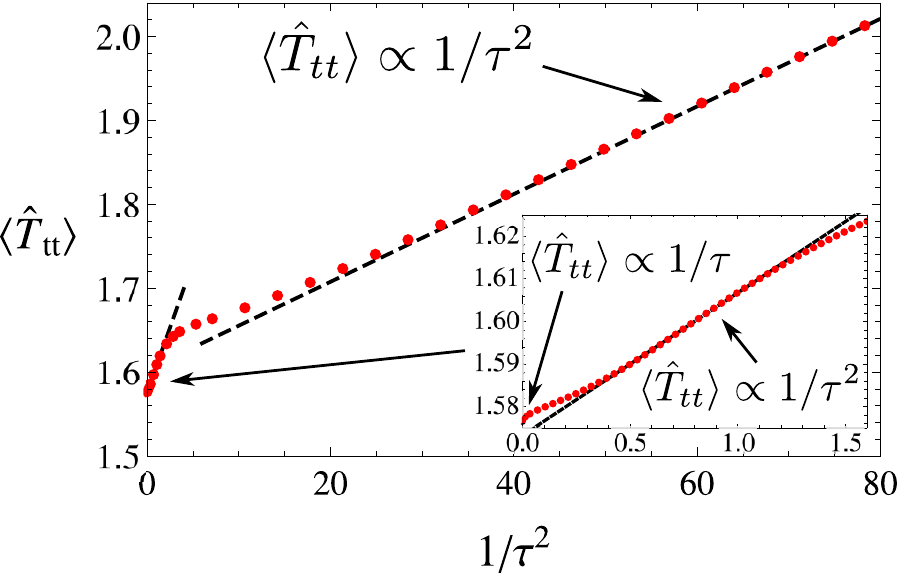}\label{fig:stTtt4s}}
\caption{Analysis of two fixed $\tau$ cases. In both cases, the small $\tau$ asymptotic behavior is given by $\langle \hat{T}_{tt} \rangle \sim 1/\tau^{2}$. In \ref{fig:stTtt4s}, there is a distinct intermediate region near $\tau \simeq 1$ where $\langle \hat{T}_{tt} \rangle \sim 1/\tau^{2}$ once again before transitioning to $\langle \hat{T}_{tt} \rangle \sim 1/\tau$ in the adiabatic limit (inset).}
\label{fig:stTtt}
\end{figure}

In  figure~\ref{fig:stTtt} we analyze the dependence  of the final state energy density on the quench speed with fixed $\tilde{\delta}$. Again, there is a pronounced scaling regime evident when the quench is fast. In this case we find that $\langle \hat{T}_{tt} \rangle$ asymptotically behaves as $1/\tau^2$ even when $\delta f_0$ is comparable in size to $\tilde{f}_0$. The scaling behavior degrades as $\tilde{\tau}$ becomes large, as is obvious from the lower left corners of the plots in figure~\ref{fig:stTtt}. This suggests that it is roughly the duration of the quench as compared to the characteristic scale of the theory $\tilde{f}_0$ that determines whether we are in this scaling regime, as opposed to the size of the width relative to the quench amplitude. In figure~\ref{fig:stTtt4s} we consider perturbations with fixed small amplitude. As shown in the inset, in this case there appears to be an intermediate region around $\tilde{\tau} \simeq 1$ where the $\langle \hat{T}_{tt} \rangle \sim 1/\tau^2$ scaling momentarily returns. However, as the quench width is further reduced we find that this scaling is not maintained, and the functional form appears to approach $\langle \hat{T}_{tt} \rangle \sim 1/\tau$ in the adiabatic limit. This adiabatic behavior at small $\tilde{\delta}$ is consistent with what was found in the absence of a confining potential in \cite{Buchel:2013lla}.

Combing the lessons of this subsection's results, we find that $\langle \hat{T}_{tt} \rangle \propto (\delta f_0/\tau)^2$ when the quench width is small compared to all other scales. This implies that a constant $\langle \hat{T}_{tt} \rangle$ line on the quench parameter plane is given by $\delta f_0 \sim \tau$. The scaling appears to be approximately respected even for small amplitude perturbations,  which is consistent with the linear behavior found in the bottom left corner of the dynamical phase diagram in figure~\ref{fig:phase}. Perhaps not surprisingly, this behavior is also consistent with a result first observed numerically under fairly different circumstances \cite{Buchel:2013lla}. It was subsequently demonstrated \cite{Buchel:2013gba,Das:2014jna,Das:2014hqa} that such a scaling is universal in the sense that it only depends on the near boundary features of the bulk theory, which is asymptotically AdS in many applications relevant for holography. Roughly, the idea is that very abrupt processes in the UV do not probe very deeply into the bulk, as the quench concludes before the disturbance has time to propagate far from the boundary. Since all the interesting features of our model appear deep in the IR, they play no role in the dynamics of this fast quench regime and it is therefore natural for our model to exhibit the same universal scaling behavior.
\subsection{Dynamical Phase Structure}

\begin{figure}[t]
\centering
\includegraphics[height=5cm]{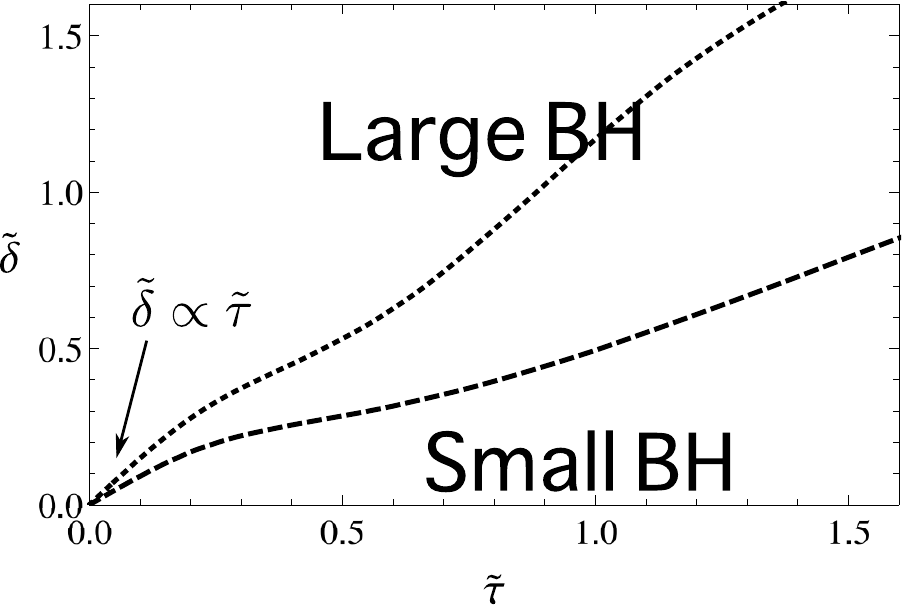}
\caption{Dynamical phase diagram for the small black brane initial state $\lambda_H/\lambda_c = 3.23$. The dashed line (lower) marks the location of a crossover between perturbations which result in a small black brane and those which produce a large black brane final state. The dotted line (upper) is the location above which the large black hole is thermodynamically favored over the thermal gas solution. Both lines behave like $\tilde{\delta} \sim \tilde{\tau}$ for $\tilde{\tau} \lesssim 0.1$.
}
\label{fig:phase}
\end{figure}

The coarse features of the quench dynamics can be efficiently summarized in a dynamical phase diagram, which we present in figure \ref{fig:phase}. To construct this diagram, we begin in the smallest black hole with which we can reliably evolve arbitrary perturbations, and scan the ($\tilde{\tau},\tilde{\delta})$ parameter plane. For each of the many perturbations, we follow the evolution until a static final state solution is obtained, and then use figure~\ref{fig:renVEV} to decide if that solution describes a large or small black hole.

The resulting phase diagram has several noteworthy features. First, the thick dashed line separating the  perturbations which result in a small black hole from those that end up in a large black hole likely marks the location of a crossover, as we have not been able to find any non-analytic behavior across this transition. To look for the sort of non-analytic behavior we have in mind, one can consider the set of susceptibilities that describe the system's dynamical response to various perturbations. For example, derivatives of the form
\begin{equation}
\chi_{\Delta \cal{E}}\big |_{\tilde{\delta}} = \left(\frac{\partial \Delta \cal{E}}{\partial{\tilde{\tau}}} \right)_{\tilde{\delta}} \qquad \textrm{or} \qquad \chi_{\Delta \cal{E}}\big |_{\tilde{\tau}}  = \left(\frac{\partial \Delta \cal{E}}{\partial{\tilde{\delta}}} \right)_{\tilde{\tau}}
\end{equation}
where $\Delta \cal{E}$ is the change in the system's energy density and the subscript is an instruction to take derivatives along the direction where that parameter is held fixed.

\begin{figure}[t]
\centering
\subfigure[$\tau$=0.168 fixed]{\includegraphics[height=4cm]{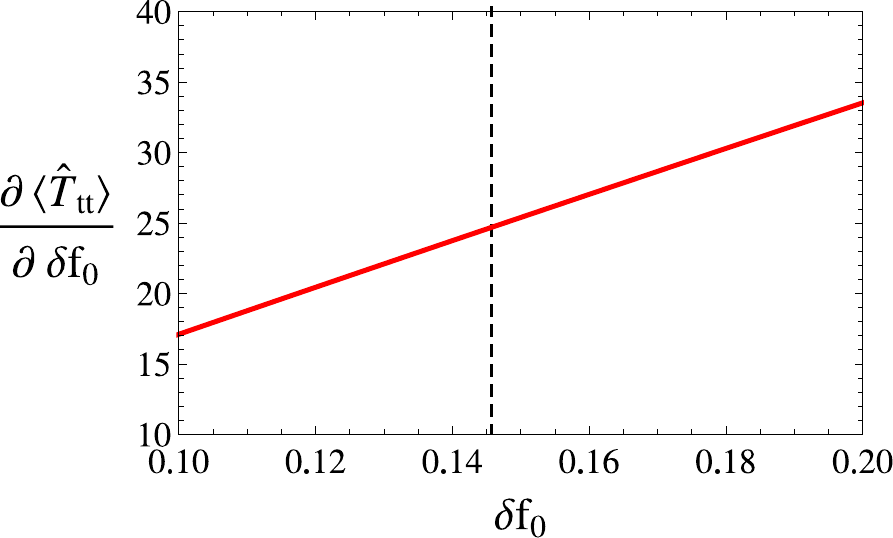}\label{dTttddf0}}
\quad
\subfigure[$\delta f_0$=0.5 fixed]{\includegraphics[height=4cm]{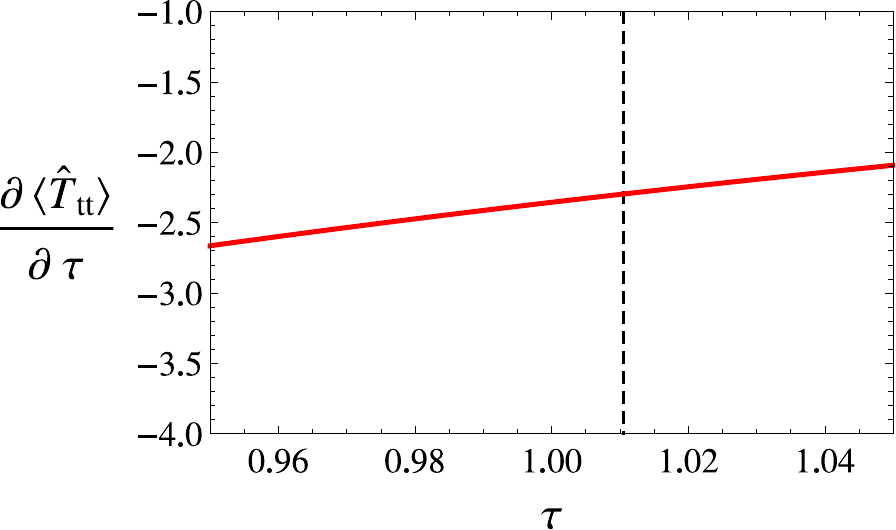}\label{fig:dTttdtau}}
\caption{Derivatives of  $\langle \hat{T}_{tt} \rangle$ in the final state with respect to the quench parameters are smooth across the line separating large and small black brane final states. The vertical dashed lines mark the location of the ``critical'' quench parameter which brings the initial bulk solution to the minimum temperature black brane. Derivatives taken at other fixed values of $\tilde{\delta}$ and $\tilde{\tau}$ show similar behavior. All quantities are measured in units of $\tilde{f}_0$.}
\label{fig:dTttdxxx}
\end{figure}

In our quenches, these susceptibilities appear to be continuous and smooth as one approaches the transition line. This can be inferred from the plots shown in figure \ref{fig:dTttdxxx} which shows an example typical of what we find when we differentiate the final state energy density with respect to the quench parameters at fixed $\tilde{\delta}$ or $\tilde{\tau}$. In general we expect that most dynamical observables in our model such as thermalization times and entropy production depend smoothly on $\Delta{\cal E}\equiv\langle \hat{T}^{tt} \rangle_\textrm{FINAL}-\langle \hat{T}^{tt} \rangle_\textrm{INIT}$, and thus we do not anticipate critical behavior to arise in other observables in the vicinity of this line. This is analogous to what happens  in weak field perturbations of global AdS, in which the line separating small and large black hole final states was also consistent with a crossover \cite{Bhattacharyya:2009uu}.

 Finally, the functional form of this dashed curve appears to be suitably described by a simple power law in the ``small-fast"  regime. Here the data can be well fit to a function of the form $\tilde{\delta} \sim \tilde{\tau}$, which is in line with general expectations as we discussed above. Independent of its precise functional form, we find that the crossover line monotonically increases in $\tilde{\delta}$ as $\tilde{\tau}$ increases for the parameter space covered by our study.

\section{Discussion}\label{sec:diss}

The line of research initiated in the present work provides another step forward in the holographic study of gauge theories that are qualitatively similar to $SU(3)$ Yang-Mills in 3+1 dimensions. Our gravitational model is characterized by a non-trivial (and carefully tuned) dilaton potential which introduces an additional level of complexity to the gravitational infall calculations dual to boundary theory thermalization. Specifically, the gravitational solutions of our model all involve a running dilaton in the bulk whose profile encodes the breaking of conformal invariance in the dual gauge theory.

The fact that these ``initial state" static solutions have nontrivial dilaton profiles requires a careful treatment of their near-boundary behavior in order to achieve stable evolution. We have applied a numerical technique based on finite difference integration to successfully evolve a broad class of perturbations, solving the fully backreacted Einstein-dilaton equations for the duration of the perturbation/equilibration process.

By studying the dependence of the final state on the parameters of the quench, we have collected several interesting lessons which we expect to prove insensitive to the precise details of the dilaton potential. Among these are the observation of a rapid transition from the perturbation (initial condition) dominated regime to the quasi-normal mode dominated regime of the quench dynamics. This behavior appears to be ubiquitous in holographic realizations of thermal quenches, and likely reflects the fact that for solutions with sufficiently large black hole horizons the thermalization time is governed by the Hawking temperature $\mathcal{T}_\textrm{therm}\sim 1/T$ (which is now the dominant length scale). The extent to which this continues to be true in the extremal limit of our model is largely an open question. From figure \ref{fig:QNM} it is clear that the decay width of the lowest lying quasi-normal mode has already departed from the simple conformal temperature scaling for final states with temperature near $T_c$. However, as we illustrate in section \ref{sec:tab}, our perturbations which remain on the small black hole branch also manifest this quick transition to the linear regime. An interesting question (intimately related to the second question concluding section \ref{sec:EX})  is then {\it where} might one expect to see deviations from this ``ubiquitous" behavior?

As we move further along the small black hole branch, the length scale introduced by the presence of the dimensionful source becomes increasingly important. This suggests that a sensible guess for where deviations from the rapid transition to the linear regime appear is for those perturbations whose final state has an energy density which is small compared to the source---in other words $\langle \hat{T}_{tt} \rangle /\tilde{f}_0{}^4\ll 1$. The smallest black hole which we  perturb in the present work has  $\langle \hat{T}_{tt} \rangle /\tilde{f}_0{}^4\sim O(1)$, and thus it is perhaps not surprising that we see no evidence for a new non-linear regime sensitive to the presence of the confining potential. Pushing the reach of our numerics closer to the extremal solution is an ongoing direction of our research which we hope to report on in the future.

Ultimately, we would like to go beyond the small black hole limit we have studied in this work and consider perturbations of the horizon-less zero temperature solution directly. The most compelling motivation for this is to determine whether or not the diverging dilaton potential characteristic of a broad class of holographic models of QCD is sufficiently repulsive to give rise to scattering type solutions which never result in black brane formation. If this is indeed the case, our dynamical phase diagram would gain a new line dividing those perturbations which result in black brane formation from those which do not. Finding the associated  scattering solutions would have interesting implications for the dual gauge theory, suggesting a class of perturbations that the strongly coupled matter can not thermalize.

Moreover, the boundary of these ``unthermalized" perturbations in the dynamical phase diagram would be interesting in its own right. By analogy with more familiar examples in asymptotically flat space, one might hope to find critical behavior akin to Choptuik phenomena \cite{Choptuik:1992jv}, in which a bulk scaling solution appears on the boundary of the perturbations that do and do not form a small black hole. Solutions of this sort are typically accompanied by various power law scalings characteristic of a second order phase transition. In our model, for example, this could manifest as the final state energy density assuming the form $\Delta{\cal E}|_{\tilde{\tau}}\sim (\tilde{\delta}-\tilde{\delta}_c)^\gamma$ where $\tilde{\delta}_c$ is the amplitude of the critical perturbation and $\gamma$ is a critical exponent quantifying the universality class of our model.
The exponent may be different from that of Choptuik, as the small black holes in our theories (unlike AdS) depart importantly from flat space black holes.

The appearance of a universal scaling regime in the fast quench limit is another noteworthy output from our calculations. As previously discussed, this scaling regime was anticipated on very general grounds, and its manifestation in our abrupt quench data is in some ways an encouraging check on our numerics. It is interesting to wonder how this scaling might be effected by the potential barrier inherent to the zero temperature solution. In so far as the universality of this scaling depends only on the UV features of the bulk solution, it is likely that quenches of the extremal solution whose width is much smaller than the scalar source will again result in a final state whose energy density satisfies (\ref{eq:uniFQ}). However, because the arguments for this scaling behavior are closely tied to the early time dynamics of the quench, it remains unclear whether the increase in energy density will manifest in black brane formation or a non-thermal scattering solution.

Investigating the properties of other probes which can be used to characterize our perturbations is another interesting future direction.  In \cite{Balasubramanian:2010ce,Buchel:2014gta} a variety of non-local probes were used to measure the approach to thermal equilibrium. These include various two point correlation functions, Wilson loops, and the system's entanglement entropy. The advantage of these non-local probes is that they are capable of moving beyond the binary thermal/non-thermal characterization of the perturbation, as they are sensitive to the scale dependence of the thermalization process. In the gravitational picture, they accomplish this by sampling the geometry away from the UV boundary.

To understand why this is true, it is instructive to consider the equal time two-point correlator for some gauge invariant boundary operator with large conformal dimension. In the semi-classical limit, this correlation function can be computed holographically by calculating the length of the bulk geodesic that connects two spatially separated points on the boundary. As the separation distance between these points increases, the bulk geodesic droops increasingly deeper into the IR. By measuring the deviation of the length of this bulk geodesic from the value one obtains in the final thermal static state as a function of boundary separation and time, one arrives at a picture of the approach to thermal equilibrium at different length scales. Constructing this picture in our model is currently in progress.

\begin{figure}[t]
\centering
\includegraphics[height=3cm]{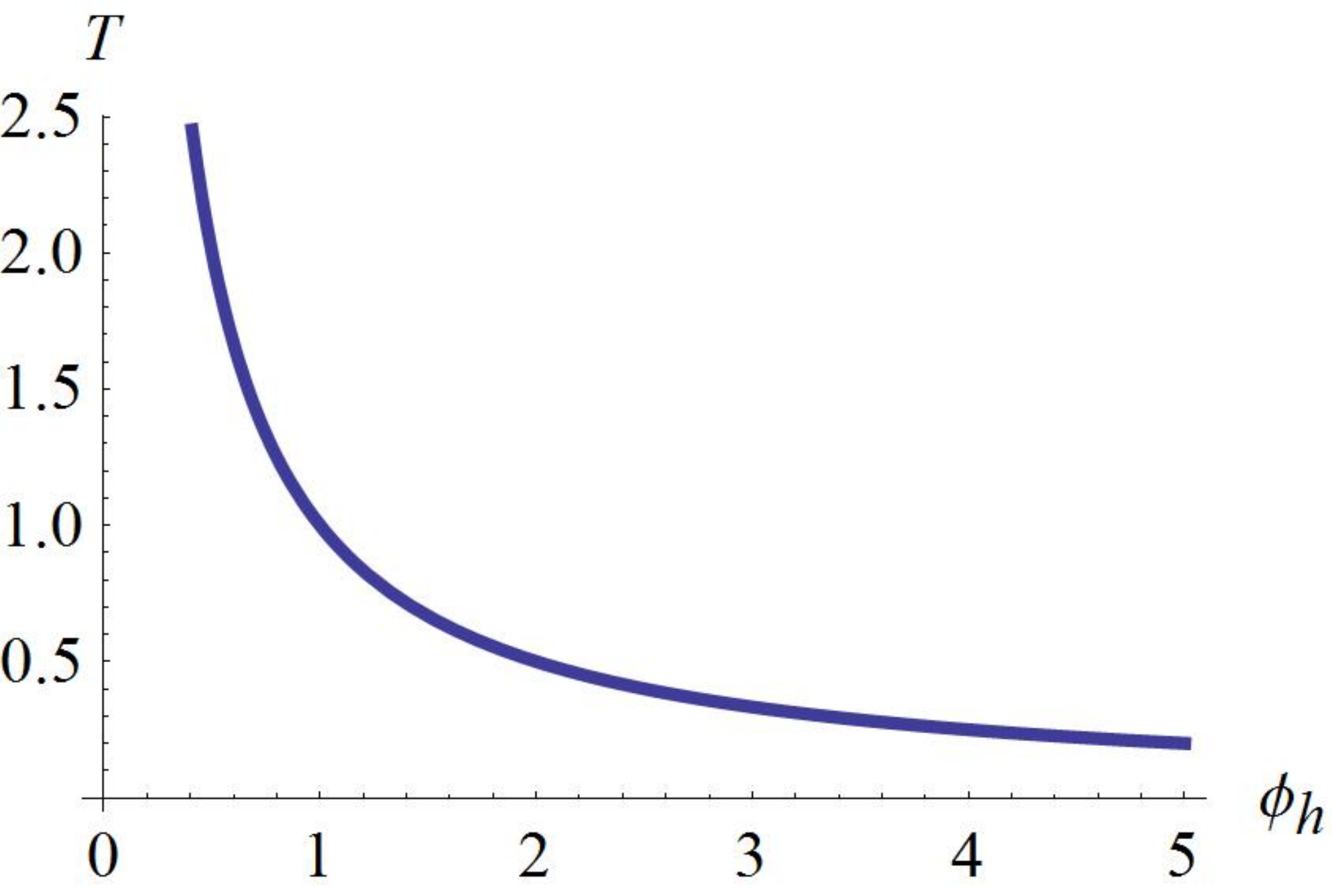}\label{plot1}
\includegraphics[height=3cm]{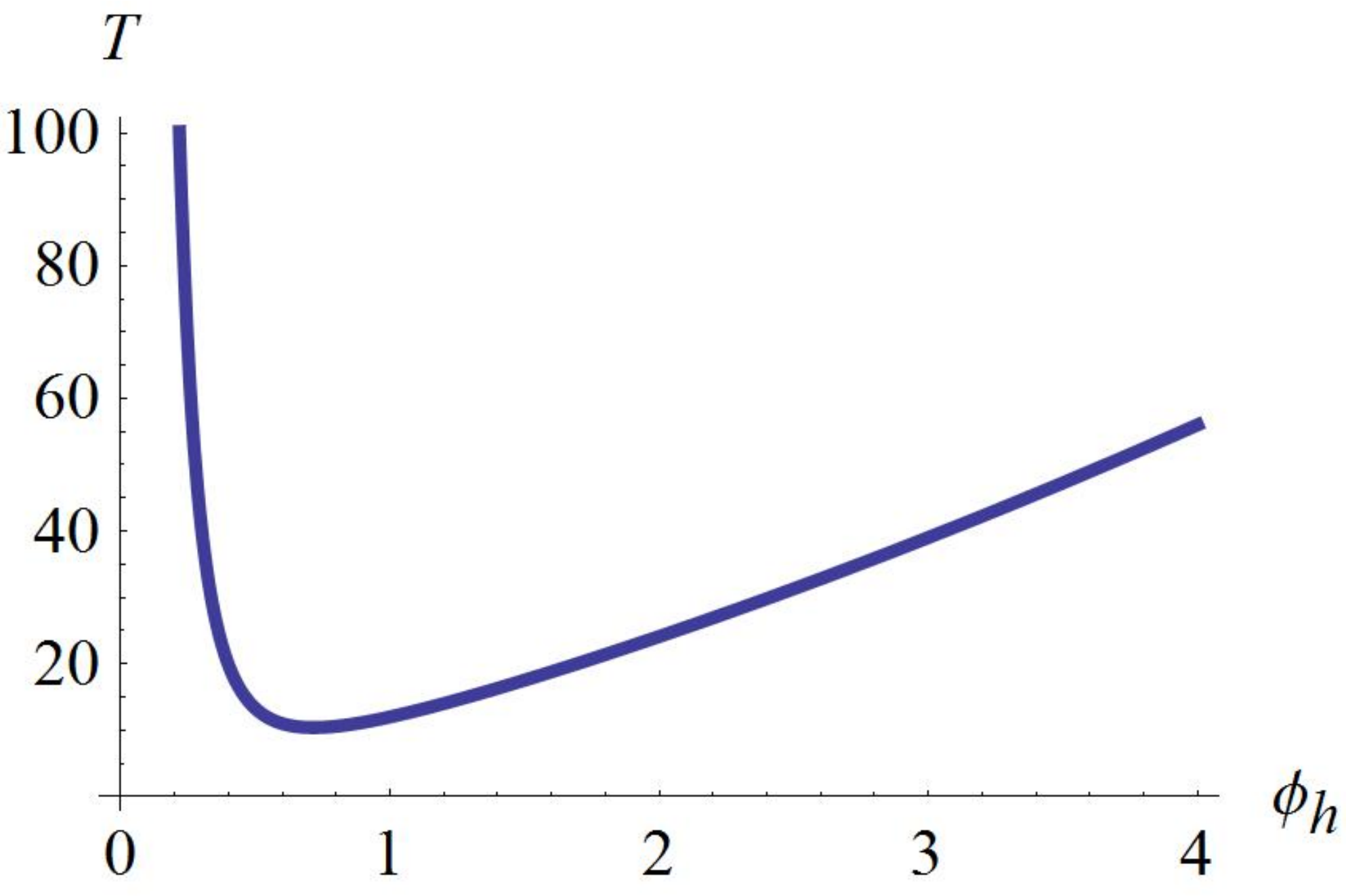}\label{plot2}
\includegraphics[height=3cm]{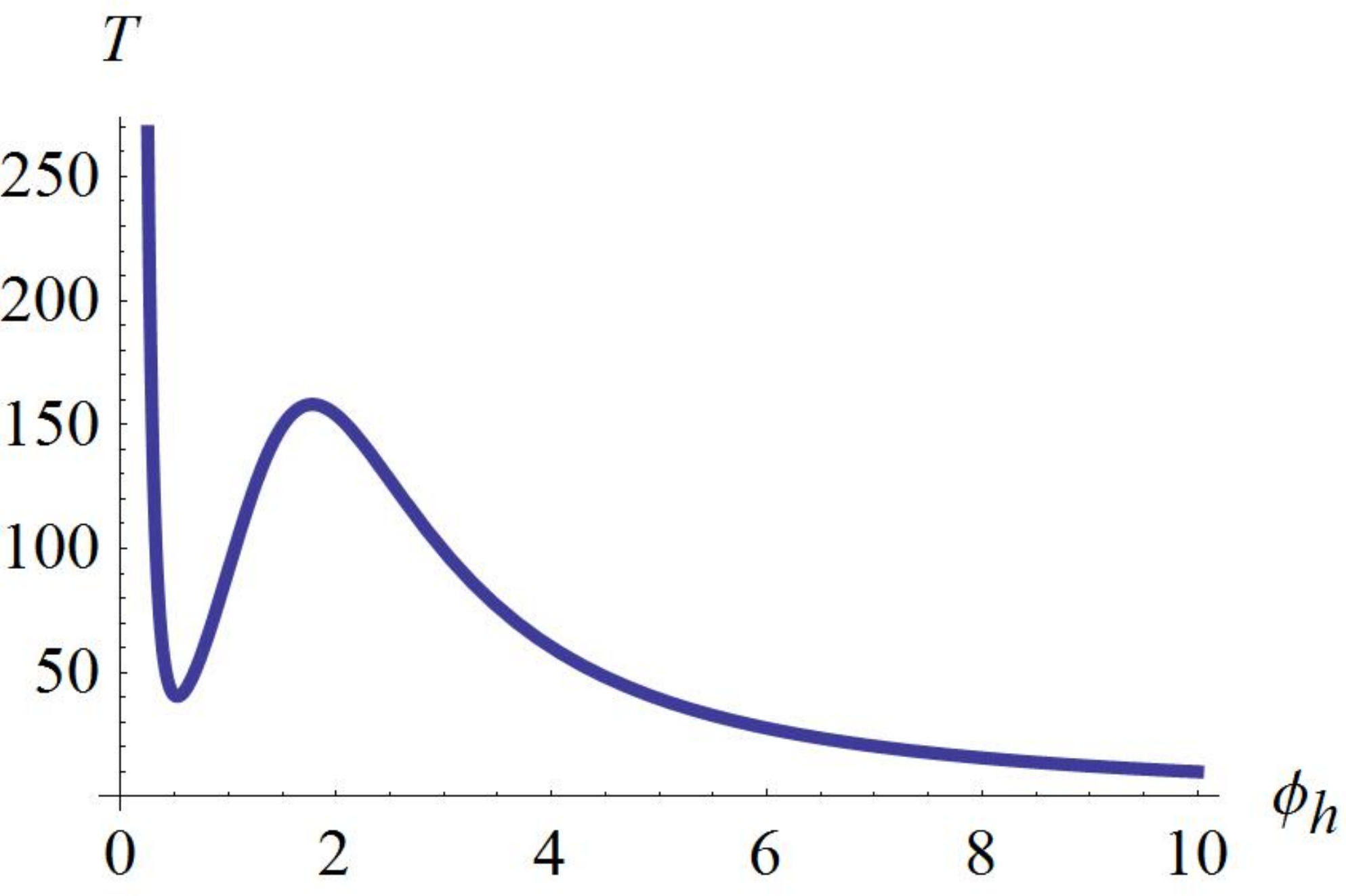}\label{plot3}
\caption{Left: Typical ($\phi_h,T$) diagram of the gapless/non-confining theory.  Center: Typical ($\phi_h,T$) diagram of a confining and gapped theory. Right: Typical ($\phi_h,T$) diagram of a non-confining and gapless theory with a first order phase transition. }
\label{plot}
\end{figure}

Following the discussion above we may put forth the following qualitative expectations, as a function of the bulk holographic theory:

\begin{itemize}

\item For theories that are simple (and smooth) RG flows between two CFTs (or hyperscaling violating geometries), the physics is characterized by a single mass scale. In such cases, the $T=0$ theory is gapless and at $T>0$, the theory is in the black hole phase. There is no phase transition or very fast crossover. The typical diagram that maps the black-holes of this theory is as shown in the left figure \ref{plot} where the temperature of the black hole solution is plotted against the value of the driving bulk scalar at the horizon, $\phi_h$.  In such cases, generically the thermalization time is expected to be proportional to the temperature $T$.

 \item For theories that are RG flows to a gapped IR theory as the case studied in this paper, the situation is different. Such theories are confining\footnote{In Einstein dilaton theory a confining theory is always gapped and vice-versa \cite{Gursoy:2008bu,Gursoy:2008za}.} and have a first-order phase transition at $T=T_c\sim \Lambda$ to the deconfined plasma phase ($\Lambda$ is the confinement scale of the $T=0$ theory). The typical diagram that maps the black-holes of this theory is as shown in the central  figure \ref{plot} where the temperature of the black hole solution is plotted against the value of the driving scalar at the horizon, $\phi_h$. There is a large stable black-hole branch to the left and a small unstable black-hole branch to the right.

      In the confined phase, $T<T_c$, the spectrum is discrete and real to leading order in $1/N_c$. The imaginary parts of various quasinormal modes are of order $1/N_c$, and therefore the thermalization time, is of order $\Lambda^{-1} N_c$  if the energy density injected into the system is much smaller than $T_c^4$. In particular, using the tree-level bulk equations of motion we do not expect the system to thermalize as the imaginary parts are zero to that order. We may therefore expect a Choptuik-like phase transition in that case.

     In the black hole phase and for energies near the phase transition the thermalization time is expected to be of order $T_c\sim \Lambda$. Finally for energy densities $T^4\gg \Lambda^4$ we expect the thermalization time to be set by the (final) temperature $T$.

\item  There are intermediate cases in which the $T=0$ theory is gapless in the IR, but with a fast crossover or phase transition to the UV regime. Such theories are always deconfined (according to the Wilson loop test) at $T=0$ \cite{Gursoy:2008bu,Gursoy:2008za}.

    The typical diagram that maps the black-holes of this theory is as shown in the right figure \ref{plot} where the temperature of the black hole solution is again plotted against the value of the driving scalar at the horizon, $\phi_h$.
    There is a large stable black-hole branch to the left and a small stable black-hole branch to the right, while there is also an unstable black-hole branch in the middle. In such a theory there is a (continuous) phase transition at $T=0^+$ to the small black hole phase, followed by a first order phase transition (or a fast crossover) to the large black hole phase\footnote{Such RG flows have two scales at $T=0$ and therefore a dimensionless parameter. An example is YM with quarks of mass $m$, and such a black hole diagram was found for example in \cite{Alho:2012mh} (see figure 22 of that paper).}.

For very small or very large final state temperatures, it is the temperature that sets the thermalization time, but in the intermediate region the characteristic scales of the theory enter into the thermalization time. Moreover, for small energy density we do not expect a Choptuik-like threshold in this case.

\end{itemize}

An important question is the implications of our results for heavy-ion collisions. The fast thermalization of sufficient high-density initial states
should be comparable to the processes studied here upon translating appropriately the quench parameters. Of course it should be kept in mind that the heavy ion collisions are anisotropic in space, but there are good reason to believe that this is not very important for initial thermalization but more important for the subsequent evolution of the plasma.
In this respect our numerical results on thermalization time should provide reliable estimates for the analogous heavy-ion thermalization process. Most importantly, the part of the physics that is not described here, namely the boundary between thermalization and non-thermalization will also provide important clues for thermalization in the less understood pp collisions, where recently CMS reported the first ever evidence for collective effects, \cite{Khachatryan:2010gv,Shuryak:2010wp}.

\section*{Note Added}
\addcontentsline{toc}{section}{Note Added}

Immediately before this work was submitted, several papers \cite{Buchel:2015saa,Fuini:2015hba,Janik:2015waa}
focusing on thermalization processes in nonconformal theories/backgrounds simultaneously appeared on the arXiv. These works have some overlap with our results on thermalization times in a non-conformal field theory.

Specifically, \cite{Buchel:2015saa} focuses on the important role played
by the low-lying quasi-normal modes in the approach to thermal equilibrium.
The fluctuation problem they consider is a linearized fluctuation problem in backgrounds like $\mathcal{N}=2^*$ which are gapless, with hyperscaling violating asymptotics in the IR. The authors emphasize the fact that even in states of non-conformal field theories with large deviations from conformality, the lowest lying quasi-normal mode approximately scales linearly with temperature. This is certainly true in many examples of holographic matter in the high temperature plasma phase, and in the corresponding states of our model as well (figure \ref{fig:QNM} ). However this approximate scaling is almost certainly violated in more phenomenologically viable models of the QGP, where the interaction measure is strongly peaked near $T_c$. Indeed this scaling is also violated noticeably in the phenomenological models investigated in \cite{Janik:2015waa}.

We believe that the subsequent claim made by the authors of \cite{Buchel:2015saa}, that ``the thermalization time is generically set by the temperature, irrespective of any other scales, in strongly coupled gauge theories" is too strong and valid only in theories with a single dynamical scale which are non-confining and gapless at $T=0$. Such systems have a phase transition to a black hole phase as soon as $T>0$.
 For reasons discussed in \cite{Craps:2013iaa}  and in the body of our text, we find it plausible that the scale introduced by the mass gap in our model introduce a new dynamical regime which is distinct from the QNM ringdown. Therefore, even the assumption made by those authors that thermalization times are approximately bounded from above by the lowest lying quasi-normal mode is called into question. Evaluating these claims necessarily requires going beyond the linearized gravitational equations, which is the approach we have adopted in the present work.

On the other hand, the authors of \cite{Janik:2015waa} focus again on linearized quasinormal  modes of a minimally coupled scalar.
The difference now is that the background is an Einstein-dilaton theory with a fast cross-over (and in one example a phase transition).
The main differences from our theory is that their theories are gapless at zero temperature while ours are gapped, and in our case the relevant scalar is the same that participates in the vacuum solution.
Some of their phenomenological formulae (like the connection of the thermalization time to the speed of sound) do not work well below $T_c$ in our case, and the reason may be the differences stated above.

Finally, the work of \cite{Fuini:2015hba} is closest in spirit to the present work, as these authors perturb their system and they follow the non-linear evolution. The main difference lies on the system to be perturbed. In their case the system is $\mathcal{N}=4$ plasma at finite charge density or in the presence of an external magnetic field.

\bigskip
\addcontentsline{toc}{section}{Acknowledgments}
\section*{Acknowledgments}
\bigskip

The authors thank Ben Craps, Matti J\"{a}rvinen, Keiju Murata, Vasilis Niarchos, Andrezj Rostworowski, Anastasios Taliotis, and Norihiro Tanahashi for helpful discussions. We additionally thank Ben Craps, Jonathan Lindgren, and Hongbao Zhang for collaboration on early stages of this work.
The numerical computations in this work were carried out in part at the Yukawa Institute Computer Facility.

This work was supported in part by European Union's Seventh Framework Programme
under grant agreements (FP7-REGPOT-2012-2013-1) no 316165, the EU program ``Thales" MIS 375734
 and was also cofinanced by the European Union (European Social Fund, ESF) and Greek national funds through
the Operational Program ``Education and Lifelong Learning" of the National Strategic
Reference Framework (NSRF) under ``Funding of proposals that have received
a positive evaluation in the 3rd and 4th Call of ERC Grant Schemes".

 \newpage
\appendix

 \renewcommand{\theequation}{\thesection.\arabic{equation}}
\addcontentsline{toc}{section}{Appendices}
\section*{APPENDIX}

\appendix
\section{Holographic Renormalization and Boundary Theory Correlators}\label{sec:AppHR}

The near boundary analysis of section \ref{sec:BA} is convenient for developing an algorithm for solving the Einstein equations, but the in-going null coordinates are ill suited to constructing the generating functional of the dual gauge theory. Following \cite{Bianchi:2001kw}, the analysis of the on shell action near the boundary is  most directly performed in Fefferman-Graham coordinates, in which the radial direction is orthogonal to the boundary directions.

In Fefferman-Graham coordinates, the metric takes the form
\begin{equation}
\dd s^2 = \frac{\dd \rho^2}{\rho^2}+\frac{1}{\rho^2}\,g_{ij}\,\dd x^i\dd x^j,
\end{equation}
and the metric and scalar permit the following expansions:
\begin{align}
\label{eq:FGxx}
g(t,\rho) =& \sum_{n=0}\left[g_{(n)}(t)+h_{(n)}(t)\log \rho +\bar{h}_{(n)}(t)\log^2 \rho + \cdots\right]\,\rho^{n},\nonumber\\
\df(t,\rho) = & \sum_{n=0}\left[\df_{(n)}(t)+\psi_{(n)}(t)\log \rho+\bar{\psi}_n(t)\log^2 \rho + \cdots\right]\,\rho^{n+1}.
\end{align}
Substituting these expansions into the Einstein equations and solving them order by order in $\rho$  allows one to determine many of the coefficients in the expansions algebraically. The primary exceptions are the leading coefficients of the normalizable modes, $g_{(4)}$ and $\df_{(2)}$ which can only be determined given the full radial profile in the bulk. Nevertheless, the near boundary analysis does constrain these undetermined coefficients, a fact realized in the boundary gauge theory by the existence of Ward identities.

The regularized on shell action (\ref{eq:action}) can be written
\begin{equation}
S_{\mathrm{R}} = -\frac{1}{3\kappa^2}\int_{\rho\ge \epsilon}\dd \rho\,\dd^4 x\,\sqrt{-g} \,V(\df)-\frac{1}{\kappa^2}\int_\epsilon \dd^4x\sqrt{-\gamma}\,\mathcal{K},
\end{equation}
where Einstein's equations have been used to eliminate the Ricci scalar. This action exhibits the following divergences in the limit $\epsilon\to 0$:
\begin{align}
S_{\mathrm{R}}=& \frac{1}{2\kappa^2}\int \dd^4 x\sqrt{-g_{(0)}}\Bigg[ \frac{6}{\epsilon^4}-\frac{4}{3\epsilon^2}\varphi_{(0)}^2-\log\epsilon\Big( \frac{1}{4}R_{ij}[g_{(0)}]R^{ij}[g_{(0)}]-\frac{1}{12}R^2[g_{(0)}] \nonumber\\
&- \frac{1}{9}\varphi_{(0)}^2 R[g_{(0)}]+\frac{2}{3}\varphi_{(0)}\Box_0\varphi_{(0)}\Big)+O(\epsilon^0)\Bigg].
\end{align}
To construct the appropriate counter terms, the regulated action must be expressed in terms of the fields living on the surface $\rho=\epsilon$. These fields are the pullback of the metric, $\gamma(t,\epsilon)$, and the scalar $\varphi(t,\epsilon)$. Performing this inversion yields
\begin{align}\label{eq:Sct}
S_{\mathrm{C}} = & -\frac{1}{2\kappa^2}\int_{\rho=\epsilon}\dd^4x\,\sqrt{-\gamma}\Bigg[6+\frac{1}{2}R[\gamma]+\frac{4}{3}\varphi^2+\log\epsilon\Bigg(F_4\,\varphi^4+\frac{2}{9}\varphi^2 R[\gamma]\nonumber\\
& -\frac{1}{2}\Big( \frac{1}{4}R_{ij}[\gamma]R^{ij}[\gamma]-\frac{1}{12}R^2[\gamma] \Big)-\frac{4}{3}\varphi\,\Box_\gamma\varphi\Bigg)+\mathcal{A}\big[ \gamma,\varphi \big]\bigg],
\end{align}
where $\mathcal{A}$ contains $O(\epsilon^0)$ finite counter terms, and the coefficient $F_4$ depends on the details of the higher order terms in the scalar potential. For the potential in (\ref{eq:Vd3}), it reads
\begin{equation}
F_4 = \left(\frac{16}{27}-\frac{V^{(4)}}{24} \right).
\end{equation}
 In the absence of an organizing principle such as supersymmetry, the finite counter terms are left unfixed and thus lead to scheme dependent ambiguities in the correlation functions.

Once the counterterms have been identified, one can form the subtracted action like
\begin{equation}
S_{\mathrm{sub}} = S_{\mathrm{R}}+S_{\mathrm{C}},
\end{equation}
and the renormalized correlation functions are then computed as follows:
\begin{align}\label{eq:corrO}
\langle \mathcal{O}\rangle = & \lim_{\epsilon\to 0}\left(\frac{1}{\epsilon^3}\frac{1}{\sqrt{-\gamma}}\frac{\delta S_{\mathrm{sub}}}{\delta\varphi}\right),\\
\langle T_{ij}\rangle = & \lim_{\epsilon\to 0}\left(\frac{1}{\epsilon^2}T_{ij}[\gamma]\right),\label{eq:corrT}
\end{align}
where $T_{ij}[\gamma]$ is the stress tensor of the theory at $\rho = \epsilon$. This boundary stress tensor is generically the sum of the contribution from the regularized action, and the contribution due to the presence of the counterterms. The regularized action gives
\begin{equation}\label{eq:Treg}
T_{ij}^R[\gamma] = -\frac{1}{\kappa^2}\left(K_{ij}-K\gamma_{ij} \right) = \frac{\epsilon}{2}\left(\partial_\epsilon\gamma_{ij}-\gamma_{ij}\gamma^{ab}\,\partial_\epsilon\gamma_{ab} \right)
\end{equation}
with $K_{ij}$ the extrinsic curvature of the regulating surface and $K = \gamma^{ij} K_{ij}$ its trace. To obtain the contribution from the counterterms, it is convenient to first catalogue the metric variations of a boundary action of the form
\begin{equation}\label{eq:Sctgen}
S^{B} = \int d^4 x \sqrt{-\gamma}\Big(A+  B\, R[\gamma]+C \,R^2[\gamma] + D\, R_{ab}[\gamma]R^{ab}[\gamma]+ E \, \varphi\,\Box_\gamma\varphi\Big),
\end{equation}
where $A,B,C,D$ and $E$ are arbitrary scalar functionals independent of the metric. Obviously this action contains (\ref{eq:Sct}) as a special case. With a bit of effort, one can show that the metric variations of $S^B$ yield
\begin{align}\label{eq:dSB}
T^B_{ij}[\gamma]\equiv-\frac{2}{\sqrt{-\gamma}}\frac{\delta S^B}{\delta \gamma^{ij}} = &\, \gamma_{ij}\left(A+BR +CR^2+DR_{ab}R^{ab}-\nabla_k(E\df)\nabla^k\df \right)\nonumber\\
&\,-  2B R_{ij}+2\nabla_i\nabla_j B - 2\gamma_{ij}\,\nabla^2 B \nonumber\\
 &\,-4CRR_{ij}+4\nabla_i\nabla_j (CR) - 4\gamma_{ij}\,\nabla^2(CR)\nonumber\\
 &\,- 4 DR_i\,^{a}R_{ja}+4\nabla_k\nabla_{( i}(DR_{j)}\,^{k})-2\nabla^2(DR_{ij})-2\gamma_{ij}\,\nabla_a\nabla_b (D R^{ab})\nonumber\\
 &\,+2\nabla_{(i}(E\df)\nabla_{j)}\df.
\end{align}
All contractions and curvatures in $T^B_{ij}$ implicitly refer to the metric $\gamma$. From the terms in (\ref{eq:dSB}) with coefficients determined by comparison to  (\ref{eq:Sct}), it is straightforward to obtain the contribution to the boundary stress tensor  resulting from the counterterms.

Performing this maneuver, summing the result with (\ref{eq:Treg})  and then inserting the on-shell near boundary expansions from (\ref{eq:FGxx}) into (\ref{eq:corrO}, \ref{eq:corrT}), one obtains the renormalized one point functions:
\begin{align}
\kappa^2\langle \mathcal{O}\rangle = &\,  \frac{2}{3}\Big(4\,\df_{(2)}+\ddot{\df}_{(0)} \Big) + \df_{(0)}^3\left(\frac{16}{27}-\frac{V^{(4)}}{24} \right)+\partial\mathcal{A}_\df,\\
\kappa^2\langle T_{tt}\rangle = & \, 2\, g_{(4)tt} \,-\frac{1}{6}\Big(8\,\df_{(2)}-\ddot{\df}_{(0)}  \Big)\df_{(0)}-\frac{1}{6}\dot{\df}_{(0)}{}^2+\frac{1}{24}\df_{(0)}^4\left(\frac{16}{27}+\frac{V^{(4)}}{24} \right)+\partial\mathcal{A}_t,\\
\kappa^2\langle T_{xx}\rangle = & \, \frac{2}{3}g_{(4)tt} \,+\frac{1}{18}\Big(8\,\df_{(2)}+5\,\ddot{\df}_{(0)}  \Big)\df_{(0)}+\frac{1}{6}\dot{\df}_{(0)}{}^2+\frac{11}{72}\df_{(0)}^4\left(\frac{208}{297}-\frac{V^{(4)}}{24} \right)+\partial\mathcal{A}_x.
\end{align}
The schematic notation $\partial\mathcal{A}$ refers to the contributions coming from the set of finite counter terms contained in $\mathcal{A}$. These contributions are scheme dependent, and will henceforth be neglected for simplicity. Note that in deriving these expressions the boundary metric has been assumed to be flat, $g_{(0)} = \eta_{ij}$. Written in terms of the near boundary expansion coefficients, it is straightforward to demonstrate that these one-point functions respect the anticipated Ward identities. For example (in units of $\kappa^2$),
\begin{align}
\langle T^{i}{}_{i}\rangle = &\,\frac{2}{3}\Big(4\,\df_{(2)}+\ddot{\df}_{(0)} \Big)\df_{(0)}+\frac{2}{3}\dot{\df}_{(0)}{}^2+\frac{1}{2}\df_{(0)}^4\left(\frac{16}{27}-\frac{V^{(4)}}{24} \right) \nonumber \\
 = & \,\df_{(0)}\,\langle \mathcal{O}\rangle+\frac{2}{3}\dot{\df}_{(0)}{}^2-\frac{1}{2}\df_{(0)}^4\left(\frac{16}{27}-\frac{V^{(4)}}{24} \right)
\end{align}
demonstrates the breaking of conformal symmetry in the presence of a dimensionful source in terms of the classical result (first term) and terms due to the matter anomaly in four dimensions. Similarly
\begin{equation}
\nabla^t\langle T_{tt}\rangle = \frac{2}{3}\Big(4\,\df_{(2)}+\ddot{\df}_{(0)} \Big) \dot{\df}_{(0)}+\df_{(0)}^3\left(\frac{16}{27}-\frac{V^{(4)}}{24} \right)\dot{\df}_{(0)}= \dot{\df}_{(0)}\langle \mathcal{O}\rangle
\end{equation}
describes the change in the system's energy in terms of the work done on it by a time dependent source. The derivation of these identities requires additional constraints which are easily obtained from the equations of motion. They relate, for example, time derivatives of the undetermined near boundary coefficients.

Because the numerical computations directly access the boundary expansion coefficients given in (\ref{eq:FUV}), it is convenient to relate these coefficients to those appearing in (\ref{eq:FGxx}). The coordinate change is given by
\begin{equation}
g_{\mu'\nu'} = \frac{\partial x^\mu}{\partial x^{\mu'}}\frac{\partial x^\nu}{\partial x^{\nu'}}g_{\mu\nu},
\end{equation}
which implies two particularly useful equations:
\begin{align}
\frac{1}{\rho^2} =&\, -v'{}^2 A-\frac{2}{z^2}v'z'\label{eq:cczz},\\
0 =&\,-\dot{v}v'A-\frac{1}{z^2}\big( \dot{v}z'+\dot{z}v'\big)\label{eq:cczt}.
\end{align}
These equations can be solved perturbatively in $\rho$ to obtain $z(t,\rho)$ and $v(t,\rho)$. One straightforward method to this end is to expand the ingoing null coordinates in powers of $\rho$, like
\begin{align}
z(t,\rho) =&\, \rho + \sum_{n=2}\rho^n\big( s_n(t)+\bar{s}_n(t)\log\rho\big),\\
v(t,\rho) =&\, t+\sum_{n=1}\rho^n \big(c_n(t)+\bar{c}_n(t)\log\rho\big),
\end{align}
and substitute these expansions into (\ref{eq:cczz}) and (\ref{eq:cczt}). The resulting system can be solved order by order in $\rho$ for the coefficients $s_n(t)$ and $c_n(t)$. This procedure yields
\begin{align}
z(t,\rho) =&\, \rho -\frac{1}{9}f_0^2\,\rho^3+\frac{4}{27}f_0\dot{f}_0\,\rho^4+\frac{1}{32}\left[4 a_4+\alpha_4+\frac{32}{81}f_0^4-\frac{16}{9}\left(\dot{f}_0^2+f_0\ddot{f}_0 \right) \right]\rho^5 \nonumber \\
&\,-\frac{1}{8}\alpha_4 \rho^5\log \rho+O(\rho^6),\\
v(t,\rho) =&\, t-\rho-\frac{1}{27}f_0^2\,\rho^3+\frac{1}{54}f_0\dot{f}_0\,\rho^4+O(\rho^5),
\end{align}
and upon inserting these expansions into (\ref{eq:AUV}) and (\ref{eq:FUV}) and regrouping terms, one directly obtains the independent coefficients $\df_{(0)}, \df_{(2)}$ and $g_{(4)tt}$ in terms of $f_0, f_2$, and $a_4$. The result is
\begin{align}
\df_{(0)} = &\,f_0,\nonumber\\
\df_{(2)} = &\, f_2-\frac{1}{2}\ddot{f}_0-\frac{1}{9}f_0^3,\nonumber\\
g_{(4)tt}=&\,-\frac{3}{4}a_4 -\frac{1}{36}\left( \dot{f}_0^2-f_0\ddot{f_0}\right)-\frac{V^{(4)}}{1152}f_0^4,\label{eq:c2c}
\end{align}
which can be used to write the one-point functions in terms of the coefficients obtained directly from the numerical routines, as in (\ref{eq:1pts1}-\ref{eq:1pts3}).

\section{Initial Data and Convergence}
\label{app:num}

The initial data we wish to perturb and evolve in time are the solutions to the static equations of motion given by setting $v$-derivatives to zero in (\ref{eq:eom1}-\ref{eq:eom5}). These solutions are constructed by integrating the static equations from the horizon to the boundary and matching to the boundary behavior of solutions in the gauge $\zeta(v)=0$. This computation can be performed at very high precisions without much effort. The results are then exported on the discretized grids desired for time evolution. Since the numerical codes used to produce and evolve the static solutions are distinct, it is an important check on our numerical package that the unperturbed initial state can be stably evolved. This procedure also helps us determine suitable grid sizes for satisfactorily suppressing numerical errors.

When the energy density of the initial state background is large, the dilaton is small throughout the bulk and it is fairly easy to achieve robust evolution. These static solutions can be evolved with only moderately small grid sizes, and numerical errors are very small. The scaling of the discretization error is second order in $\Delta z$, as desired.

\begin{figure}[t]
\centering
\subfigure[Static $\langle \hat{O} \rangle$.]{\includegraphics[height=4.5cm]{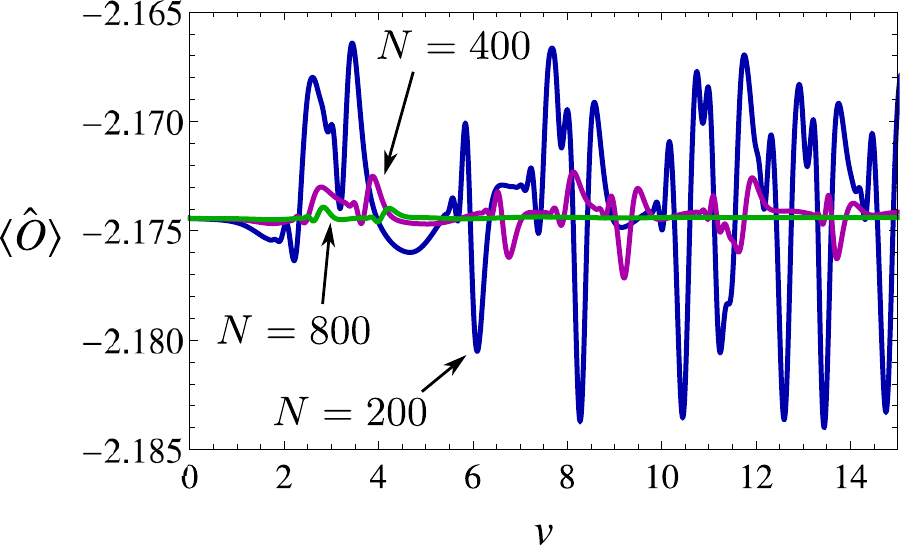}\label{fig:bg2a}}
\quad
\subfigure[Log plot of static $\delta \langle \hat{T}_{tt} \rangle$.]{\includegraphics[height=4.5cm]{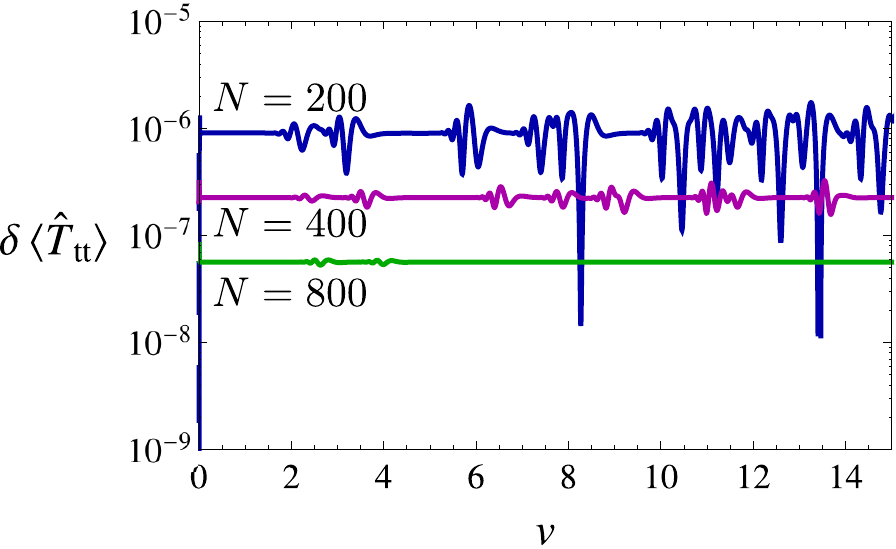}\label{fig:bg2b}}
\caption{Convergence of static time evolution by changing the grid size for an initial data set with $\lambda_H/\lambda_c = 3.23$. In~\ref{fig:bg2b}, the magnitude of the difference of $\langle \hat{T}_{tt}(v) \rangle$ from the input value is shown, $\delta \langle \hat{T}_{tt} \rangle \equiv |\langle \hat{T}_{tt}(v) \rangle - \langle \hat{T}_{tt}(0) \rangle|/|\langle \hat{T}_{tt}(0) \rangle|$. The blue, purple and green lines correspond to $N=200, \, 400, \, 800$, respectively. In the late time, the green lines do not oscillate and converge to the static value exponentially.}
\label{fig:bg2}
\end{figure}

As the background scalar becomes larger, however, we find that even simply maintaining the static initial state becomes a challenge. In particular, the difficulty drastically increases as the black hole size decreases as it heads towards the small black hole branch. In figure~\ref{fig:bg2}, time evolution of a static solution with $\lambda_H/\lambda_c = 3.23$ is shown. This corresponds to almost the smallest black brane in which we can reliably perform many distinct perturbations. We divide the domain between the boundary and the initial black hole horizon $z_H |_\textrm{stat}$ into $N$ intervals, and hence the grid size is $\Delta z= z_H|_{\textrm{stat}}/N$. In the left panel is the time evolution of $\langle \hat{\mathcal{O}} \rangle$. When $N=200$ and $400$, there are some numerical oscillations which persist on the order of $\Delta z$, which is larger than the desired numerical error. When the grid size is decreased by one half, although oscillations remain, the decrease of their magnitude is demonstrably second order in $\Delta z$. This implies that the presence of these oscillations is not indicating any real numerical instability, but is instead noise. We observe that this noise is largely insensitive to the size of $\Delta v$. For the initial data shown in figure~\ref{fig:bg2}, we find that the noise is heavily suppressed and the late time behaviors of both panels converge to constant values when $N=800$. These results suggest that for large background scalars, taking a very small grid size is necessary in order to rid the computation of unwanted noise. We have also made several tests with larger $\lambda_H/\lambda_c$, and in those cases the required smallness of the grid size for suppressing the noise continues to quickly increase. The static evolution of $\langle \hat{T}_{xx} \rangle$ behaves similarly to $\langle \hat{\mathcal{O}} \rangle$.

These oscillations are far less pronounced in $\langle \hat{T}_{tt} \rangle$, as indicated by \eqref{eq:dW}. In the right panel of figure~\ref{fig:bg2}, we plot the magnitude of the difference of $\langle \hat{T}_{tt}\rangle$ for the unperturbed solution evolved in time from the input value. Notice the difference of scale compared to the left panel. In this logarithmic plot, quadratic convergence against the grid size of the finite discretization is evident. We also checked the bulk constraint equation \eqref{eq:eom5}, and verified that this too converges quadratically.

One consequence of evolving our unperturbed initial state solutions in time is that any numerical irregularities in the solution are damped by the presence of the horizon as the numerical solution ``rings down" to a numerically stable configuration. This procedure thus cleans the numerical data for static initial states from the discrepancy between using different numerical codes for static and dynamical computations. In practice, before we perform any sort of quench, we typically allow the numerical data for the initial state to evolve unperturbed for a bit to clean the noise, and then compute the evolution of the perturbed state with suitably small (quench dependent) grid sizes.

\section{Geometrical Aside}\label{sec:AppAH}
\subsection{The Apparent Horizon}
The formulation of our gravitational problem relies crucially on the notion of an apparent horizon. This horizon is primarily important because unlike its more familiar sibling the event horizon, it is not teleological. This is to say that the location of the apparent horizon can be determined on each time slice, whereas the event horizon can only be deduced once the final state of the geometry is known.

An operational definition of an apparent horizon is a spacelike surface on which an outgoing null congruence normal to the surface has zero expansion. Following \cite{poisson} one may study this expansion $\theta$ via a null vector tangent to an outgoing null geodesic, $k$. In the basis provided by the coordinates of (\ref{eq:gans}), such a vector is given by
\begin{equation}\label{eq:nullvec}
k = k^\mu \partial_\mu = \partial_v - \frac{z^2}{2}A\,\partial_z.
\end{equation}
This vector is not affinely parametrized, which means that it satisfies the geodesic equation
\begin{equation}
k^\mu\nabla_\mu k_\nu = \kappa\, k_\nu
\end{equation}
for non-zero $\kappa$. It is easy to show that in the present case,
\begin{equation}\label{eq:kap}
\kappa = -\frac{1}{2}z^2 A' ,
\end{equation}
where the prime denotes differentiation with respect to $z$. Because $\kappa$ is non-zero, the expansion equation is modified as follows:
\begin{equation}
\theta  = e^\Gamma\left(\nabla_\mu k^\mu - \kappa \right).
\end{equation}
In this expression, the exponential pre-factor is necessary to convert to an affine parametrization. As it is manifestly positive and non-zero, it will play no role in the present discussion.

From (\ref{eq:gans}), (\ref{eq:nullvec}) and (\ref{eq:kap}) a short calculation reveals that in this background ansatz
\begin{equation}
\theta = \partial_v \Sigma -\frac{z^2}{2}A\,\partial_z\Sigma,
\end{equation}
and combining this result with the definition of the outwards directed derivative $\dd_+$ from (\ref{eq:ddp}), and the requirement that the expansion vanish at the location of the apparent horizon $z_H$ leads to this section's main result:
\begin{equation}
\left. \dd_+\Sigma \right|_{z_H} = 0.
\end{equation}
Apparent horizons have several interesting properties that make them particularly well suited to the studying of dynamical processes in gravity. Specifically, for static spacetimes with black holes in the bulk, the apparent, event, and Killing horizons all coincide. Moreover, it is possible to show on general grounds that in the non-static case an apparent horizon will always lie within the event horizon. This suggests that an apparent horizon provides an IR cutoff for the numeric problem that is both natural and practical, as well as available on each time slice.

\subsection{The Event Horizon}
When the dynamics is such that the gravitational system settles back into a static state, as is the case in the holographic thermalization processes we study here, it is also straightforward to compute the location and area of the event horizon over time.  For this one needs only the observation (mentioned above) that the event and apparent horizons coincide in static spacetimes, and that the event horizon necessarily travels along a geodesic in the bulk.

The line element traversed by a light ray in the ingoing null coordinates satisfies
\begin{equation}
0 = -A\,\dd v^2 -\frac{2}{z^2}\dd v\dd z ,
\end{equation}
which implies the following geodesic equation for the location of the event horizon, $z_\mathrm{EH}$:
\begin{equation}\label{eq:zhgeo}
\dot{z}_\mathrm{EH} = -\frac{1}{2}z_\mathrm{EH}^2\,A(z_\mathrm{EH},v),
\end{equation}
to be solved subject to the boundary condition $z_\mathrm{EH}(v\to\infty) = z_\mathrm{AH}(v\to\infty)$, where $z_\mathrm{AH}$ is the location of the apparent horizon. In practice, equation (\ref{eq:zhgeo}) can be integrated backwards in time along the geodesic using the numerically determined metric function $A$ at each step.

\bibliography{stringsBib}

\providecommand{\href}[2]{#2}\begingroup\raggedright\begin{thebibliography}{10}

\bibitem{Chesler:2008hg}
P.~M. Chesler and L.~G. Yaffe, {\it {Horizon formation and far-from-equilibrium
  isotropization in supersymmetric Yang-Mills plasma}},  {\em Phys.Rev.Lett.}
  {\bf 102} (2009) 211601, [\href{http://arxiv.org/abs/0812.2053}{{\tt
  arXiv:0812.2053}}].

\bibitem{Chesler:2009cy}
P.~M. Chesler and L.~G. Yaffe, {\it {Boost invariant flow, black hole
  formation, and far-from-equilibrium dynamics in N = 4 supersymmetric
  Yang-Mills theory}},  {\em Phys.Rev.} {\bf D82} (2010) 026006,
  [\href{http://arxiv.org/abs/0906.4426}{{\tt arXiv:0906.4426}}].

\bibitem{Heller:2011ju}
M.~P. Heller, R.~A. Janik, and P.~Witaszczyk, {\it {The characteristics of
  thermalization of boost-invariant plasma from holography}},  {\em
  Phys.Rev.Lett.} {\bf 108} (2012) 201602,
  [\href{http://arxiv.org/abs/1103.3452}{{\tt arXiv:1103.3452}}].

\bibitem{Bizon:2011gg}
P.~Bizon and A.~Rostworowski, {\it {On weakly turbulent instability of anti-de
  Sitter space}},  {\em Phys.Rev.Lett.} {\bf 107} (2011) 031102,
  [\href{http://arxiv.org/abs/1104.3702}{{\tt arXiv:1104.3702}}].

\bibitem{Maliborski:2013jca}
M.~Maliborski and A.~Rostworowski, {\it {Time-Periodic Solutions in an Einstein
  AdS-Massless-Scalar-Field System}},  {\em Phys.Rev.Lett.} {\bf 111} (2013),
  no.~5 051102, [\href{http://arxiv.org/abs/1303.3186}{{\tt arXiv:1303.3186}}].

\bibitem{Buchel:2013uba}
A.~Buchel, S.~L. Liebling, and L.~Lehner, {\it {Boson stars in AdS spacetime}},
   {\em Phys.Rev.} {\bf D87} (2013), no.~12 123006,
  [\href{http://arxiv.org/abs/1304.4166}{{\tt arXiv:1304.4166}}].

\bibitem{Craps:2014eba}
B.~Craps, E.~Lindgren, A.~Taliotis, J.~Vanhoof, and H.-b. Zhang, {\it
  {Holographic gravitational infall in the hard wall model}},  {\em Phys.Rev.}
  {\bf D90} (2014), no.~8 086004, [\href{http://arxiv.org/abs/1406.1454}{{\tt
  arXiv:1406.1454}}].

\bibitem{Bhattacharyya:2009uu}
S.~Bhattacharyya and S.~Minwalla, {\it {Weak Field Black Hole Formation in
  Asymptotically AdS Spacetimes}},  {\em JHEP} {\bf 0909} (2009) 034,
  [\href{http://arxiv.org/abs/0904.0464}{{\tt arXiv:0904.0464}}].

\bibitem{Balasubramanian:2013rva}
V.~Balasubramanian, A.~Bernamonti, J.~de~Boer, B.~Craps, L.~Franti, et~al.,
  {\it {Inhomogeneous Thermalization in Strongly Coupled Field Theories}},
  {\em Phys.Rev.Lett.} {\bf 111} (2013) 231602,
  [\href{http://arxiv.org/abs/1307.1487}{{\tt arXiv:1307.1487}}].

\bibitem{Caceres:2014pda}
E.~Caceres, A.~Kundu, J.~F. Pedraza, and D.-L. Yang, {\it {Weak Field Collapse
  in AdS: Introducing a Charge Density}},
  \href{http://arxiv.org/abs/1411.1744}{{\tt arXiv:1411.1744}}.

\bibitem{Craps:2013iaa}
B.~Craps, E.~Kiritsis, C.~Rosen, A.~Taliotis, J.~Vanhoof, et~al., {\it
  {Gravitational collapse and thermalization in the hard wall model}},  {\em
  JHEP} {\bf 1402} (2014) 120, [\href{http://arxiv.org/abs/1311.7560}{{\tt
  arXiv:1311.7560}}].

\bibitem{Bhaseen:2012gg}
M.~Bhaseen, J.~P. Gauntlett, B.~Simons, J.~Sonner, and T.~Wiseman, {\it
  {Holographic Superfluids and the Dynamics of Symmetry Breaking}},  {\em
  Phys.Rev.Lett.} {\bf 110} (2013), no.~1 015301,
  [\href{http://arxiv.org/abs/1207.4194}{{\tt arXiv:1207.4194}}].

\bibitem{Buchel:2012gw}
A.~Buchel, L.~Lehner, and R.~C. Myers, {\it {Thermal quenches in N=2*
  plasmas}},  {\em JHEP} {\bf 1208} (2012) 049,
  [\href{http://arxiv.org/abs/1206.6785}{{\tt arXiv:1206.6785}}].

\bibitem{Buchel:2013lla}
A.~Buchel, L.~Lehner, R.~C. Myers, and A.~van Niekerk, {\it {Quantum quenches
  of holographic plasmas}},  {\em JHEP} {\bf 1305} (2013) 067,
  [\href{http://arxiv.org/abs/1302.2924}{{\tt arXiv:1302.2924}}].

\bibitem{Buchel:2014gta}
A.~Buchel, R.~C. Myers, and A.~van Niekerk, {\it {Nonlocal probes of
  thermalization in holographic quenches with spectral methods}},  {\em JHEP}
  {\bf 1502} (2015) 017, [\href{http://arxiv.org/abs/1410.6201}{{\tt
  arXiv:1410.6201}}].

\bibitem{Buchel:2013gba}
A.~Buchel, R.~C. Myers, and A.~van Niekerk, {\it {Universality of Abrupt
  Holographic Quenches}},  {\em Phys.Rev.Lett.} {\bf 111} (2013) 201602,
  [\href{http://arxiv.org/abs/1307.4740}{{\tt arXiv:1307.4740}}].

\bibitem{Das:2014jna}
S.~R. Das, D.~A. Galante, and R.~C. Myers, {\it {Universal scaling in fast
  quantum quenches in conformal field theories}},  {\em Phys.Rev.Lett.} {\bf
  112} (2014) 171601, [\href{http://arxiv.org/abs/1401.0560}{{\tt
  arXiv:1401.0560}}].

\bibitem{Das:2014hqa}
S.~R. Das, D.~A. Galante, and R.~C. Myers, {\it {Universality in fast quantum
  quenches}},  {\em JHEP} {\bf 1502} (2015) 167,
  [\href{http://arxiv.org/abs/1411.7710}{{\tt arXiv:1411.7710}}].

\bibitem{Gursoy:2007cb}
U.~Gursoy and E.~Kiritsis, {\it {Exploring improved holographic theories for
  QCD: Part I}},  {\em JHEP} {\bf 0802} (2008) 032,
  [\href{http://arxiv.org/abs/0707.1324}{{\tt arXiv:0707.1324}}].

\bibitem{Gursoy:2007er}
U.~Gursoy, E.~Kiritsis, and F.~Nitti, {\it {Exploring improved holographic
  theories for QCD: Part II}},  {\em JHEP} {\bf 0802} (2008) 019,
  [\href{http://arxiv.org/abs/0707.1349}{{\tt arXiv:0707.1349}}].

\bibitem{Gubser:2000nd}
S.~S. Gubser, {\it {Curvature singularities: The Good, the bad, and the
  naked}},  {\em Adv.Theor.Math.Phys.} {\bf 4} (2000) 679--745,
  [\href{http://arxiv.org/abs/hep-th/0002160}{{\tt hep-th/0002160}}].

\bibitem{Kinar:1998vq}
Y.~Kinar, E.~Schreiber, and J.~Sonnenschein, {\it {Q anti-Q potential from
  strings in curved space-time: Classical results}},  {\em Nucl.Phys.} {\bf
  B566} (2000) 103--125, [\href{http://arxiv.org/abs/hep-th/9811192}{{\tt
  hep-th/9811192}}].

\bibitem{Kiritsis:2011qv}
E.~Kiritsis and A.~Taliotis, {\it {Mini-Black-Hole Production at RHIC and
  LHC}},  {\em PoS} {\bf EPS-HEP2011} (2011) 121,
  [\href{http://arxiv.org/abs/1110.5642}{{\tt arXiv:1110.5642}}].

\bibitem{Kiritsis:2011yn}
E.~Kiritsis and A.~Taliotis, {\it {Multiplicities from black-hole formation in
  heavy-ion collisions}},  {\em JHEP} {\bf 1204} (2012) 065,
  [\href{http://arxiv.org/abs/1111.1931}{{\tt arXiv:1111.1931}}].

\bibitem{Gubser:2000mm}
S.~S. Gubser and I.~Mitra, {\it {The Evolution of unstable black holes in
  anti-de Sitter space}},  {\em JHEP} {\bf 0108} (2001) 018,
  [\href{http://arxiv.org/abs/hep-th/0011127}{{\tt hep-th/0011127}}].

\bibitem{DeWolfe:2013cua}
O.~DeWolfe, S.~S. Gubser, C.~Rosen, and D.~Teaney, {\it {Heavy ions and string
  theory}},  {\em Prog.Part.Nucl.Phys.} {\bf 75} (2014) 86--132,
  [\href{http://arxiv.org/abs/1304.7794}{{\tt arXiv:1304.7794}}].

\bibitem{Girardello:1999hj}
L.~Girardello, M.~Petrini, M.~Porrati, and A.~Zaffaroni, {\it {Confinement and
  condensates without fine tuning in supergravity duals of gauge theories}},
  {\em JHEP} {\bf 9905} (1999) 026,
  [\href{http://arxiv.org/abs/hep-th/9903026}{{\tt hep-th/9903026}}].

\bibitem{Gubser:2008yx}
S.~S. Gubser, A.~Nellore, S.~S. Pufu, and F.~D. Rocha, {\it {Thermodynamics and
  bulk viscosity of approximate black hole duals to finite temperature quantum
  chromodynamics}},  {\em Phys.Rev.Lett.} {\bf 101} (2008) 131601,
  [\href{http://arxiv.org/abs/0804.1950}{{\tt arXiv:0804.1950}}].

\bibitem{Bourdier:2013axa}
J.~Bourdier and E.~Kiritsis, {\it {Holographic RG flows and nearly-marginal
  operators}},  {\em Class.Quant.Grav.} {\bf 31} (2014) 035011,
  [\href{http://arxiv.org/abs/1310.0858}{{\tt arXiv:1310.0858}}].

\bibitem{Gursoy:2010fj}
U.~Gursoy, E.~Kiritsis, L.~Mazzanti, G.~Michalogiorgakis, and F.~Nitti, {\it
  {Improved Holographic QCD}},  {\em Lect.Notes Phys.} {\bf 828} (2011)
  79--146, [\href{http://arxiv.org/abs/1006.5461}{{\tt arXiv:1006.5461}}].

\bibitem{Boyd:1996bx}
G.~Boyd, J.~Engels, F.~Karsch, E.~Laermann, C.~Legeland, et~al., {\it
  {Thermodynamics of SU(3) lattice gauge theory}},  {\em Nucl.Phys.} {\bf B469}
  (1996) 419--444, [\href{http://arxiv.org/abs/hep-lat/9602007}{{\tt
  hep-lat/9602007}}].

\bibitem{Bianchi:2001kw}
M.~Bianchi, D.~Z. Freedman, and K.~Skenderis, {\it {Holographic
  renormalization}},  {\em Nucl.Phys.} {\bf B631} (2002) 159--194,
  [\href{http://arxiv.org/abs/hep-th/0112119}{{\tt hep-th/0112119}}].

\bibitem{Papadimitriou:2011qb}
I.~Papadimitriou, {\it {Holographic Renormalization of general dilaton-axion
  gravity}},  {\em JHEP} {\bf 1108} (2011) 119,
  [\href{http://arxiv.org/abs/1106.4826}{{\tt arXiv:1106.4826}}].

\bibitem{Chesler:2013lia}
P.~M. Chesler and L.~G. Yaffe, {\it {Numerical solution of gravitational
  dynamics in asymptotically anti-de Sitter spacetimes}},  {\em JHEP} {\bf
  1407} (2014) 086, [\href{http://arxiv.org/abs/1309.1439}{{\tt
  arXiv:1309.1439}}].

\bibitem{Murata:2010dx}
K.~Murata, S.~Kinoshita, and N.~Tanahashi, {\it {Non-equilibrium Condensation
  Process in a Holographic Superconductor}},  {\em JHEP} {\bf 1007} (2010) 050,
  [\href{http://arxiv.org/abs/1005.0633}{{\tt arXiv:1005.0633}}].

\bibitem{Kovtun:2005ev}
P.~K. Kovtun and A.~O. Starinets, {\it {Quasinormal modes and holography}},
  {\em Phys.Rev.} {\bf D72} (2005) 086009,
  [\href{http://arxiv.org/abs/hep-th/0506184}{{\tt hep-th/0506184}}].

\bibitem{DeWolfe:2011ts}
O.~DeWolfe, S.~S. Gubser, and C.~Rosen, {\it {Dynamic critical phenomena at a
  holographic critical point}},  {\em Phys.Rev.} {\bf D84} (2011) 126014,
  [\href{http://arxiv.org/abs/1108.2029}{{\tt arXiv:1108.2029}}].

\bibitem{Nunez:2003eq}
A.~Nunez and A.~O. Starinets, {\it {AdS / CFT correspondence, quasinormal
  modes, and thermal correlators in N=4 SYM}},  {\em Phys.Rev.} {\bf D67}
  (2003) 124013, [\href{http://arxiv.org/abs/hep-th/0302026}{{\tt
  hep-th/0302026}}].

\bibitem{Choptuik:1992jv}
M.~W. Choptuik, {\it {Universality and scaling in gravitational collapse of a
  massless scalar field}},  {\em Phys.Rev.Lett.} {\bf 70} (1993) 9--12.

\bibitem{Balasubramanian:2010ce}
V.~Balasubramanian, A.~Bernamonti, J.~de~Boer, N.~Copland, B.~Craps, et~al.,
  {\it {Thermalization of Strongly Coupled Field Theories}},  {\em
  Phys.Rev.Lett.} {\bf 106} (2011) 191601,
  [\href{http://arxiv.org/abs/1012.4753}{{\tt arXiv:1012.4753}}].

\bibitem{Gursoy:2008bu}
U.~Gursoy, E.~Kiritsis, L.~Mazzanti, and F.~Nitti, {\it {Deconfinement and
  Gluon Plasma Dynamics in Improved Holographic QCD}},  {\em Phys.Rev.Lett.}
  {\bf 101} (2008) 181601, [\href{http://arxiv.org/abs/0804.0899}{{\tt
  arXiv:0804.0899}}].

\bibitem{Gursoy:2008za}
U.~Gursoy, E.~Kiritsis, L.~Mazzanti, and F.~Nitti, {\it {Holography and
  Thermodynamics of 5D Dilaton-gravity}},  {\em JHEP} {\bf 0905} (2009) 033,
  [\href{http://arxiv.org/abs/0812.0792}{{\tt arXiv:0812.0792}}].

\bibitem{Alho:2012mh}
T.~Alho, M.~J{\"a}rvinen, K.~Kajantie, E.~Kiritsis, and K.~Tuominen, {\it {On
  finite-temperature holographic QCD in the Veneziano limit}},  {\em JHEP} {\bf
  1301} (2013) 093, [\href{http://arxiv.org/abs/1210.4516}{{\tt
  arXiv:1210.4516}}].

\bibitem{Khachatryan:2010gv}
{\bf CMS} Collaboration, V.~Khachatryan et~al., {\it {Observation of Long-Range
  Near-Side Angular Correlations in Proton-Proton Collisions at the LHC}},
  {\em JHEP} {\bf 1009} (2010) 091, [\href{http://arxiv.org/abs/1009.4122}{{\tt
  arXiv:1009.4122}}].

\bibitem{Shuryak:2010wp}
E.~Shuryak, {\it {Comments on the CMS discovery of the 'Ridge' in High
  Multiplicity pp collisions at LHC}},
  \href{http://arxiv.org/abs/1009.4635}{{\tt arXiv:1009.4635}}.

\bibitem{Buchel:2015saa}
A.~Buchel, M.~P. Heller, and R.~C. Myers, {\it {Equilibration rates in a
  strongly coupled nonconformal quark-gluon plasma}},
  \href{http://arxiv.org/abs/1503.07114}{{\tt arXiv:1503.07114}}.

\bibitem{Fuini:2015hba}
J.~F. Fuini and L.~G. Yaffe, {\it {Far-from-equilibrium dynamics of a strongly
  coupled non-Abelian plasma with non-zero charge density or external magnetic
  field}},  \href{http://arxiv.org/abs/1503.07148}{{\tt arXiv:1503.07148}}.

\bibitem{Janik:2015waa}
R.~A. Janik, G.~Plewa, H.~Soltanpanahi, and M.~Spalinski, {\it {Linearized
  nonequilibrium dynamics in nonconformal plasma}},
  \href{http://arxiv.org/abs/1503.07149}{{\tt arXiv:1503.07149}}.

\bibitem{poisson}
E.~Poisson, {\em {A Relativist's Toolkit: The Mathematics of Black-Hole
  Mechanics}}.
\newblock Cambridge University Press, 2004.

\end{thebibliography}\endgroup

\end{document}